\documentclass[usenatbib]{mn2e}
\usepackage{amsmath}
\usepackage{amssymb}
\usepackage{longtable}
\usepackage{lscape}
\usepackage{rotating}
\usepackage{natbib}
\usepackage{epsfig}
\newcommand{\lsim}{\lower 2pt \hbox{$\, \buildrel {\scriptstyle<}\over {\scriptstyle \sim}\,$}}  
\newcommand{\gsim}{\lower 2pt\hbox{$\, \buildrel {\scriptstyle >}\over {\scriptstyle \sim}\,$}}

\newcommand{\oxythr}{{[OIII]}}
\newcommand{\oxythrw}{{[OIII]}$\lambda\lambda$4959,5007}

\newcommand{\oxytwow}{{[OII]}$\lambda\lambda$3726,3729}

\newcommand{\sulphurtwo}{{[SII]}}
\newcommand{\sulphurtwow}{{[SII]}$\lambda\lambda$6716,6731}
\newcommand{\nitrogen}{{[NII]}}

\newcommand{\kms}{km s$^{-1}$ }
\newcommand{\ha}{H$\alpha$}
\newcommand{\hb}{H$\beta$}

\newcommand{\lala}{$\lambda\lambda$}

\bibpunct{(}{)}{;}{a}{,}{,} \title[The nature of the nuclear outflows in Sy-ULIRGs]{The importance of warm, AGN-driven outflows in the nuclear regions of nearby ULIRGs}

\author[J.Rodr\'iguez Zaur\'in,  C.N.Tadhunter ]{ J.Rodr\'iguez
  Zaur\'in$^{1,2}$\thanks{E-mail:javierr@iac.es}, C.N.
  Tadhunter$^{3}$, M. Rose$^{3}$, J. Holt$^{4}$\\ 
  $^{1}$Insituto de Astrof\'isca de Canarias, E-38205, La Laguna, Tenerife, Spain\\
  $^{2}$Departamento de Astrof\'isica, Universidad de La Laguna, Tenerife, Spain\\
  $^{3}$Department of physics and Astronomy, University of Sheffield, Sheffield S3 7RH\\
  $^{4}$Leiden Observatory, Leiden University, PO Box 9513, 2300 RA Leiden, the Netherlands}
\begin{document}

\pagerange{\pageref{firstpage}--\pageref{lastpage}} \pubyear{2002}

\maketitle

\label{firstpage}

\begin{abstract}
We present an optical spectroscopic study of a 90\% complete sample of nearby
ULIRGs ($z < 0.175$) with optical Seyfert nuclei, with the aim of investigating
the nature of the near-nuclear ($r \la 3.5$~kpc) warm gas outflows. A high
proportion (94\%) of our sample show disturbed emission line kinematics in the
form of broad ($FWHM > 500$ km~s$^{-1}$) and/or strongly blueshifted ($\Delta V
< -150$ km s$^{-1}$) emission line components. This proportion is significantly
higher than found in a comparison sample of nearby ULIRGs that lack optical
Seyfert nuclei (19\%). We also find evidence that the emission line kinematics
of the Sy-ULIRGs are more highly disturbed than those of samples of non-ULIRG
Seyferts and PG quasars in the sense that, on average, their
[OII]$\lambda\lambda$5007,4959 emission lines are broader and more asymmetric.

The Sy-ULIRG sample encompasses a wide diversity of emission line profiles. In
most individual objects we are able to fit the profiles of all the emission
lines of different ionization with a kinematic model derived from the strong
[OIII]$\lambda\lambda$4959,5007 lines, using between 2 and 5 Gaussian
components. From these fits we derive diagnostic line ratios that are used to
investigate the ionization mechanisms for the different kinematic components. We
show that, in general, the line ratios are consistent with gas of super-solar
abundance photoionized by a combination of AGN and starburst activity, with an
increasing contribution from the AGN with increasing FWHM of the individual
kinematic components, and the AGN contribution dominating for the broadest
components. However, shock ionization cannot be ruled out in some cases. Our
derived upper limits on the mass outflows rates and kinetic powers of the
emission line outflows show that they can be as energetically significant as the
neutral and molecular outflows in ULIRGs -- consistent with the requirements of
the hydrodynamic simulations that include AGN feedback. However, the
uncertainties are large, and more accurate estimates of the radii, densities and
reddening of the outflows are required to put these results on a firmer footing.
\end{abstract}


\section{Introduction}

AGN-induced outflows are increasingly recognized as a key element in the overall
galaxy formation process, regulating the correlations between black hole mass
and host galaxy properties \citep{Silk98,Fabian99}, and influencing the
luminosity functions of galaxies, particularly at the high luminosity end
\citep{Benson03}. Indeed, AGN feedback is now routinely incorporated into
numerical simulations of the hierarchical growth of galaxies through major
galaxy mergers \citep{Springel05, Johansson09}. However, there remain
considerable uncertainties about the exact nature of the symbiosis between black
holes and their host galaxies. Consequently, the importance of AGN-induced
outflows relative to those driven by the supernovae associated with the
merger-induced starbursts has yet to be established observationally.

In order to test the models it is important to identify samples of actively
evolving galaxies in which both the black holes and the galaxy bulges are
growing rapidly. Although much attention has been paid recently to sub-mm
galaxies at high redshifts that frequently show signs of AGN activity in X-ray
observations \citep{Alexander05,Alexander10,Harrison12}, there is a limit to what can be
learnt in detail about the co-evolution of black holes and galaxy bulges in such
objects because of their faintness. Moreover, for high redshift galaxies many
important diagnostic emission lines are shifted out of the optical/near-IR, and
their outflows are difficult to resolve spatially.

Fortunately, the hierarchical evolution of galaxies is also continuing in the
local Universe, albeit at a reduced rate. The Ultra Luminous Infrared Galaxies
(ULIRGs: $L_{IR} > 10^{12}$~L$_{\odot}$) represent a class of objects at
relatively low redshifts in which the prodigious far-IR radiation represents the
dust re-processed light of major starbursts and/or AGN buried in the nuclei of
the galaxies \citep{Sanders96}. ULIRGs almost invariably show morphological
evidence (e.g. tidal tails, double nuclei) consistent with triggering of the
activity in major galaxy mergers
\citep[e.g.][]{Sanders88a,Surace98,Surace00a,Surace00b,Kim02,Veilleux02}. Based
on their mid- to far-IR colours they can be sub-divided into ``warm'' ($f_{25\mu
  m}/f_{60\mu m} > 0.20$) and ``cool'' ($f_{25\mu m}/f_{60\mu m} < 0.20$)
sub-types, with the warm sub-type frequently associated with visible Seyfert or
quasar nuclei \citep[e.g.][]{Surace99}. In terms of evolutionary scenarios, it
has been proposed that cool, starburst-dominated ULIRGs evolve into warm,
AGN-dominated ULIRGs as the circum-nuclear dust is dispersed by starburst and
AGN-induced outflows \citep{Sanders88,Surace98}. Such scenarios are broadly
consistent with current simulations of major, gas-rich galaxy mergers
\citep{Hopkins05}. Therefore, the ULIRGs represent local analogues of the
rapidly evolving galaxies now being detected in large numbers in the distant
Universe, but they have the considerable advantage over the high redshift
objects that they are close enough to study in detail. In this sense the nearby
ULIRGs are key objects for understanding the detailed physical mechanisms --
particularly the outflows -- involved in the co-evolution of super-massive black
holes and their host galaxies.

As the most rapidly evolving galaxies in the local Universe, ULIRGs represent
just the situation modelled in many of the most recent hydrodynamic simulations
of gas-rich mergers \citep{diMatteo05,Johansson09}. In the final stages of such
mergers, as the nuclei coalesce, the simulations predict that the super-massive
black holes accrete at sufficiently high rates to produce luminous, quasar-like
AGN; the outflows associated with the AGN are potentially capable of ejecting
the remaining gas and halting star formation in the bulges of the merger
remnants. However, fine tuning of the models is required in order for them to
reproduce the observed M$_{bh}$ vs. $\sigma$ relationship. In particular it is required
that the outflows carry a relatively large proportion of the available accretion
power of the AGN \citep[$\sim$5 -- 10\%:][but see Hopkins \& Elvis 2010]{Fabian99,diMatteo05}.  
{\it Clearly it is important to use direct observations to determine whether this fine
tuning is justified: are the AGN-driven outflows in ULIRGs truly as important
as many of the simulations require?}

In line with the expectations of the hydrodynamic simulations, spectroscopic
evidence for outflows has now been found in several ULIRGs using optical
observations of neutral NaID absorption lines \citep{Rupke05a,Rupke05b,Rupke11},
and far-IR/sub-mm observations of the molecular OH and CO lines
\citep{Feruglio10,Fischer10,Sturm11}. However, the evidence for a clear link
between these outflows and nuclear AGN activity is mixed.  For example, although
the mass outflow rates and kinetic powers associated with the neutral NaID
absorption line outflows are significant, comparisons between the properties of
the neutral outflows in objects with and without optical AGN show surprisingly
little difference \citep{Rupke05c}. Moreover, while the results of
\cite{Sturm11}, based on Herschel data, do suggest that the molecular outflows
may be faster and more powerful in ULIRGs hosting powerful AGN \citep{Sturm11},
the sample is small (only 6 objects) and incomplete.

It is important to recognize that the gas associated with any AGN-driven outflow
may be highly ionized and therefore not adequately sampled by NaID absorption
and molecular line studies. In addition, the quantitative estimates of the
properties of the neutral and molecular outflows are compromised by the fact
that, in most cases, the radial extents of these outflows are not well
constrained by the observations, which sample gas along the line of sight.

Results based on the optical studies of the large-scale ($>$5~kpc) warm ionized
gas in ULIRGs \citep{Heckman90,Westmoquette12,Soto12a,Soto12b} reveal a similar
picture to the molecular and neutral gas studies: energetically significant
outflows, but no clear differences between the outflows in ULIRGs with and
without AGN; again it is not clear whether the large-scale emission line gas is
truly sampling the AGN-driven outflows.  However, optical and mid-IR
spectroscopic observations have recently revealed highly ionized outflows in the
{\it near-nuclear regions} of several ULIRGs with Seyfert-like nuclei in the
form of broad, blueshifted emission line components
\citep{Wilman99,Lipari03,Holt03,Holt06,Spoon09a,Spoon09b}. These observations
also suggest that the near-nuclear outflows are highly stratified
\citep{Holt03,Spoon09b}, and that relativistic jets may play an
important role in accelerating the gas \citep{Holt09,Spoon09b}. Given that they
isolate the AGN-driven outflow components, the near-nuclear high ionization
emission lines thus provide a powerful tool for addressing the key questions
surrounding AGN-driven outflows in ULIRGs. These questions include the
following.

\begin{itemize}
\item {What is the dominant outflow driving mechanism in ULIRGS: AGN or
    starbursts?} As discussed above, previous studies based on observations of
  the molecular, neutral and large-scale ionized gas outflows have not provided a
  clear-cut answer to this question. Clearly it is important to consider this
  question in the context of the near-nuclear warm outflows.

\item {Are AGN-driven outflows more significant in Seyfert-like ULIRGs than in
  other types of AGN?} Given that the ULIRGs are among the most actively
  evolving galaxies in the local Universe, with (predicted) high nuclear nuclear
  accretion rates, we might expect the AGN-induced outflows in such objects to
  be unusually strong. However, a large proportion of AGN of all types show
  evidence for outflows at some level, in the form of blue wings to the [OIII]
  emission lines
  \citep[e.g.][]{Heckman81,Whittle85a,Veilleux91b,Nelson95}. Therefore, it is
  important to establish whether the outflows in Seyfert-like ULIRGs are
  significant more extreme than those detected in the general population of
  non-ULIRG AGN.

\item {What is the nature of the near-nuclear outflows in ULIRGs?}  Comparisons
  of spectroscopic observations of emission lines of different ionisation have
  provided evidence that the near-nuclear emission lines outflows in some ULIRGs
  are stratified. However, it is currently uncertain whether the outflows
  comprise a small number of distinct outflow velocity components, each
  associated with a particular spatial location and set of physical condition
  \citep{Holt03}, or rather encompass a continuous gradient of velocities and
  physical conditions across the full extent of the emission line region
  \citep[e.g.][]{DeRobertis86}. At the same time it is important to confirm
  using a wide range of emission lines that the outflows are indeed
  decelerating, as suggested by observations of the mid-IR fine structure neon
  lines \citep{Spoon09b}.

\item {Do relativistic jets play an important role in accelerating the ULIRG
  outflows?} Although it is certainly the case that some of the most extremely
  blueshifted emission line components are are found in ULIRGs that contain
  powerful relativistic jets \citep{Holt03,Holt06}, the issue of the jet
  contribution to the acceleration and ionisation of the outflows warrants
  further investigation using a complete sample of ULIRGs that is unbiased
  towards jet properties.

\item {How energetically significant are the AGN-induced outflows in ULIRGs?} In
  terms of testing the hydrodynamical merger simulations that incorporate AGN
  feedback as a key element, it is crucial to accurately quantify the mass
  outflow rates and kinetic powers of the near-nuclear outflows. In
  this way it will also be possible to compare the properties of the
  near-nuclear AGN outflows with those of the neutral and large-scale ionised
  gas outflows, to determine whether the AGN-induced outflows are truly energetically
  significant.

\end{itemize}

In this paper we directly address the these questions using detailed optical
spectroscopic observations of a complete sample of 17 nearby ($z < 0.175$)
ULIRGs with Seyfert-like nuclear spectra. Throughout this paper we assume a
cosmology with H$_{0}$ = 71 km s$^{-1}$ Mpc$^{-1}$, $\Omega_{0} = 0.27$,
$\Omega_{\Lambda} = 0.73$.

\section{Sample selection, observations and data reduction}

In order to achieve the aims of this project we require a sample of ULIRGs with
Seyfert-like nuclear spectra that are close enough to study in detail and have
high quality nuclear spectra; the sample also needs to be complete as possible.
As a starting point we took the ``Extended Sample'' (ES) of \cite{Rodriguez-Zaurin09},
which comprises 36 nearby ($z < 0.175$) ULIRGs of all spectral types from the 1Jy
sample of \cite{kim98} that have deep WHT spectra. We then selected all the
objects with Seyfert-like nuclear spectra from this sample. To make this
selection we use the criteria of \citet{yuan10}, which in turn are based on the
emission line diagnostic diagrams of \cite{Kewley06} for SDSS
galaxies. Specifically, we selected all the objects classified as Sy2 in at
least 2 out of 3 of the \cite{Kewley06} diagnostic diagrams, as well as the
three objects with broad-line Seyfert 1 spectra. This results in a sample of 19
nearby ULIRGs with Seyfert-like spectra, redshifts $z < 0.175$, declinations $\delta
> -25$, and right ascensions $12 < RA < 02$~hr (see Table 1 for details). In
comparison, a total of 22 ULIRGs from the full 1Jy sample of \cite{kim98} meet
the same selection criteria in optical classification, redshift, declination and
right ascension, and of these only 3 objects -- F12265+0219 (3C273), F13443+08,
F14121-0126 -- are not included in our sample of ULIRGs with Seyfert-like
nuclear spectra and deep WHT spectra. Therefore our sample is 86\%
complete\footnote{This rises to 90\% if we exclude from consideration the
powerful quasar F12265+0219 (3C273), which is dominated by non-thermal jet
emission, rather than thermal dust emission, at infrared wavelengths.}, and is
certainly representative of the subset of ULIRGs with optical Seyfert nuclei in
the local Universe.

As described in \cite{Rodriguez-Zaurin09}, deep spectra of all of the 19 objects
in our sample were taken with the ISIS dual-beam spectrograph on the 4.2-m
William Herschel Telescope (WHT), on La Palma, Spain. The observations were
carried out using the R300B grating with the EEV12 CCD, and the R316R grating
with the MARCONI2 CCD on the blue and the red arm respectively. The instrumental
setup resulted in a spatial scale of 0.4 arcsec/pix for both arms, and a
dispersion of 1.72~\AA/pix for the blue and 1.65~\AA/pix for the red arm.  The
useful wavelength range is $\sim$ 3300 -- 7800~\AA, in the observed frame.  A
slit of width 1.5 arcsec was used for all the objects.  In order to minimize the
effects of differential atmospheric refraction, the objects were observed either
with the slit aligned along the parallactic angle, or at small airmass (AM $<$
1.1). Although we lack accurate seeing estimates for individual objects, because
ISIS/WHT does not provide imaging observations, measurements by the DIMM seeing
monitor at the observatory suggest that the seeing was in the range $0.6 < FWHM
< 2.6$ arcsec, and usually less than 2.0 arcsec, for the nights of the
observations.

The data were reduced (bias subtracted, flat field corrected, cleaned of cosmic
rays, wavelength calibrated and flux calibrated) and straightened before
extraction of the individual spectra using the standard packages in {\it IRAF}
and the {\it STARLINK} packages {\it FIGARO} and {\it DIPSO}. The wavelength
calibration accuracy, measured as the mean shift between the measured and
published \citep{Osterbrock96} wavelength of night-sky emission lines, is $\sim$
0.35~\AA\, for the blue spectra and $\sim$ 0.25 for the red spectra. The spectral
resolutions, calculated using the widths of the night-sky emission lines (FWHM),
are in the ranges 4.9 -- 5.6~\AA\, and 5.1 -- 5.5~\AA\, for the blue and red
arms, respectively. Finally, the estimated uncertainty for the relative flux
calibration is $\pm$5\%, based on comparison of the response
curves of several spectrophotometric standard stars observed during 
all the observing runs. 

Following reduction, inspection of the spectra revealed that in 2 of the 3
objects with Seyfert 1 spectra (Mrk231, F21219-1775), the strength the broad
Balmer and FeII emission features and non-stellar continuum prevented analysis
of their narrow line emission. Therefore we excluded these two objects from
further analysis. Note however that, although it has not proved possible to
study the narrow emission line kinematics in the nucleus of the two excluded
Seyfert 1 objects, one of these objects -- Mrk231 -- is well known to show
evidence for outflows based on absorption line studies of the neutral and
molecular gas, and the extended emission line regions \citep[e.g.][]{Rupke11}.
Overall our final sample comprises 17 Seyfert-like ULIRGs with viable spectra.

\begin{table*}
\centering
\begin{tabular}{lllrlllll}
\hline\hline
Name & & & & &&&\\
IRAS & z & RA        & DEC       & L$_{\rm IR}$   & L$_{\rm [OIII]}$ &
L$_{1.4\rm GHz}$\\ 
FSC  &   & (J2000.0) & (J2000.0) & (W) & (W)     & (W Hz$^{-1}$)\\ 
(1)  & (2)& (3)      & (4)       & (5)       &  (6)     & (7)\\
\hline
F00188--0856 & 0.128  & 00 21 26.5  & -08 39 26  & 9.2E+38 & 2.1E+33  & 6.8E+23\\
F01004--2237 & 0.117  & 01 02 51.2  & -22 21 51  & 7.5E+38 & 1.6E+34  & 4.4E+23\\
F12072--0444 & 0.129  & 12 09 45.4  & -05 01 14  & 9.4E+38 & 5.4E+34  & 3.3E+23\\
F12112+0305  & 0.073  & 12 13 46.0  &  02 48 38 & 8.2E+38 & 1.6E+33  & 5.7E+23 \\
F13305--1739 & 0.148  & 13 33 15.2  & -17 55 01  & 6.8E+38 & 4.4E+35  & 2.6E+24\\
F13428+5608  & 0.037  & 13 44 41.8  &  55 53 14  & 5.3E+38 & 9.0E+33  & 4.6E+23 \\
F13451+1232E  & 0.122  & 13 47 33.3  &  12 17 24  & 8.0E+38 & 1.2E+35  & 1.9E+26 \\
F14394+5332E  & 0.104  & 14 41 04.3  &  53 20 08  & 4.7E+38 & 4.1E+34  & 1.0E+24 \\
F15130--1958 & 0.109  & 15 15 55.6  & -20 09 18  & 5.3E+38 & 3.7E+34  & 2.7E+23\\
F15462--0450 & 0.100  & 15 48 56.8  & -04 59 34  & 6.2E+38 & 3.4E+34  & 3.1E+23\\
F16156+0146NW  & 0.132  & 16 18 08.2  &  01 39 21  & 4.7E+38 & 6.0E+34  & 3.6E+23 \\
F17044+6720  & 0.135  & 17 04 28.4  &  67 16 23  & 5.8E+38 & 1.9E+34  & 4.9E+23 \\
F17179+5444  & 0.147  & 17 18 55.1  &  54 41 50  & 6.8E+38 & 5.0E+34  & 1.7E+25 \\
F23060+0505  & 0.173  & 23 08 34.2  &  05 21 29  & 1.2E+39 & 1.0E+35  & 5.0E+23 \\
F23233+2817  & 0.114  & 23 25 48.7  &  28 34 19  & 4.3E+38 & 7.8E+34  & 2.1E+23 \\
F23327+2913S  & 0.107  & 23 35 11.9  &  29 30 00  & 4.9E+38 & 5.5E+33  & 2.1E+23 \\
F23389+0300N  & 0.145  & 23 41 31.1  &  03 17 31  & 5.3E+38 & 2.7E+34  & 4.3E+25 \\
\end{tabular}
\caption []{The sample of 17 ULIRGs discussed in this papers. These objects were
  selected from the \cite{Rodriguez-Zaurin09} ES sample of 36 ULIRGs and are
  classified as Sy2 in at least two of the 3 of the \cite{Kewley06} diagnostic
  diagrams. Col (1): object designation in the IRAS Faint Source Catalogue
  Database (FSC). For those sources with multiple nuclei, the individual nucleus
  considered for this study is indicated in the table (N = north, S = south,
  E = east and W = west). Col (2): optical redshifts from Kim \& Sanders
  (1998). Cols (3) and (4): right ascension (hours, minutes and seconds) and
  declination (degrees, arcminutes and arcseconds) of the IRAS source positions
  as listed in the Faint Source Database (FSDB). Col (5): IR luminosity from Kim
  and Sanders (1998) adapted to our cosmology. Col (6): total
  [OIII]$\lambda$5007 emission line luminosities obtained from our modelling
  results (see text for details). Col (7): 1.4 GHz monochromatic radio
  luminosities (L$_{\rm 1.4GHz}$). }
\label{Sample}
\end{table*}

\section{Results}

In this paper we are interested in the AGN-induced outflows in the near-nuclear
regions of ULIRGs. Therefore we concentrate on the 5 kpc-diameter nuclear
apertures described in the \cite{Rodriguez-Zaurin09,Rodriguez-Zaurin10} stellar
population study of the galaxies. In this section we describe the results
obtained by fitting the profiles of the emission lines
detected in the nuclear spectra. 

\subsection{Fitting the emission line profiles}

\begin{figure*}
\begin{tabular}{cc}
\hspace{-1.0 cm}\psfig{figure=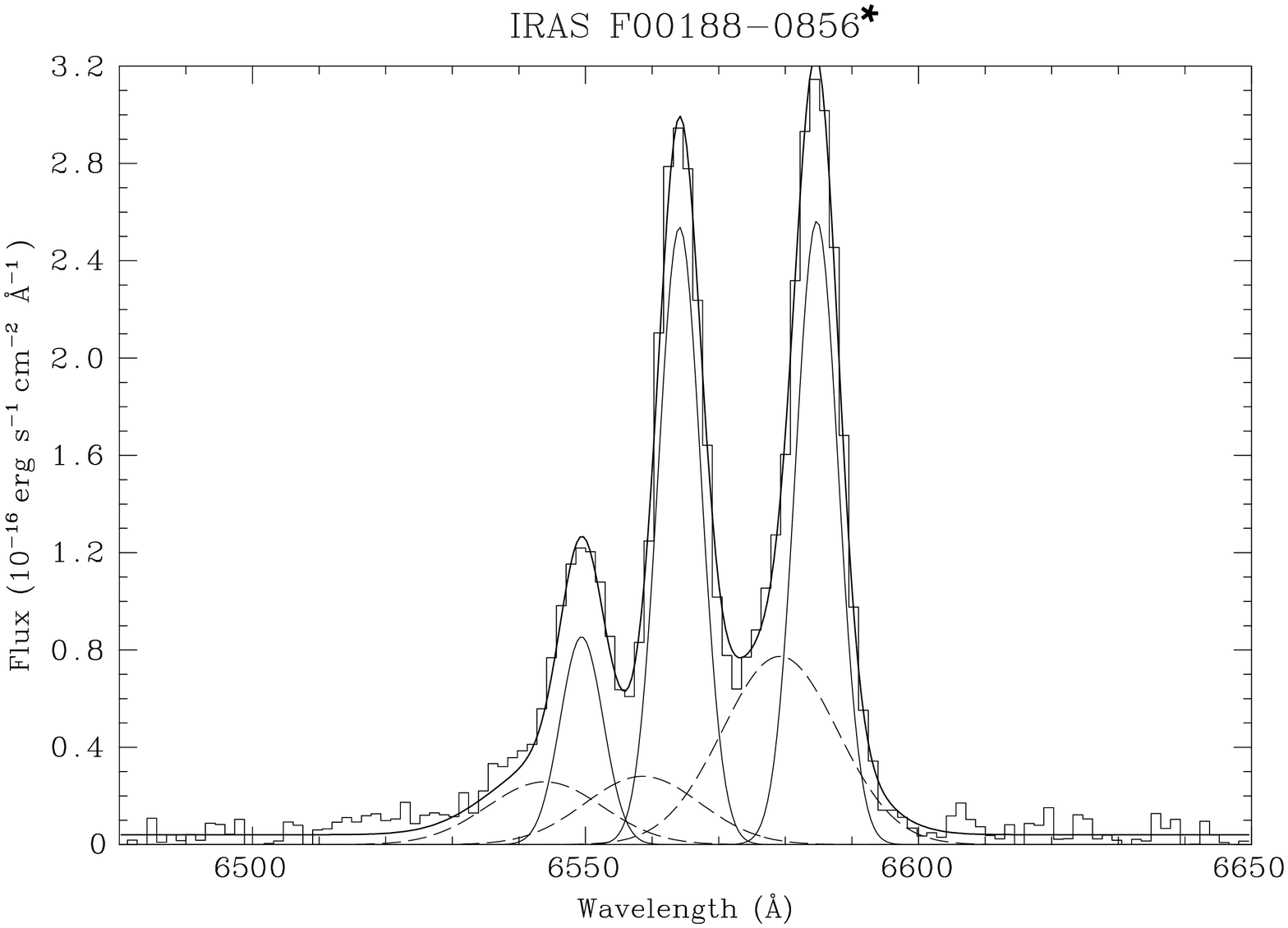,width=5.cm,angle=-90.}&
\psfig{figure=IRASF01004_profile.ps,width=5.cm,angle=-90.}\\
\hspace{-1.0 cm}\psfig{figure=IRASF12072_profile_2nd.ps,width=5.cm,angle=-90.}&
\psfig{figure=IRASF12112_profile_2nd.ps,width=5.cm,angle=-90.}\\
\hspace{-1.0 cm}\psfig{figure=IRASF13305_profile_2nd.ps,width=5.cm,angle=-90.}&
\psfig{figure=IRASF13428_profile_2nd.ps,width=5.cm,angle=-90.}\\
\end{tabular}
\caption{Fits to the [OIII] emission lines profile of all the objects in our
  sample, with the exception of IRAS F13451+1232, IRAS F00188-0856 and IRAS
  F23327+2913. In the case of IRAS F13451+1232 the fit to the [OIII] profile can
  be found in Holt et al. (2003) and therefore is not included in the figure. In
  the latter two cases the \oxythr~emission lines have low equivalent widths and
  we used the H$\alpha$+[NII] complex to find one model that reproduces all
  emission lines. In each case the best fitting model is over-plotted (bold
  line) on the extracted spectrum (faint line). The different kinematic
  components are also plotted in the figure. The narrow, intermediate, broad and
  very broad (if present) components correspond to the the solid line, dashed
  line, dotted line and dot-dashed line. The exceptions are IRAS F13305--1739,
  IRAS F15130--1958 and IRAS 23060+0505. In the first case the dashed line
  corresponds to one of the two broad components detected in this object,
  refereed as B1 in Table 2. In the case of IRAS F15130--1958, the solid line
  corresponds to the intermediate component referred as I1 in Table 2. Finally,
  in the case of IRAS 23060+0505, the dashed line corresponds to the narrow
  component refereed as N2 in Table 2. In addition, IRAS F14394+5332 shows a
  spectacular [OIII] profile that is not observed in all the other emission
  lines detected in its spectrum. See Notes on individual sources for details of
  the modelling for this and the other ULIRGs in our sample. }
\label{OIII-profiles}
\end{figure*}
\addtocounter{figure}{-1}
\begin{figure*}
\begin{tabular}{cc}
\hspace{-1.0 cm}\psfig{figure=IRASF14394_profile_2nd.ps,width=5.cm,angle=-90.}&
\psfig{figure=IRASF15130_profile_2nd.ps,width=5.cm,angle=-90.}\\
\hspace{-1.0 cm}\psfig{figure=IRASF15462_profile.ps,width=5.cm,angle=-90.}&
\psfig{figure=IRASF16156_profile_2nd.ps,width=5.cm,angle=-90.}\\
\hspace{-1.0 cm}\psfig{figure=IRASF17044_profile_2nd.ps,width=5.cm,angle=-90.}&
\psfig{figure=IRASF17179_profile_2nd.ps,width=5.cm,angle=-90.}\\
\hspace{-1.0 cm}\psfig{figure=IRASF23060_profile_2nd.ps,width=5.cm,angle=-90.}&
\psfig{figure=IRASF23233_profile_2nd.ps,width=5.cm,angle=-90.}\\
\end{tabular}
\caption{Continued}
\label{OIII-profiles}
\end{figure*}
\addtocounter{figure}{-1}
\begin{figure*}
\begin{tabular}{cc}
\hspace{-1.0 cm}\psfig{figure=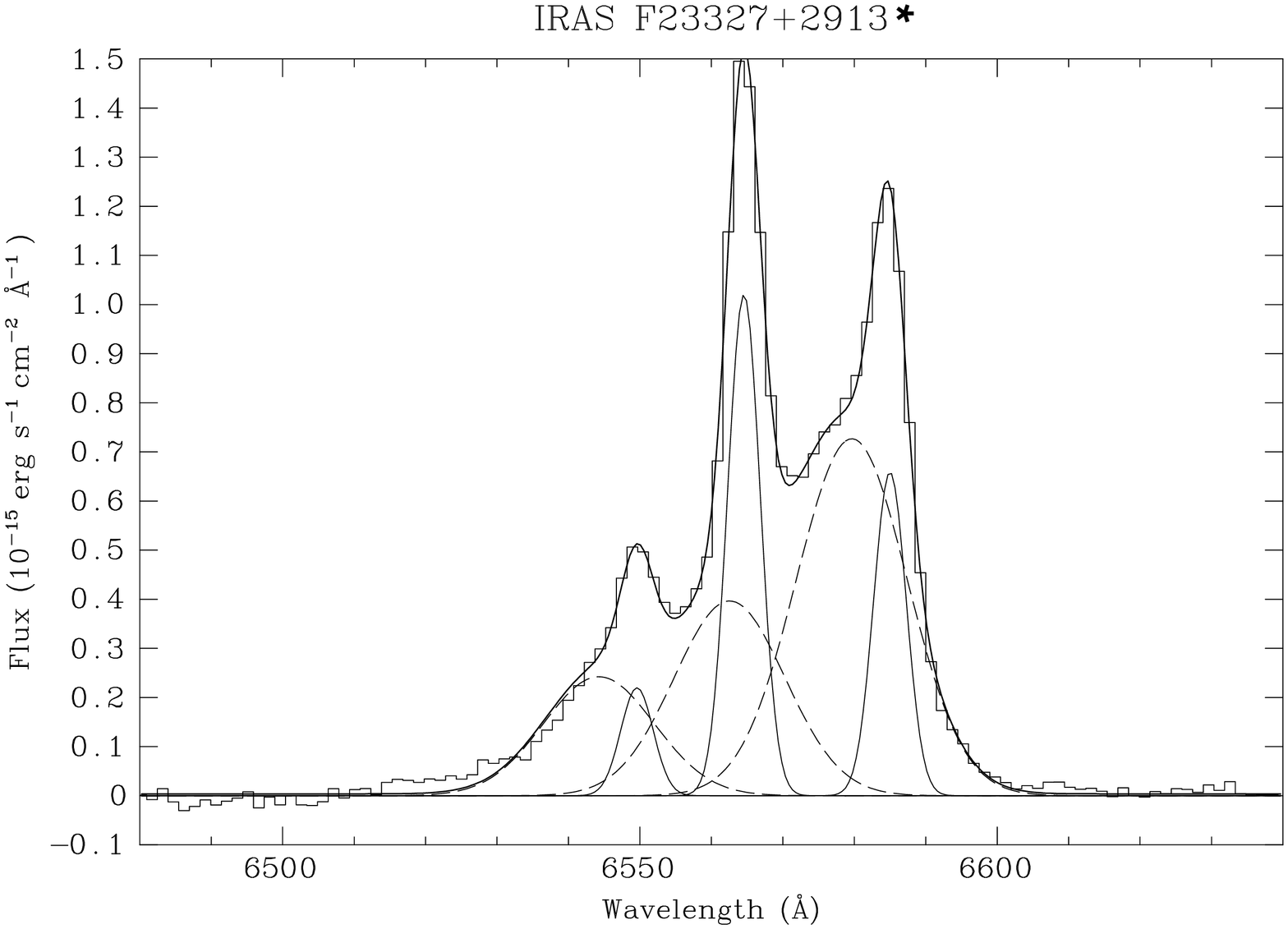,width=5.cm,angle=-90.}&
\psfig{figure=IRASF23389_profile_2nd.ps,width=5.cm,angle=-90.}\\
\end{tabular}
\caption{Continued}
\label{OIII-profiles}
\end{figure*}

Prior to the modelling of the emission line profiles in the nuclear regions, the
spectra were shifted to the galaxy rest frame and a continuum model was
subtracted from the data. In each case the continuum model was selected among
the best fitting models to the stellar continuum emission from the galaxy
\citep[see][for details on the stellar continuum modelling
technique]{Rodriguez-Zaurin09}.

Once the continuum emission was subtracted, we used the $DIPSO$ package to fit
Gaussian profiles to the emission lines. Our general modelling philosophy is to
fit each emission line with the minimum number of Gaussian components required
to produce an acceptable fit\footnote{We define a ``acceptable'' fit to a
  emission line based on visual inspection, i.e. one that has acceptable low
  residuals and adequately represents the overall shape of the line
  profiles.}. The velocity widths derived from the fits were quadratically
corrected for the instrumental profile, and all linewidths and radial velocity
shifts were corrected to the rest frame of the objects.

One of key aspects we want to investigate is the nature of the outflows: whether
they comprise a small number of distinct outflow velocity components, each
associated with a particular spatial location and set of physical conditions
\citep[e.g.][]{Villar-Martin99a,Holt08}, or on the contrary, there is a
continuous gradient of velocities and physical conditions across the full extent
of the emission line region \citep[e.g.][]{Holt03,Spoon09b}. In the first
scenario one kinematic model would provide an adequate fit to all the different
emission lines, while in the second scenario each line would have different
kinematic components.

To test these scenarios it is first necessary to select a prominent emission line that
is suitable to generate a model that will be used to fit all the other
emission lines. In this context the strong [OIII]$\lambda$$\lambda$4959,5007
lines usually have high S/N in our spectra,  are in a region clean from any
atmospheric absorption, and  are not blended with other emission lines (as in
the case of H$\alpha$ and [NII]$\lambda$$\lambda$6548,6583 or the
[OII]$\lambda$$\lambda$3727,3729 doublet). Therefore, where possible, we started
the individual analysis of each source by fitting the
[OIII]$\lambda$$\lambda$4959,5007 emission lines. When fitting these lines we
used three constraints in accordance with atomic physics: i) the flux ratio
between [OIII]$\lambda$4959 and [OIII]$\lambda$5007 was set at 2.99:1 (based
on the transition probabilities); ii) the widths of the corresponding kinematic
components of each line were forced to be equal; iii) the shifts between the
corresponding components of each line were fixed to be 48.0\AA. We will refer to
this as `the {[OIII]} model' hereafter.

We then attempted to model the other prominent emission lines in the spectra
with the same kinematic model (velocity widths and shifts) as {[OIII]}, leaving
the relative fluxes in the kinematic sub-components to vary.  As well as the
constraints derived from the {[OIII]} model, it is possible to further constrain
the fits to other doublets in accordance with atomic physics: the shifts between
the corresponding components of each line in all doublets (e.g.
{[NeIII]}\lala3868,3968, {[NeV]}\lala3346,3425, {[NII]}\lala6548,6583,
{[OI]}\lala6300,6363 and {[SII]}\lala6716,6731) were set, and for some doublets
(e.g. {[Ne III]}\lala3868,3968, {[Ne V]}\lala3346,3425, {[NII]}\lala6548,6583,
and {[OI]}\lala6300,6363) the flux ratios were set based on the transition
probabilities. For {[SII]}\lala6716,6731, the flux ratio was required to be
within the range 0.44 $<$ {[SII](}6716/6731) $<$ 1.42, the ratios corresponding
to the high and low density limits respectively. For simplicity, The
[OII]\lala3726,3728~lines were treated as a single line for the majority of the
objects in our sample\footnote{There are 3 cases (F12072-0444, F17044+6720 and
  F17179+5444) for which accounting for the doublet nature of at least one of
  the different kinematic components, was necessary to adequately model the
  \oxytwow~emission lines.}. Therefore, in those cases we assumed 3727\AA~as the
rest frame wavelength.

The {[OIII]} modelling technique was successful for all but 4 of the 16
objects in the sample discussed in this paper (75\%). This is consistent with
the idea that, in the majority of cases, there exist a discrete number kinematic
components, each with its own line width, velocity shift, physical conditions
and ionization. However, in the cases of F13451+1232, F14394+5332, F16156+0146
and F23389+0300 (25\% of the objects in our sample) it was not possible to use a
single model to fit all the emission lines observed in their optical spectra,
suggesting that there may be a continuous gradient of density, ionization and
kinematics through the Narrow Line Region (NLR), leading to each line having a
different profile.

Figure \ref{OIII-profiles} shows the fits to the [OIII] emission lines profile
for all but 3 of the objects in our sample. In the cases of F00188-0856 and
F23327+2913, the [OIII]\lala4959,5007 emission lines have a low equivalent width
and the profiles are affected by residual structure in the underlying continuum;
hence it is not possible to precisely fit these lines. However, it is still
possible in these cases to find one kinematic model that fits all emission
lines, based on the stronger H$\alpha$+[NII] lines. Figure \ref{OIII-profiles}
shows the fits to the H$\alpha$+[NII] complex for these two objects.  In
addition, a detailed study of the kinematics for F13451+1232 is presented in
\cite{Holt03} and \cite{Holt11}, including the fit to the [OIII] emission
line. Therefore, no fit for this source is shown in Figure
\ref{OIII-profiles}. Table \ref{widths_shifts} presents the widths and shifts of
the different kinematic components for the [OIII] model (or H$\alpha$+[NII]
model in two cases) for the central 5kpc aperture for all galaxies in our
sample, whereas Table \ref{fluxes} presents the line fluxes and emission line
ratios.

We emphasize that the fits to some emission line blends have the potential to
suffer from degeneracies, in the sense that it is possible for the fitting
programme to allocate the fluxes between the different kinematic components in
various ways and still produce acceptable fits to the overall emission line
profile. This is a particular problem in cases where there are multiple, broad
kinematic components, and the components of the blend are relatively close
together in wavelength (e.g. H$\alpha$+[NII]$\lambda\lambda$6548,6584 and
[SII]$\lambda\lambda$6717,6731). To check whether the degeneracy is a
consequence of the particular fitting programme (DIPSO) we have used, we have
also modelled the emission line profiles of those objects for which the
degeneracies were more important (e.g. F01004--2237 or F15130--1958) using the
MPFIT code \citep{Markwardt09}. The main difference between DIPSO and MPFIT is
that, whereas DIPSO fits the kinematic components in each individual emission
line or blend individually, we have used MPFIT to fit all the kinematic
components in all the lines/blend in the spectrum simultaneously. Overall, the
modelling results were consistent with those obtained using DIPSO and the
degeneracies were still present. The impact of these potential degeneracies on
the fits to emission line blends in individual objects is discussed in the
Appendix.

As well as fitting the 17 objects in our sample of ULIRGs with optical Seyfert
nuclei, for the purposes of comparison, we used identical techniques to fit the
[OIII]$\lambda$$\lambda$4959,5007 and/or H$\alpha$+[NII] profiles in the 5kpc
aperture nuclear spectra of the 16 nearby ULIRGs in the ES ($z < 0.175$) of
\citet{Rodriguez-Zaurin09} that do not show evidence for Seyfert nuclei at
optical wavelengths. 

For ease of reference in the following sections, we use the following scheme to
label kinematic components, based on line widths (FWHM):

\begin{itemize}
\item narrow: FWHM $<$ 500 \kms ;\\ 
\item intermediate: 500 $<$ FWHM $<$ 1000 \kms ; \\
\item broad: 1000 $<$ FWHM $<$ 2000 \kms;  \\
\item very broad 2000 \kms $>$ FWHM. \\
\end{itemize}

Detailed descriptions of the line profile fits for individual ULIRGs are
described in the Appendix, while in the following sections we describe the
general results derived from these fits.

\begin{table*}
\begin{tabular}{llll}
\hline\hline
Object &    & FWHM &$\Delta V$ \\
IRAS   &    &      &           \\
       &    & \kms &  \kms    \\
(1)    & (2)&  (3) &   (4)    \\
\hline 
F00188--0856 & N & 269$\pm$14   &               \\ 
             & I & 904$\pm$79   & -253$\pm$36   \\
F01004--2237 & N & unres        &               \\
             & I & 849$\pm$79   & -229$\pm$32   \\ 
             & B & 1590$\pm$94  & -999$\pm$80   \\
F12072--0444 & N & 275$\pm$18   &               \\
             & I & 525$\pm$36   & -276$\pm$30   \\
             & B & 1343$\pm$84  & -446$\pm$35   \\
F12112+0305  & N1 & 162$\pm$73  & -             \\
             & N2 & 473$\pm$66  & -171$\pm$14   \\
F13305--1739 & N & 435$\pm$ 29  &               \\
             & B1 & 1275$\pm$31 & -36$\pm$14    \\
             & B2 & 1685$\pm$286& -281$\pm$29   \\
F13428+5608  & N & 450$\pm$ 13  &               \\
             & B & 1368$\pm$66  & +14$\pm$11    \\
F13451+1232  & N & 340$\pm$23   &               \\
             & B1& 1255$\pm$41  & -402$\pm$9    \\
             & B2& 1944$\pm$65  & -1980$\pm$36  \\
F14394+5332  & N1& 335$\pm$26   &               \\
             & N2& 390$\pm$32   & -700$\pm$10   \\
             & N3& 435$\pm$52   & -1358$\pm$18  \\
             & B1& 1272$\pm$67  & -156$\pm$8    \\
             & B2& 1401$\pm$131 & -1574$\pm$19  \\
F15130--1958 & I1 & 545$\pm$85  &               \\
             & I2 & 700$\pm$222 & -350$\pm$281  \\
             & B & 1630$\pm$42  & -725$\pm$131  \\
F15462--0450 & N & 147$\pm$57   &               \\
             & B & 1426$\pm$39  & -822$\pm$25   \\
F16156+0146  & N & unres        &               \\
             & I & 804$\pm$26   & -186$\pm$10   \\ 
             & B & 1535$\pm$63  & -374$\pm$26   \\
F17044+6720  & N & 290$\pm$20   &               \\
             & B & 1765$\pm$103 & -553$\pm$65   \\
F17179+5444$^a$& N & 358$\pm$75 & -123$\pm$33   \\
               & I & 515$\pm$33 & 123$\pm$34    \\
               & B & 1562$\pm$43& -242$\pm$61   \\
F23060+0505  & N1 & 376$\pm$10  & -152$\pm$6    \\
             & N2 & unres       & 152$\pm$4     \\
             & B & 1001$\pm$23  & -463$\pm$20   \\
             & VB & 2150$\pm$125& -1073$\pm$122 \\
F23233+2817  & N & 178$\pm$37   &               \\
             & I & 640$\pm$24   & -161$\pm$15   \\
             & B & 1511$\pm$118 & -447$\pm$40   \\
             & VB & 3433$\pm$903& -2706$\pm$422 \\
F23327+2913  & N & 109$\pm$12   &             \\
             & I & 804$\pm$14   & -94$\pm$4     \\
F23389+0300  & N & 289$\pm$12   &             \\
             & VB & 2223$\pm$30 & 47$\pm$13     \\
\end{tabular}
\caption {The FWHM and velocity shifts ($\Delta V$) for the different kinematic
  components of the galaxies in our sample. Col (1): object name. Col (2): The
  label of the different components as defined in the previous section [N
    (narrow), I (intermediate), B (broad) and VB (very broad)]. For those object
  with two components within the same FWHM range, these are indicated with
  numbers (e.g. B1 and B2 corresponding to the two broad components detected in
  IRAS F13305--1739). Col (3) and (4): rest-frame widths (FWHM) and shifts
  relatively to galaxy rest frame ($\Delta V$) of the different
  components. \newline $^{a}$ Two models adequately reproduced the emission
  lines in the case of F17179+5444. One model comprises a narrow, an
  intermediate and a broad components while the second model includes two narrow
  components plus a broad one. The results show in the table for this galaxy
  correspond to the first model. For details on the modelling for this and all
  the other objects in our sample, see Section 3.2}
\label{widths_shifts}
\end{table*}

\begin{landscape}
\centering
\begin{table}
\begin{tabular}{llcccccccccccccc}
\hline\hline
Object &&  \hb & [NeV]$\lambda$3426 & [OII]$\lambda3727$ & [NeIII]$\lambda$3869 & [OIII]$\lambda$5007 & [OI]$\lambda$6300 & \ha & $\Delta$(\ha/\hb) & [NII]$\lambda$6583 & [SII]$\lambda\lambda$6716,6731 \\
IRAS   & &erg cm$^{-2} $s$^{-1}$ \\
(1) & (2) & (3) &  (4) &  (5) &  (6) &  (7) &  (8) &  (9) &  (10) & (11) & (12)\\   
\hline 
F00188--0856 & N & 2.92E-16 & -    & 1.96 &  -   & 0.40  & 0.97 & 6.97 & 0.93 & 7.03 & -\\ 
             & I & 1.76E-16 & -    & 3.01 &  -   & 2.00  & 0.38 & 3.46 & 1.07 & 9.54 & -\\
F01004--2237 & N & 4.03E-16 & -    & 1.21 & 0.25 & 2.66 & 0.29 & 4.01 & 0.24 & 0.99 & 1.00\\ 
             & I & 4.60E-16 & -    & 1.85 & 0.55 & 2.07 & 0.42 & 4.45 & 0.72 & 1.90 &0.46\\  
             & B & 3.65E-16 & -    & 0.62 & 0.94 & 6.47 & 0.40 & 2.51 & 0.51 & 7.47 &1.17\\
F12072--0444 & N$^{a}$  & 1.23E-15 & 0.02 & 2.78 & 0.27 & 3.50 & 0.70 & 5.91 & 0.13 & 3.71 & -\\  
             & I           & 4.82E-16 & 0.69 & 1.70 & 0.67 & 8.47 & 0.47 & 5.48 & 0.61 & 6.46 & -\\
             & B           & 4.68E-16 & 0.09 & 1.26 & 0.79 & 6.92 &1.07 & 6.75 & 0.92 & 6.30 & -\\
F13305--1739 & N & 1.67E-15 & 0.46 & 0.65 & 0.20 & 4.61 & 0.06 & 3.37 & 0.46 & 1.59 & 1.24\\
             & B1 &4.57E-15 & 0.62 & 1.67 & 0.95 & 10.15& 0.71 & 3.74 & 0.70 & 5.83 & 1.88\\ 
             & B2 &9.69E-16 & -    & 2.03 & 1.97 & 14.06& 0.85 & 8.35 & 5.60 & 4.98 & 1.20\\
F13428+5608  & N & 8.38E-15 & -    & 1.37 & -    & 1.98 & 0.66 & 5.67 & 0.19 & 4.09 & 3.36 \\
             & B & 9.02E-16 &      &10.34 & -    & 11.80& 2.65 & 9.81 & 3.80 & 29.36& 6.94 \\
F13451+1232$^{b}$  & N & 5.85E-16 &  0.02& 3.24 & 0.30 & 2.34 & 0.89 & 3.32 & 0.33 & 4.76 & 3.79\\
             & B1& 2.11E-15 &  0.01& 1.34 & 0.65 & 9.37 & 1.10 & 5.16 & 0.28 & 6.27 & 1.83\\
             & B2& 3.24E-16 &  1.22& 0.55 & 4.81 &24.42 & 7.31 & 18.81& 4.74 & 1.98 & 6.50\\
F14394+5332$^{b}$  & N & 2.88E-15 & - & 1.32 & 0.02 & 0.63 & 0.49 & 6.61 & 0.17 & 3.14 & 2.69\\
             & B & 8.06E-16 & - & 16.81& 2.88 & 15.12 & 6.25 & 17.59 & 5.02 & 27.82 & 16.14 \\
F15130--1958 & I1 & 6.06E-16& 0.13 & 2.91 & 0.30 & 2.91 & 1.03 & 4.24 & 0.42 & 8.04 & 3.21\\
             & I2 & 2.47E-16& 1.21 & 3.05 & 1.97 & 12.73& 0.85 & 3.99 & 0.89 & 11.96& 0.27\\
             & B &  4.51E-16& 3.69 & 1.57 & 3.09 & 15.05& 1.41 & 4.01 & 0.85 & 12.04& 2.97\\
F15462--0450$^{c}$& N    & 1.11E-15 & 0.38 & 3.59 & 0.38 & 1.23 &  - & 7.47 &  0.75 & 2.74 & 2.81  \\
                  & B    & 3.56E-15 & 0.58 & 0.85 & 0.95 & 3.25 & -  & -    & & -     & 0.85 \\
                  & BLR  & 2.20E-14 & -    &  -   &  -   &  -   &  - & 4.79 & 0.20 & -     &  - \\
F16156+0146$^{b}$  & N  & 5.62E-16& 0.49 & 0.21 & 0.84 & 9.06 & 0.17 & 4.04 & 0.48 & 0.51 & - \\
                   & N/I&4.66E-16 & 0.15 & 6.30 & 0.05 & 14.43 &1.65 & 6.26 & 1.54 & 3.93 & - \\
                   & B  &9.05E-16 & 0.28 & 1.29 & 0.78 & 5.05 & 0.88 & 5.71 & 0.80 & 4.33 & - \\
F17044+6720  & N$^{a}$  & 1.25E-15 & -    & 2.98 & 0.24 & 1.86 & 0.65 & 5.05 & 0.17 & 2.50 & -  \\ 
             & B$^{a}$  & 2.49E-16 & -    & 4.26 &  -   & 5.27 & 1.48 & 5.33 & 2.89 & 14.13& - \\
F17179+5444$^{d}$ & N$^{a}$  &7.82E-16& 0.37 & 1.16 & 0.23 & 1.15 & 0.79 & 4.72 & 0.50 & 5.17 & 2.97\\
             & I$^{a}$  &3.31E-16& -    & 6.42 & 1.32 & 10.24 & 0.43 & 5.89 & 2.27 & 4.36 & 0.84\\
             & B           &4.64E-16& -    & -    & 0.18 & 7.40 & 0.77 & 7.71 & 2.04 & 9.37 & 2.46\\
F23060+0505$^{e}$  & N1  & 1.83E-15& 0.21 & 2.05& 0.45 & 5.42 & 0.42 & 4.23 & 0.14 & 2.55 & 1.75\\
                   & N2 & 6.23E-16& 0.23 & 1.24 & 0.97 & 6.76 & 0.24 & 2.30 & 0.26 & 0.84 & 0.50\\ 
                   & B  & 8.39E-16& 1.28 & 2.19 & 1.00 & 11.96& 0.36 & 8.11 & 0.85 & 7.40 & 2.83\\

\hline
\end{tabular}
\caption{}
\label{fluxes}
\end{table}
\end{landscape}
\addtocounter{table}{-1}
\begin{landscape}
\centering
\begin{table}
\begin{tabular}{llcccccccccccccc}
\hline\hline
Object &&  \hb & NeV$\lambda$3426 & [OII]$\lambda3727$ & NeIII$\lambda$3869 &
[OIII]$\lambda$5007 & [OI]$\lambda$6300 & \ha & $\Delta$(\ha/\hb) & [NII]$\lambda$6583 & [SII]$\lambda\lambda$6716,6731 \\
IRAS   & &erg cm$^{-2} $s$^{-1}$   \\
(1) & (2) & (3) &  (4) &  (5) &  (6) &  (7) &  (8) &  (9) &  (10) & (11) & (12)\\   
\hline 
F23233+2817$^{e}$  & N & 8.10E-16 & 0.45 & 1.28 & 0.42 & 1.98 & 0.33 & 5.12 & 0.69 & 5.66 & 2.37\\  
                   & I & 8.86E-16 & 0.88 & 2.59 & 1.26 & 10.68& 0.63 & 4.46 & 1.12 & 13.37 & 1.41\\
                   & B & 1.58E-15 & 1.10 & 0.41 & 1.14 & 7.34 & 0.40 & 2.10 & 0.33 & 4.99 & 1.04\\
F23327+2913$^{f}$  & N &  -       & -   &  0.14   &  -   &  0.05  &  0.08   &  5.82E-15   &  0.65   & 0.47  \\
                   & I & 3.02E-15 &  -   & 0.91 &  -    & 0.62 & 0.26 & 2.55 & 0.14 & 4.68 & 0.75\\ 
F23389+0300$^{b}$ & N/I & 4.12E-16 & -  & 3.33 & 0.28 & 2.28 & 4.88 & 7.87 & 1.02 & 4.71 & 2.52\\
                  & VB  & 9.88E-16 & - & 1.91 & 0.43 & 3.63 & 7.34 & 8.35 & 1.26 & 25.88 & 9.46\\
\hline
\end{tabular}
\caption{Table showing the fluxes of various emission lines detected in the
  optical spectra of the sources. Col (1): object name. Col (2): the label of
  the different components (Same as Col (2) in Table 2). Col (3): \hb~flux in
  erg s$^{-1}$ cm$^{-2}$. From Col (4) to Col (9) and columns (11) and (12):
  ratios of the fluxes of the main emission lines to the \hb~line. The
  uncertainties in these ratio are typically $\lsim$15\%, as estimated
  accounting for the 5\% flux calibration uncertainty, and the uncertainties in
  the model fits themselves. However, there are a few particular cases, mainly
  the broadest component in some emission lines, or those kinematic components
  that make a small contribution to the flux, for which the uncertainty can be
  as high as 50\% (for example, the I2 component of the
  [SII]$\lambda$$\lambda$6716,6731 emission lines in the case of
  IRASF15130--1958). Col (10): uncertainties in the \ha/\hb~ratio (we discuss in
  the text a possible correlation between the broadness of the different
  kinematic components and their corresponding reddening. Therefore, it is
  particularly important to give the uncertainties associated with this ratio
  explicitly).
\newline $^{a}$ As a first approach during the modelling, \oxytwow~ was treated
as a singlet rather than a doublet. However, there are cases for which
accounting for the doublet nature of at least one of the different kinematic
components is necessary to adequately model the \oxytwow~emission lines. The
fluxes presented in the table for these cases are are the sum of the two lines
of the doublet. \newline $^{b}$ For these four galaxies it was not possible to
find one model that adequately reproduces all the emission lines. In the cases
of IRAS F13451+1232, F16156+0146 and F23389+0300 the number of components is the
same for all the emission lines, although with different widths and
shifts. Therefore, it is still possible to use the H$\beta$ line as a reference
and give line fluxes relative to that line. In the case of F14394+5332 the
modelling results reveal the presence of a narrow component at rest frame found
in {\it all} the emission lines, plus a number of different blueshifted
components with different widths and shifts for the different emission
lines. Therefore, to give an idea of the fluxes from the narrow component and
the shifted components, we define two kinematic components in this case: narrow
(N), corresponding to the narrow component found for all emission lines, and
broad (B), which comprises the sum of all the kinematic components for each line
that are not the narrow component. \newline $^{c}$ This is the only object with
a strong broad-line Sy1 nucleus the in our sample. Due to the large contribution
of the broad component from the broad line region (BLR) in the case of the
\ha~emission, it is not possible to constrain the flux in the broad, blueshifted
component detected in [OIII] (referred as B in the table) for this particular
emission line. Therefore, no flux value for the B component is presented in the
table for this galaxy (see appendix A for details). \newline $^{d}$ Two models
adequately modelled the emission lines in the case of F17179+5444. One model
comprises a narrow, an intermediate and a broad components, while the second
model includes two narrow components of the same width plus a broad one. The
results shown in the table for this galaxy correspond to the first
model.\newline $^{e}$ The very broad (VB) kinematic component for these two
ULIRGs is only detected in high ionization emission lines (i.e. not in the case
of \hb) and therefore not included in the table. \newline $^{f}$ The \hb~
emission line for this galaxy falls close to the region of the spectrum that is
heavily affected by the dichroic. Therefore, the narrow component, which makes a
relative small contribution for most of the emission lines, is not detected in
the case of \hb. For this components line fluxes are given relative to \ha.}
\label{fluxes}
\end{table}
\end{landscape}

\subsection{The incidence of kinematically disturbed components in Seyfert and non-Sy-ULIRGs}

Kinematic disturbance of the emission line gas does not necessarily involve
large line shifts, but can also be indicated by large line widths in components
that are relatively unshifted. Therefore we define the term ``kinematic
disturbance'' in this paper to encompass kinematic components that have
intermediate or broad line widths ($FWHM > 500$ km s$^{-1}$) and/or large blue
shifts relative to the galaxy rest frame ($\Delta V < -150$ km s$^{-1}$).  By
this definition, the overwhelming majority (16/17 or 94\%) of the Sy-ULIRGs
for which we have been able to measure the emission line profiles in this study
show evidence for kinematic disturbance in their nuclear regions.  In contrast,
only a small minority (3/16 or $\sim$19\%) of the non-Sy-ULIRGs from
Rodriguez Zaurin et al. (2009) present evidence for kinematic disturbance in the
nuclear 5 kpc apertures\footnote{These three ULIRGs are F14348--1447 and
  F23234+0946 and F15327+2340.}. This difference is significant at the
$\sim$3$\sigma$ level, strongly suggesting that the kinematically disturbed
components are driven by the AGN in most cases, rather than by the
circum-nuclear starbursts.  Significantly, the one Sy-ULIRG that shows no
clear signs of nuclear kinematic disturbance --- F12112+0305 --- is also the
object that presents the weakest evidence for Seyfert-like AGN activity based on
its line ratios --- falling close to the AGN/HII composite region in the
diagnostic diagrams of \cite{yuan10}; it also has the lowest [OIII] emission
line luminosity of all the Sy-ULIRGs in our sample (see Table 1).

At first sight these results may seem in contrast with those of \cite{Soto12a}
and \cite{Soto12b} who studied a sample of 39 ULIRGs and found evidence for
disturbed kinematics in the majority of the objects in their sample, which is
dominated by non-Sy ULIRGs. However, the \cite{Soto12a} spectroscopic dataset
has higher spectral and spatial resolution than ours. In addition, the sizes of
the extraction apertures are different from those used in this paper. Therefore,
it is likely that they detect some broad components that are not observed in our
spectra. Finally, \cite{Soto12a} and \cite{Soto12b} used a different modeling
approach from the one we use here (i.e. a ``simplest model approach''). As a
result, their kinematic models for the nuclear emission lines in some of the
sources in their sample comprise more than one kinematic component, regardless
of whether the additional components are actually required to produce an
adequate fit.

\subsection{The diversity of emission line profiles}

Figure 1 and Table 2 demonstrate a striking diversity in the emission line
profiles of Sy-ULIRGs. The profiles range from the complex, multi-component
profile visible in F14394+5332, through the case of IRAS F17044+6720, in which
there is a single broad, blueshifted component along with a strong narrow
component, to the case of F23389+0300, which has a strong, and extremely broad
component that makes up a large proportion of the flux, yet is barely shifted
relative to the narrow component.  The minimum number of Gaussian components
required to fit the individual emission lines ranges from 2 to 5, with 7 objects
($\sim$41\%) requiring two, 7 objects ($\sim$41\%) requiring three, 2 objects
($\sim$12\%) requiring four, and 1 object (F14394+5332) requiring five
components for an adequate fit.  However, it is important to note that, in the
two cases -- F23060+0505, F23233+2817 -- where a fourth component is required to
adequately model the emission lines, the relative flux contribution of this
fourth component is relatively small. Therefore, its FWHM and shift with respect
to the narrow component are relatively unconstrained.

In terms of determining the direction and magnitude of the velocity shift of any
kinematically disturbed component, it is important to carefully determine the
rest frame velocities of the ULIRG host galaxies (see Holt et al., 2003, 2006,
2008). In the absence of other indicators (e.g. accurate stellar kinematics) the
best way to do this is to compare the kinematics of the extended narrow
components along the slit with the various kinematic components detected in the
spatially-integrated emission line profile of the 5~kpc aperture. For example,
if the undisturbed gas shows a rotation curve across the nucleus, the component
in the spatially-integrated emission line profile that corresponds to the galaxy
rest frame will have a wavelength corresponding to the centroid of the blue and
redshifted components on either side of the nucleus. In the majority of the
cases in which we have spatially-resolved information on the narrow line
kinematics, we find that the redmost narrow component in the spatially
integrated profile coincides with the galaxy rest frame. However, in two cases
--- F17179+5444 and F23060+0505 --- a narrow and a relatively low-FWHM
intermediate component (FWHM = 515 $\pm$ 33 \kms), or two narrow components, are
present in the nuclear line profiles, each corresponding to the velocity shift
of the spatially resolved gas on one or other side of the nucleus. In these
cases, the observed narrow line splitting is likely to represent unresolved
rotation, or to trace a large-scale bipolar outflow in the gas \citep[see, for
  example][]{Holt08}. In these two cases we used the average of the wavelengths
of the two narrow components to define the galaxy rest frame.

With the exceptions of F17179+5444 and F23060+0505, the velocity shifts listed
in Table 2 are defined relative to the redmost narrow component fitted in the
5kpc aperture spectrum. It is clear that the broadest kinematic components in
each object are significantly blueshifted relative to the galaxy rest frames in
all cases except F13428+5608 and F23389+0300. In the case that the emission line
clouds suffer extinction by a general field of dust in the nuclear regions of
the galaxies, rather than by dust in the clouds themselves, this implies that in
most cases the gas is undergoing a radial outflow from the nucleus.

\begin{figure}
\psfig{figure=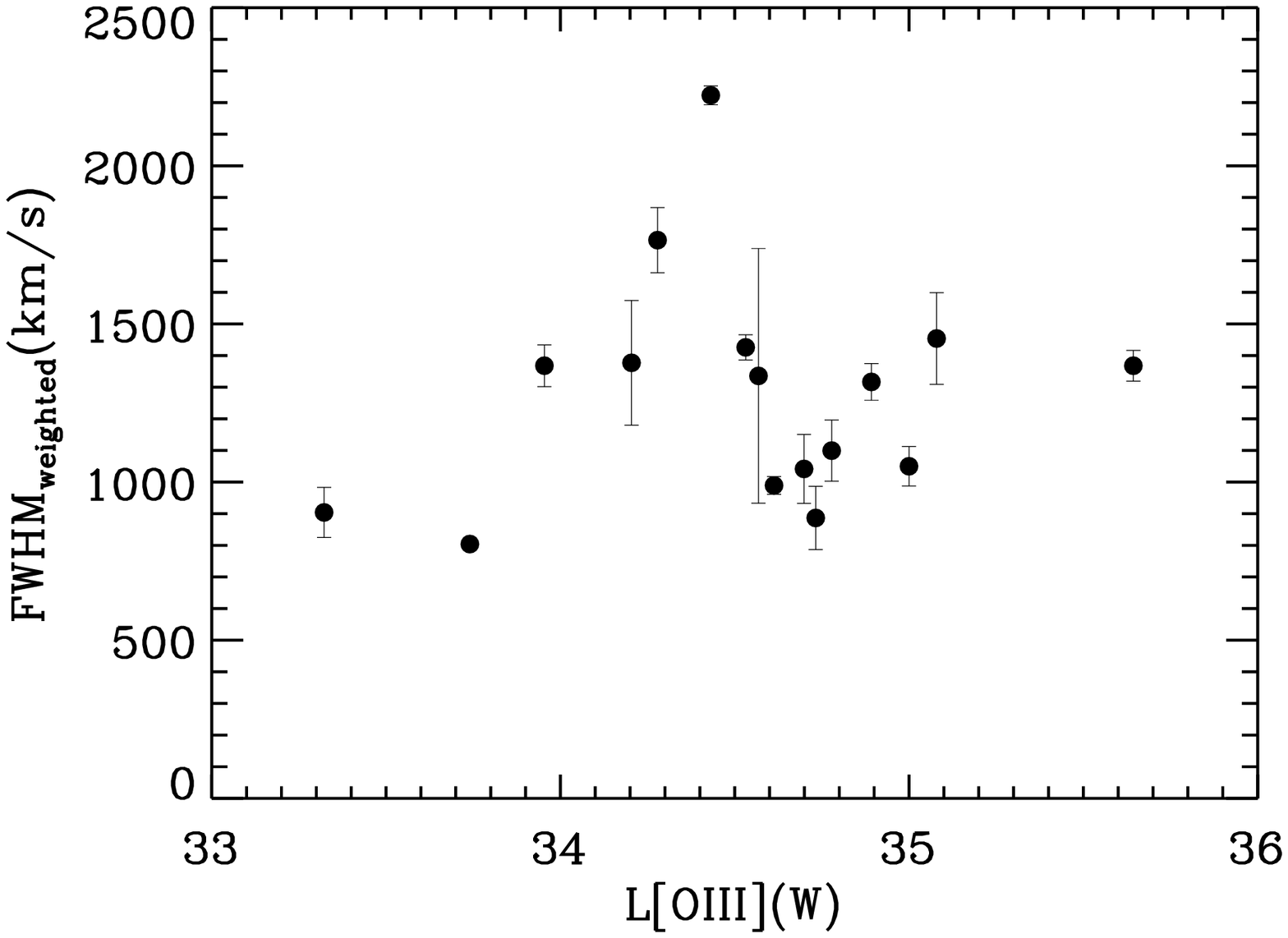,width=9cm}\\
\psfig{figure=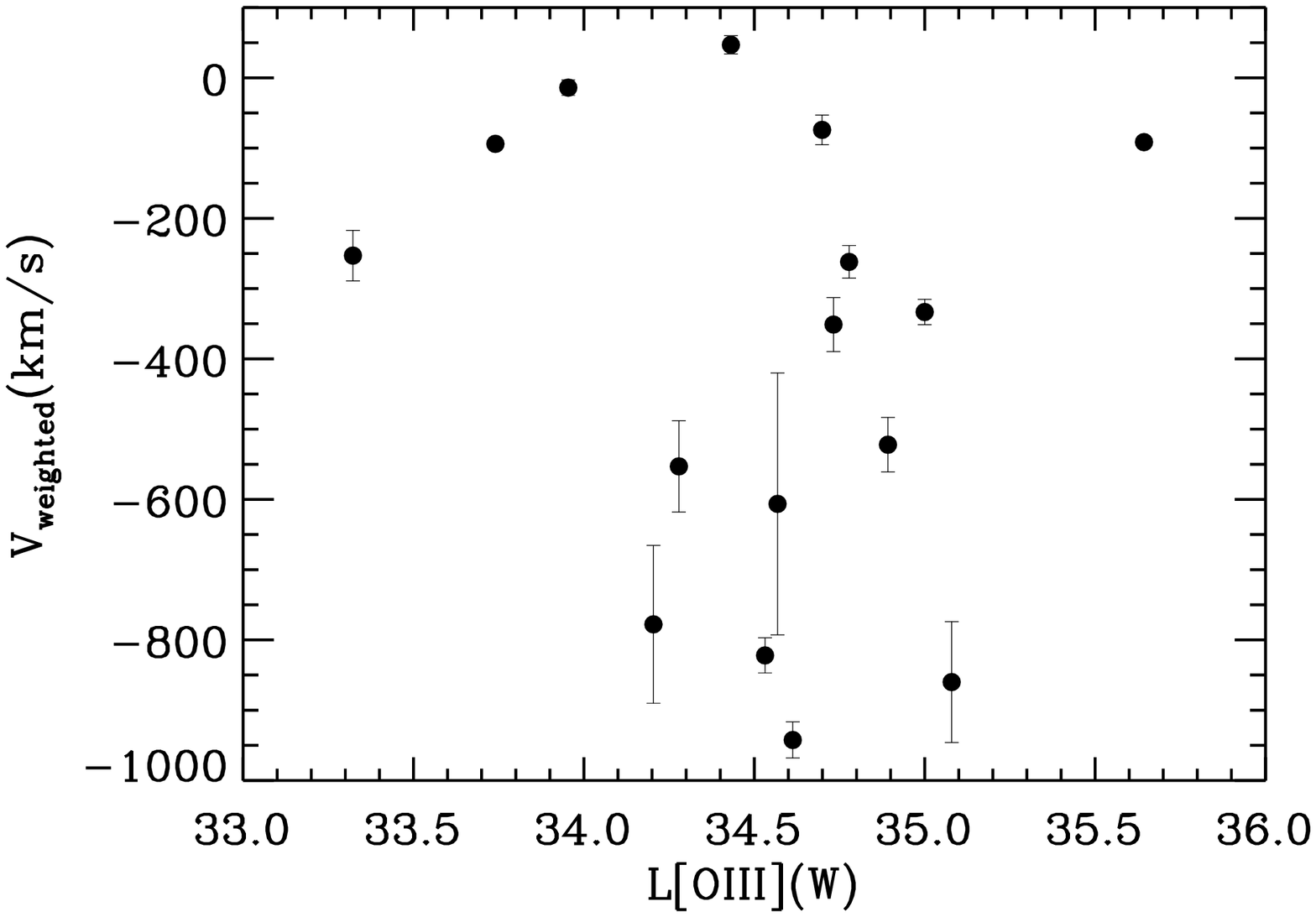,width=9cm}
\caption{Flux weighted mean FWHM (FWHM$_{\rm w}$) and shifts
  ($\Delta$V$_{\rm w}$) plotted against the [OIII]$\lambda$5007
  luminosities (L$_{\rm [OIII]}$) of the sources.}
\label{FWHMvsLOIII}
\end{figure}

\begin{figure}
\psfig{figure=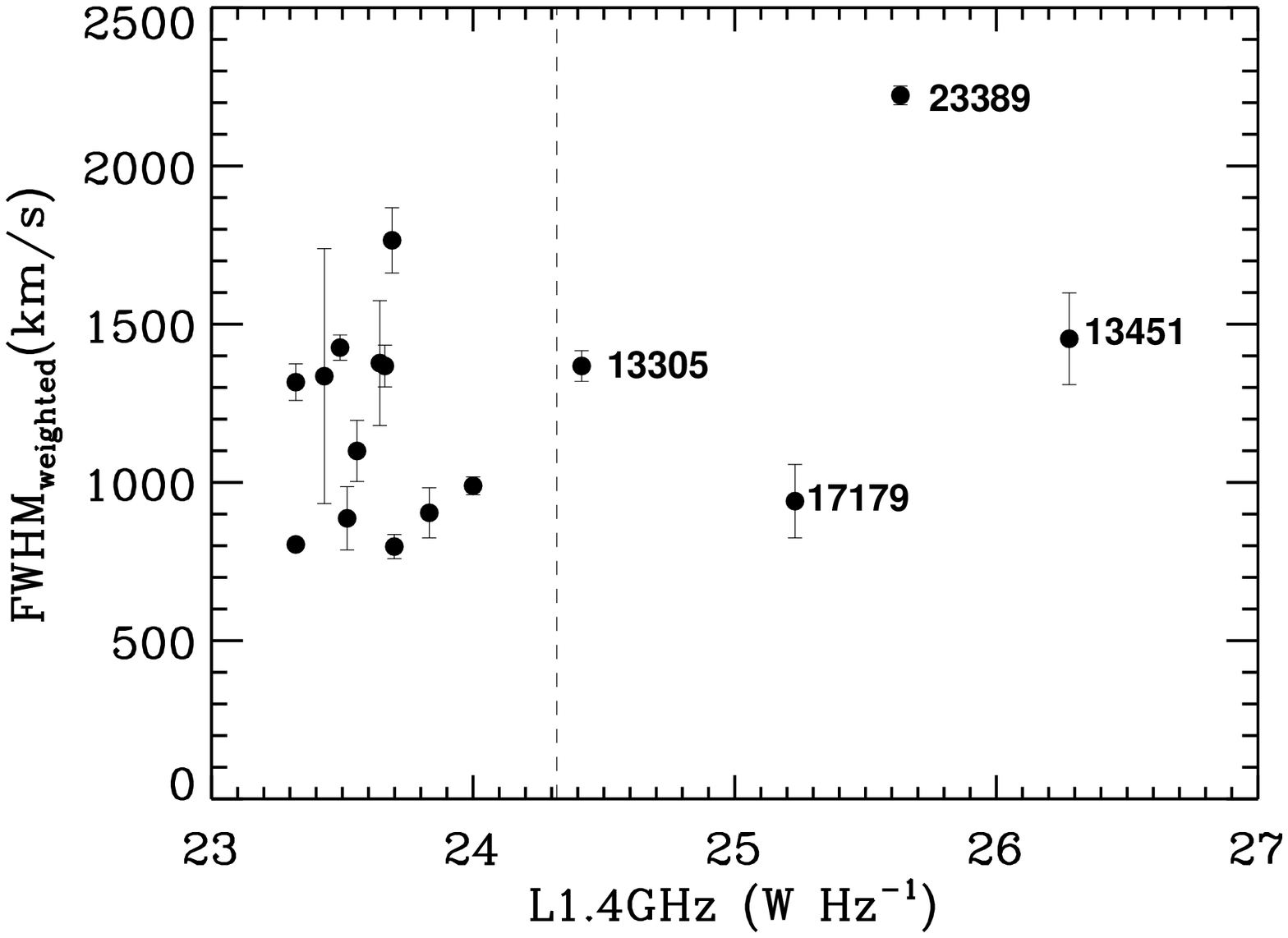,width=9cm}\\
\psfig{figure=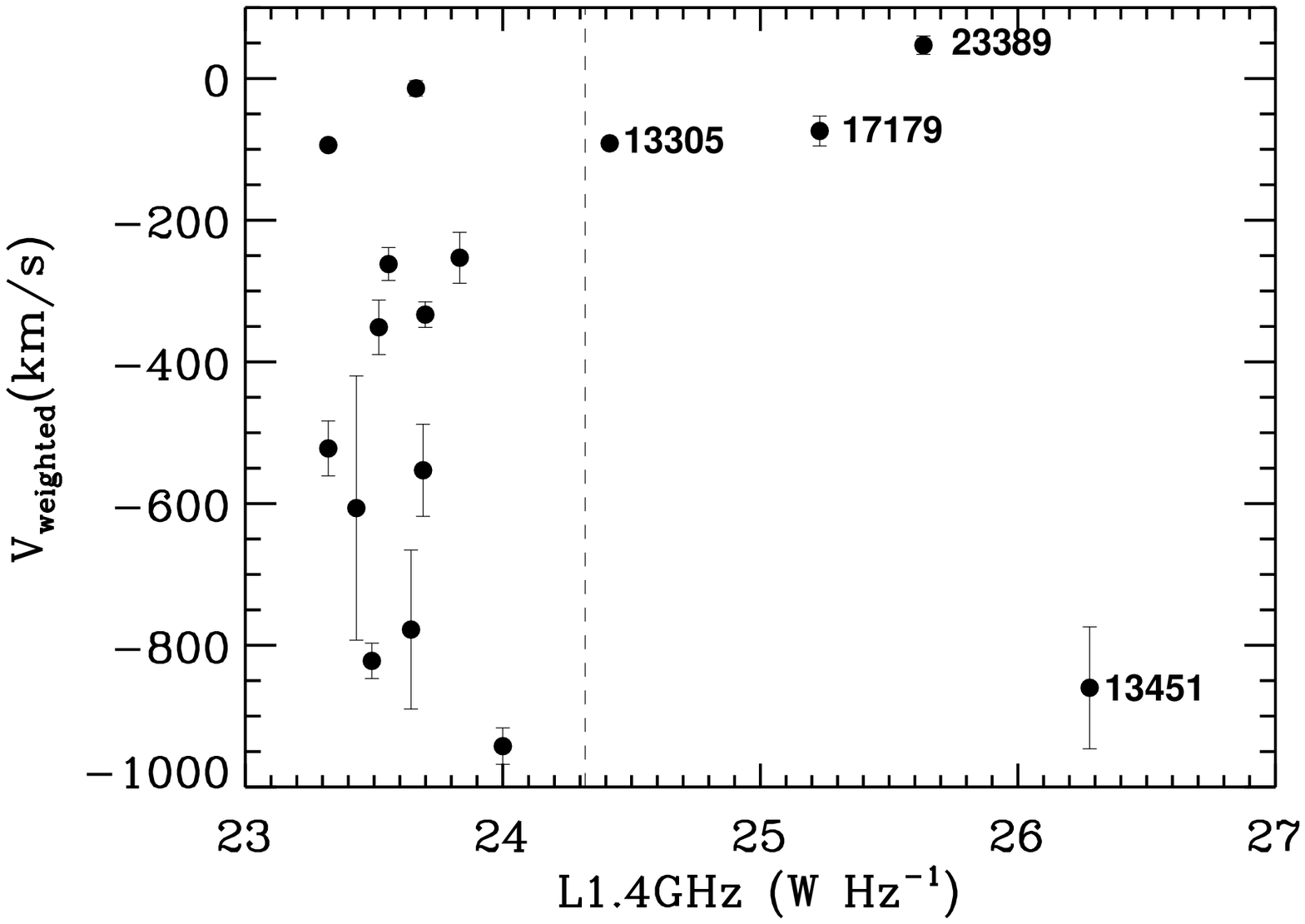,width=9cm}
\caption{Same as Figure \ref{FWHMvsLOIII} but with the 1.4 GHz monochromatic
  radio luminosities (L$_{\rm 1.4 GHz}$). The dashed line indicates the maximum
  1.4 GHz luminosity for the galaxies in our sample calculated using their
  L$_{\rm IR}$ along with the Condon et al. (1991) IR-radio luminosity relation
  for starburst galaxies. Four objects show an excess of radio emission for
  their corresponding L$_{\rm IR}$, which reveals the presence of a strong AGN
  component. These objects are indicated in the Figure using the first 5 digits
  of their designation in the IRAS Faint Source Catalogue Database. }
\label{FWHMvsL1.4}
\end{figure}

Finally, we can define the flux weighted mean line width and
velocity shift  of the disturbed kinematic components as:

\begin{equation} 
{FWHM_{\rm w}} = \frac{\sum_{\rm i} (F_{i} \times
  FWHM_{\rm i})}{\sum_{i} {F_{\rm i}}}
\end{equation}

\begin{equation} 
{\Delta{V}_{\rm w}} = \frac{\sum_{\rm i} (F_{\rm i} \times \Delta {V}_{
    \rm i})}{\sum_{\rm i} {\rm F_{\rm i}}}
\end{equation}

\noindent
where $F_{\rm i}$, $FWHM_{\rm i}$ and $\Delta V_{\rm i}$ are the fluxes, FWHM and velocity
shifts of all the intermediate and broad kinematic components with respect to
the rest frame. The results are shown in Table \ref{weighted}. For our sample of
16 ULIRGs, we find values in the range of 797 $\leq FWHM_{\rm w} \leq$ 2223~\kms
with a mean and median values of 1253 \kms and 1336 \kms respectively. In the
case of the $\Delta V_{\rm w}$ we find vaues in the range -942 $< \Delta V_{\rm
  w} < $ 47~\kms with mean and median values of -406 \kms and -333 \kms
respectively.

\begin{table}
\begin{tabular}{lllll}
\hline\hline
Object &$FWHM_{\rm w}$ & $\Delta V_{\rm w}$ &${R_{\rm [OIII]}}$\\
IRAS   &     &              & \\
       &\kms          &\kms    &(kpc)     \\       
(1)    &  (2)     &  (3)     &   (4)     \\   
\hline 
F00188--0856 & 904$\pm$79     & -253$\pm$36  &2.2 \\		
F01004--2237 & 1377$\pm$197    & -778$\pm$112  &3.5 \\
F12072--0444 & 886$\pm$100    & -351$\pm$38 &2.0 \\
F13305--1739 & 1368$\pm$49     & -92$\pm$5 &3.2 \\
F13428+5608 & 1368$\pm$66     & +14$\pm$11  &1.5 \\
F13451+1232 & 1454$\pm$145    & -860$\pm$86 &0.2 \\
F14394+5332 & 989$\pm$28     & -942$\pm$26 &1.4 \\
F15130--1958 & 1336$\pm$402    & -606$\pm$186 &1.7 \\
F15462--0450 & 1426$\pm$40     & -822$\pm$25 &1.4 \\
F16156+0146 & 1099$\pm$97     & -262$\pm$23 &2.4 \\
F17044+6720 & 1765$\pm$103    & -553$\pm$65 &3.1 \\
F17179+5444 & 941$\pm$116    & -74$\pm$16 &1.6 \\
F23060+0505 & 797$\pm$38     & -333$\pm$18 &1.8 \\
F23233+2817 & 1317$\pm$58     & -522$\pm$39 &1.4 \\
F23327+2913 & 804$\pm$14     & -94$\pm$4 &1.5 \\
F23389+0300 & 2223$\pm$30     & +47$\pm$13 &1.4\\
\end{tabular}
\caption{Weighted mean FWHM and velocity shifts ($FWHM_{\rm w}$ and $\Delta
  V_{\rm w}$) for the objects in our sample.  The final column gives the
  estimated radii of the [OIII] outflow regions. For most of the objects these
  have been estimated by fitting a Gaussian to the core of the
  [OIII]$\lambda\lambda$5007,4959 spatial profile along the slit, with the
  radius estimated as $R_{\rm [OIII]}=FWHM(kpc)/2$. However, in the case of
  F13451+1232 the radius was estimated from the narrow-band images published in
  \citet{batcheldor07}. }
\label{weighted}
\end{table}

Figure \ref{FWHMvsLOIII} shows $FWHM_{\rm w}$ and $\Delta V_{\rm w}$ plotted
against the total [OIII]$\lambda$5007 emission line luminosity ($L_{\rm
  [OIII]}$) from column 6 of Table 1. This figure is designed to investigate any
possible correlation between the properties of the outflows and the AGN power,
as indicated by the $L_{\rm [OIII]}$ luminosity, which is often considered to be
a reliable AGN bolometric indicator
\citep[e.g.][]{Dicken10,LaMassa09,LaMassa10}.  However, we do not find any clear
correlation between $L_{\rm [OIII]}$ and either the shifts or the widths of the
outflows observed at optical wavelengths, for the galaxies in our sample. Our
results suggest that the kinematic properties of the AGN-induced outflows
present in Sy-ULIRGs are not strongly dependent in the instantaneous AGN
radiative power as indicated by the [OIII] luminosity.  A possible reason for
this lack of a correlation is that the [OIII] luminosity represents the AGN
power on the timescale of the light crossing time of the narrow line region
($t_{\rm NLR} \la 10^4$~yr), but the AGN activity may be highly variable within
the $\sim10^8$~yr timescale of the final stages of gas-rich mergers represented
by the ULIRGs. In this case, the instantaneous [OIII] luminosity might not
provide a good indication of average AGN power over the entire activity
cycle. Alternatively, because of the substantial extinction on a kpc-scale in
the NLRs of ULIRGs, the [OIII] luminosity may not be as reliable an indicator of
the AGN bolometric luminosity as it is in other types of active
galaxies. Indeed, many ULIRGs show compelling evidence for AGN activity at
mid-IR wavelengths \citep{Veilleux09}, but no sign of such activity at optical
wavelengths. 

Unfortunately, estimating the reddening is challenging based on our existing
data. H$\alpha$ is in a blend, and there are potentially degeneracies involved
in the fits that may affect the accuracy of the line ratio measurements,
especially in cases with highly complex, multiple-component line
profiles. Nonetheless, we used the H$\alpha$/H$\beta$ ratios in Table 3 to
calculate a ``crude'' estimate of the de-redenned [OIII] luminosities. We then
used these de-reddened luminosities to reproduce Figure 2 and obtained identical
results, i.e. no trends and correlations are found. Furthermore, we carried out
the same excercise but using the [OIV]25.89 luminosities from \cite{Veilleux09},
which are a more reliable AGN luminosity indicator and are available for 9
ULIRGs in our sample. The results obtained are identical to those of the
previous attempt, i.e., no trends and correlations are found.

\subsection{The spatial scales of the kinematically disturbed regions}

Our analysis of the emission line profiles has concentrated on spatially
integrated, 5~kpc diameter aperture, data. However, in order to properly
quantify the outflows it is important to attempt to estimate their spatial
extents within the apertures.  Spatially resolved, kpc-scale emission line
outflows are found in many Seyfert galaxies, and directly detected in one low-z
Sy-ULIRG (Mrk231: see Rupke \& Veilleux 2011). Moreover, \cite{Alexander10}
and \cite{Harrison12} have recently found evidence for outflows on a scale of 4
-- 15~kpc in a sample of high-z ULIRGs. In contrast, HST emission line imaging
of the outflows in two radio-loud ULIRGs -- PKS1345+12 and PKS1549-79 -- shows
that they are significantly more compact, with radial extents of only $\sim$100
-- 200~pc \citep{batcheldor07}. Therefore, the radial extents of the AGN-induced
outflows in the Sy-ULIRGs in the local Universe are highly uncertain.

Based on visual inspection of our 2D long-slit data, we find that the dominant
broad/intermediate kinematic components are generally confined to the central
few pixels -- on a similar scale to the seeing FWHM estimated by the DIMM seeing
monitor at the time of the observations. More quantitatively, we have extracted
the continuum-subtracted spatial profiles of the [OIII] emission from the 2D
frames and fitted Gaussians to the central cores of the profiles. We show the
results in column 4 of Table 4, quantified as $FWHM/2$ in units of kpc. We find
that the [OIII] FWHM estimates are comparable with the DIMM seeing estimates at
the times of the observations for $\sim$56\% of the objects in our sample.
Moreover, in some objects the spatial profiles may be dominated by spatially
resolved narrow components. Therefore, in most cases the numbers presented in
column 4 of Table 4 are likely to represent upper limits on the true spatial
extents of the near-nuclear warm outflows.

Overall, based on our measurements of the spatial profiles of the [OIII]
emission in our sample of Sy-ULIRGs, and the direct HST imaging estimates of the
scales of the outflow regions in PKS1345+12 and PKS1549-79 \citep{batcheldor07},
it is likely that the radial extents of the near-nuclear outflows fall in the
range $0.1 < r < 3.5$~kpc.

\section{DISCUSSION}

\subsection{Are the outflows jet-driven?}

\cite{Spoon09b} found evidence for a correlation between the width of the mid-IR
     [NeIII] line at 15.56~$\mu$m and the 1.4~GHz radio luminosity ($L_{\rm
       1.4GHz}$), based on Spitzer IRS observations of a combined sample of
     ULIRGs and local Seyfert galaxies. The correlation is in the sense that
     some of the objects with the highest radio powers have the most extreme
     emission line kinematics, suggesting that the outflows might be jet-driven.
     Further evidence for jet-driven outflows in ULIRGs is provided by HST
     images which show that the outflow regions have similar scales to the
     relativistic radio jets in two of the most extreme radio-loud ULIRGs with
     high velocity outflows \citep{batcheldor07}.
 
In Figure \ref{FWHMvsL1.4} we show the weighted mean velocity shifts and FWHM
plotted against the radio powers listed in column 7 of Table 1. Note that an
important caveat with interpreting this plot in terms of jet-driven outflows is
that the radio emission does not necessarily have an AGN origin in the ULIRGs,
but may instead be associated with the prodigious starburst activity. The
vertical dashed line in \ref{FWHMvsL1.4} show the maximum radio power expected
for the starburst-related radio emission in the Sy-ULIRGs, as predicted on
from their measured infrared luminosities and the well-known correlation between
radio power and infrared luminosity for starburst galaxies
\citep{Condon91b}. From this it is clear that only four of the Sy-ULIRGs --
F13305-1739, F13451+1232, F17179+5444, F23389+03000 -- can be considered to have
genuinely radio-loud AGN with $L_{\rm 1.4GHz} > 10^{24.3}$~W Hz$^{-1}$.

On the basis of Figure \ref{FWHMvsL1.4} there is no evidence for a strong
correlation between radio power and emission line kinematics.  Although it is
certainly true that the two most radio-powerful objects in our sample --
F13451+1232 and F23389+03000 -- also display some of the most extreme emission
line kinematics, the other two radio-loud Sy-ULIRGs show relatively
quiescent kinematics.  Moreover, the radio-quiet majority of objects in our
sample ($L_{\rm 1.4GHz} \la 10^{24}$~W Hz$^{-1}$) encompass the same range of
kinematic disturbance as the radio-loud objects, and some radio-quiet objects
show emission line kinematics as extreme as those found in the radio-loud
objects, with [OIII] emission line profiles dominated by broad, shifted
components (e.g. F01004-2237, F14394+5332, F15130-1958, F15462-0450).  Therefore
we argue that, while the emission line outflows {\it can} be driven by the jets
in {\it some} Sy-ULIRGs, especially in cases where powerful relativistic
jets have the same spatial scales as the emission line structures
\citep[e.g. F13451+1232][]{Holt03,batcheldor07}, in most Sy-ULIRGs there is
no clear evidence for such jet-driven outflows.

\subsection{Ionization mechanisms}

\subsubsection{AGN and starburst photoionization}

\begin{figure*}
\begin{tabular}{cc}
\hspace{-1.0 cm}\psfig{figure=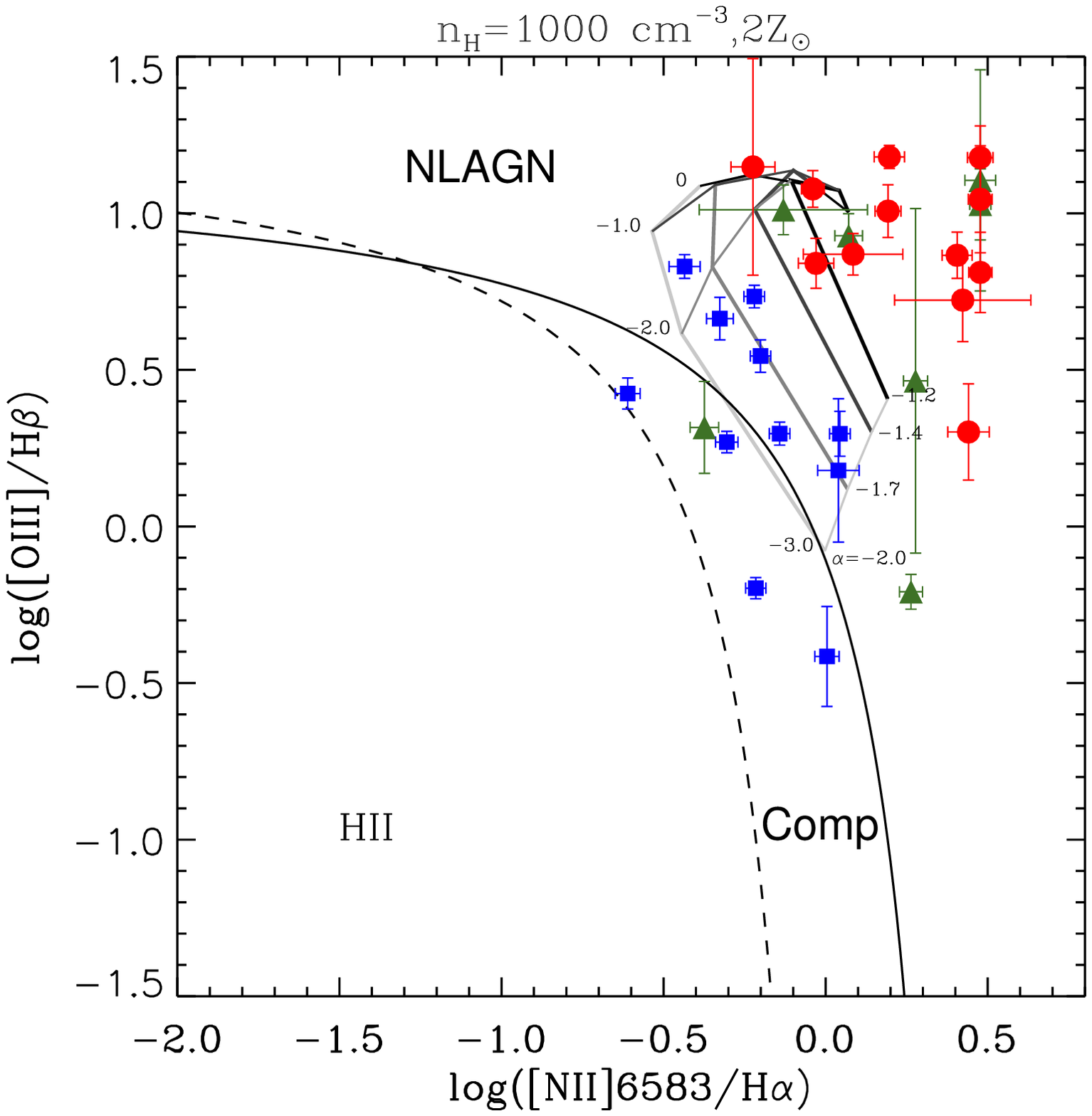,width=8.cm}&
\psfig{figure=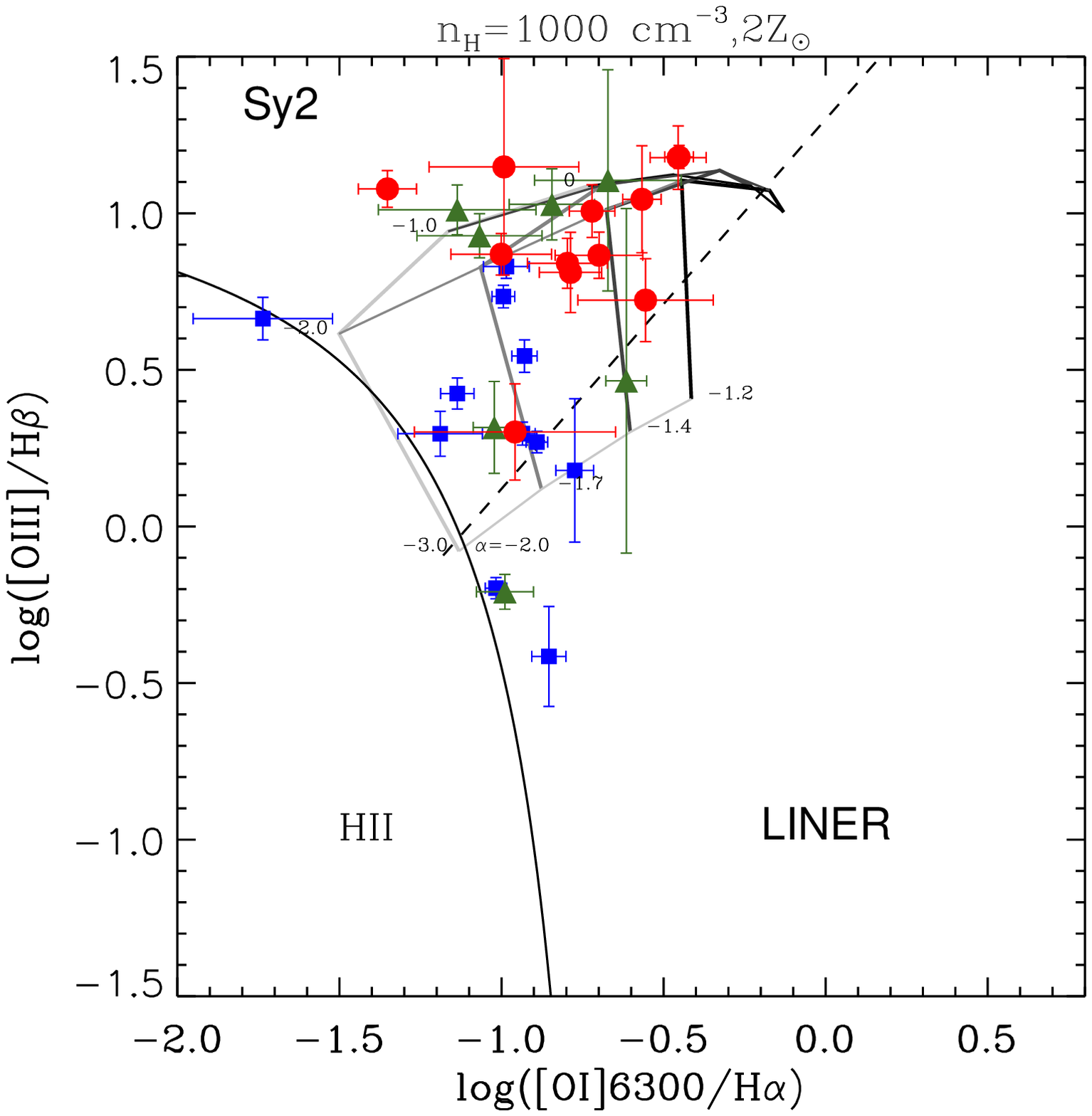,width=8.cm}\\
\hspace{-1.0 cm}\psfig{figure=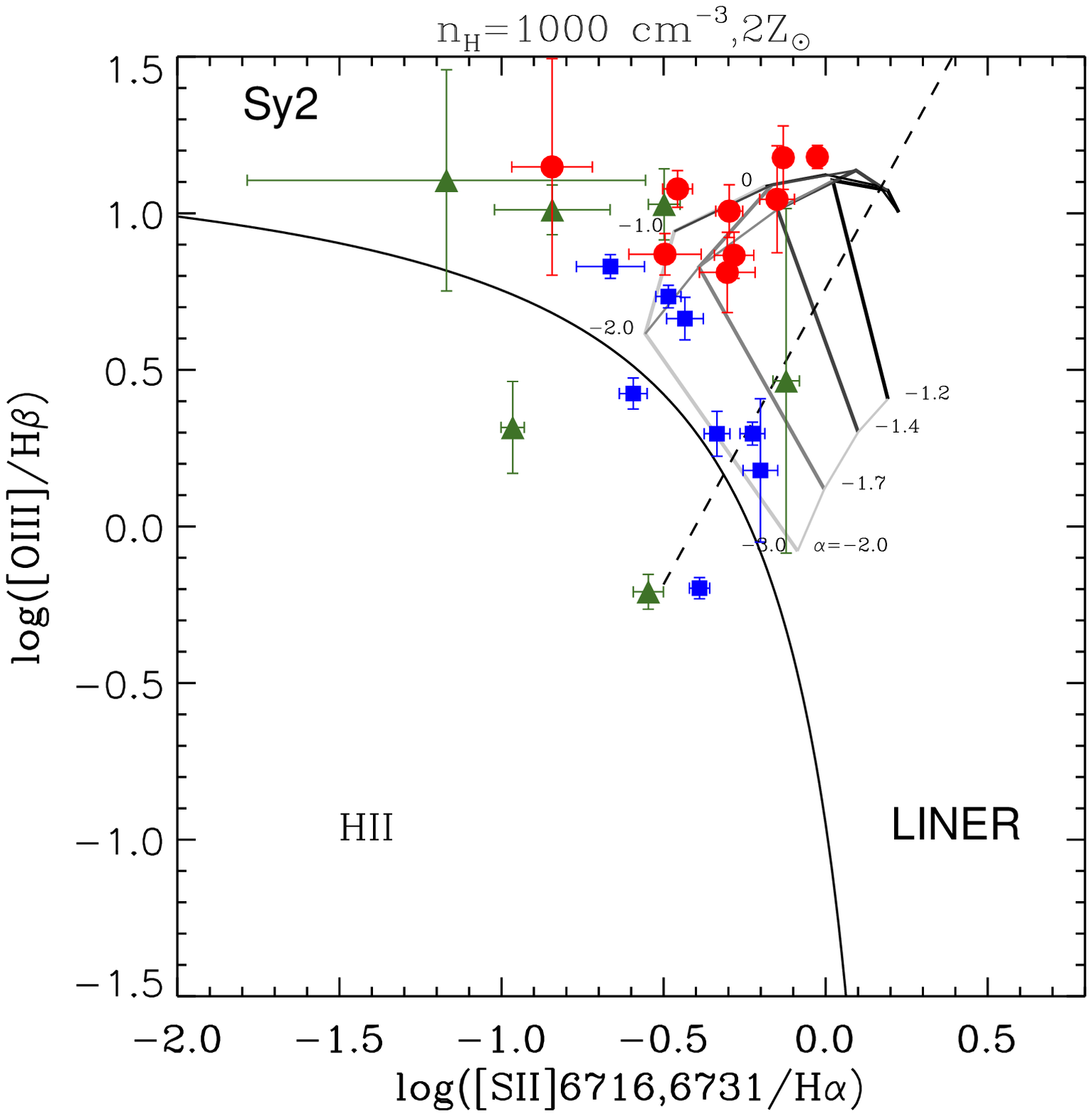,width=8.cm}&
\end{tabular}
\caption[]{Standard optical diagnostic diagrams showing the classification
  scheme by \cite{Kewley06}. The solid curves are the theoretical ``maximum
  starburst line'' derived by \cite{Kewley01} as an upper limit for star-forming
  galaxies and the empirical boundary lines between Seyfert 2 galaxies and
  LINERs. The dashed curve on the [NII] diagram is the \cite{Kauffmann03}
  semi-empirical lower boundary for the star forming galaxies. The
  \cite{Groves04a} grids of dusty, radiation pressure-dominated models are also
  plotted in the Figure. These grids have been generated assuming twice solar
  metallicity (2Z$_{\odot}$) and hydrogen density of n$_H$ = 1000
  cm$^{-3}$. Gridlines corresponding to five values of ionizing parameter ($U_0$
  = 0,-1,-2,-3,-4) and four values of power law index (F$_{\nu} \propto$
  ${\nu}^{\alpha}$, $\alpha$ = -1. 2, -1.4, -1.7, -2.0) are shown in the
  figure. To help the reader follow the gridlines these are grey-coded from
  ``light-grey'' to black, with light-grey and black corresponding to the lowest
  and highes values of $U_0$ and $\alpha$ respectively. Over-plotted on the
  diagrams are the results of our kinematic study. Blue squares are the line
  ratios corresponding to the narrow components, green triangles correspond to
  the intermediate component and red circles represent the broadest kinematic
  component.}
\label{Diagnostics}
\end{figure*}

It is well known that starburst and AGN activity can co-exist in the nuclear regions of
ULIRGs, both energy sources potentially contributing to the ionization the surrounding gas
\citep[e.g.][]{Genzel98,Farrah07,Armus07,Veilleux09}. Therefore  it is interesting to
determine whether there there is correlation between the emission line kinematics and 
the nature of the ionizing source(s). 

Figure \ref{Diagnostics} shows the standard, optical BPT diagnostic line ratio
diagrams \citep{Baldwin81}. The lines drawn in the diagrams correspond to the
the optical classification scheme of \cite{Kewley06}. Overplotted on the figure
are the results from our kinematic study. Blue squares, green triangles and red
circles correspond to the narrow, intermediate and broad components
respectively. The error bars have been calculated by combining the 5\%
uncertainty of the relative flux calibration in quadrature with the uncertainty
associated with the model fits.

It is notable the line ratios obtained for most, if not all, the ULIRGs in our
sample are consistent with a Sy2 classification. However, the figure
demonstrates a clear trend with emission line kinematics: whereas the line
ratios for the broad components (red circles) in most cases fall squarely in
Seyfert part of the diagram, suggesting AGN photoionization as the dominant
ionization mechanism, the line ratios for the narrow components (blue squares)
measured in many objects fall in, or close to, the transition zone between HII
region and Seyfert classifications -- more consistent with a combination of the
AGN and stellar photoionization; the ratios measured for the intermediate
components (green triangles) tend to fall in between those of the narrow and the
broad.

We emphasise that the trends in the emission line ratios with linewidth do not
necessarily provide evidence for an ionization stratification in the AGN outflow
regions themselves. Rather, they are naturally explained in terms of the varying
contribution of the starburst activity to the ionization of the gas associated
with each of the different kinematic components. For example, it is likely that
starburst activity plays an important role in ionizing the gas associated with
the narrow components in many of the objects. The importance of the starburst
activity as an ionizing source gradually decreases for the intermediate
component. Finally, in the case of the broadest kinematic component AGN activity
is the dominant source of ionization. Note that the association between the
ionization state of the gas and the line width provides further evidence that
the broader (generally blueshifted) components are driven by the AGN rather than
by the starburst components.

With the aim of better understanding the nature of ionization mechanisms
responsible for the measured line ratios, Figure \ref{Diagnostics} also shows
the \cite{Groves04a} dusty, radiation pressure-dominated photoionization models
for NLG in AGN. For these models we assume a hydrogen density n$_H$ = 1000
cm$^{-3}$ (the only hydrogen density used in the Groves et al. 2004a paper) and
twice solar abundance (2Z$_{\odot}$). This later value was found by
\citet{Groves04b} to provide a better match to the line ratios measured in
Seyfert NLR \citet{Groves04b}. A grid of models for various values of the
ionizing parameter ($U_0$ = 0, -1, -2, -3, -4) and ionizing continuum SED
power-law indicies (F$_{\nu} \propto$ ${\nu}^{\alpha}$, $\alpha$ = -1.2, -1.4,
-1.7, -2.0) is shown in the figure.

\begin{figure}
\psfig{figure=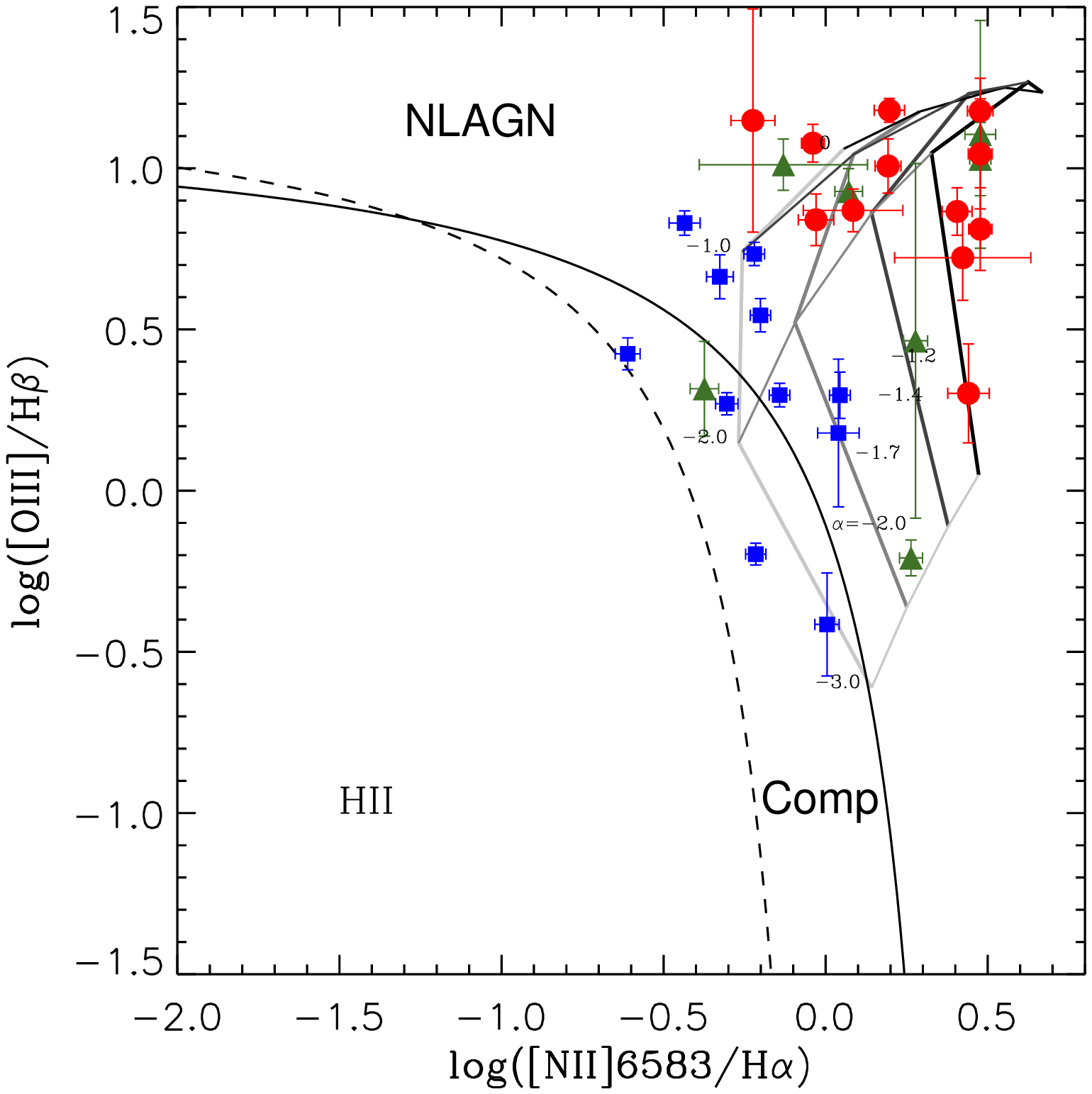,width=8.cm}
\caption[]{Same as Figure \ref{Diagnostics} but only for the [NII]6583/H$\alpha$
  diagram and for 4Z$_{\odot}$. The other two diagrams remain relatively
  unchanged by the increase in metallicity and therefore are not shown in the
  figure.}
\label{Diagnostics_2}
\end{figure}

At first sight it is apparent that, with few exceptions, the line ratios of the
intermediate and the broad kinematic components fall within the grids of models
in the case of the diagram involving [OI]/H$\alpha$. However, these models fail
to reproduce most of the line ratios for such components in the case of the
[NII]/H$\alpha$ diagram (and to a lesser extent in the case of the
[SII]/H$\alpha$ diagram). A possible explanation for this finding is that the
emission line regions are more metal rich than we have assumed. In this context
Figure \ref{Diagnostics_2} shows the [OIII]5007/H$\beta$ vs [NII]6583/H$\alpha$
diagram assuming n$_H$ = 1000 cm$^{-3}$ and 4Z$_{\odot}$ (the other two diagrams
remain relatively unchanged after increasing the metallicity and therefore are
not shown in the figure). It is clear from the figure that a 4Z$_{\odot}$
abundance certainly improves the match between the models and our
data. Therefore, the line ratios derived for the Sy-ULIRGs in our sample are
consistent with gas of super-solar abundances that has been photoionized by a
combination of starburst and AGN components, with the AGN photoionization
dominating for the broad/shifted kinematic components.
 
\subsubsection{Shock ionization}

As well as photoionization by AGN and starburst activity, it is possible that
line ratios plotted in Figure \ref{Diagnostics} can also be explained in terms
of ionization by fast radiative shocks
\citep[][]{Dopita95,Dopita96,Groves04a,Allen08}. The shock models predict a
series of line ratios for a range of magnetic field ($B$), electron densities
(n$_e$), abundances and shock velocities ($v_s$). In addition, the predicted
line ratios depend on the geometry of the shock, i.e. the presence of a
photoionized precursor of the shock \citep[see][for a detail
  discussion]{Allen08}.

\begin{figure*}
\begin{tabular}{cc}
\hspace{-1.0 cm}\psfig{figure=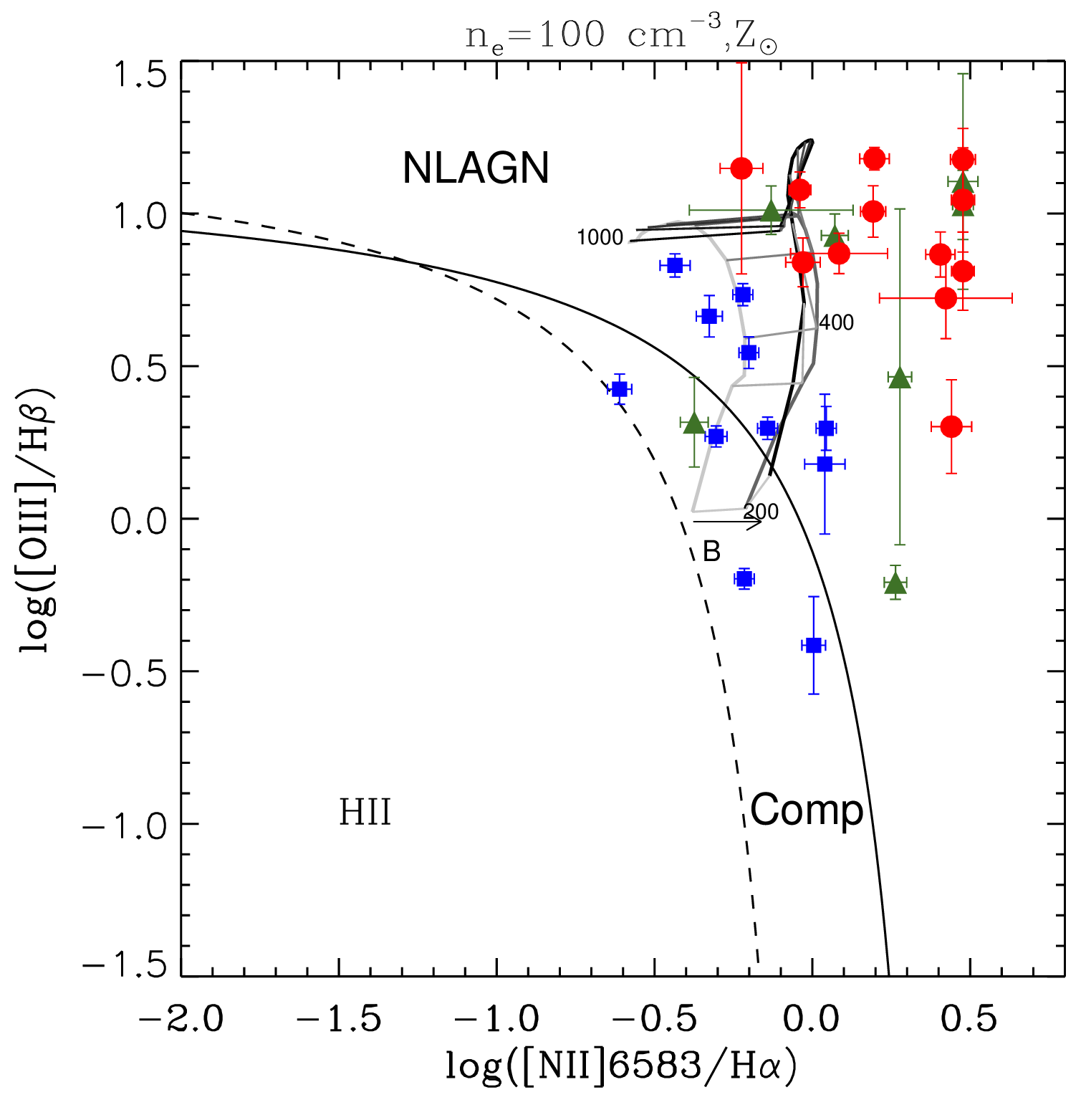,width=8.cm}&
\psfig{figure=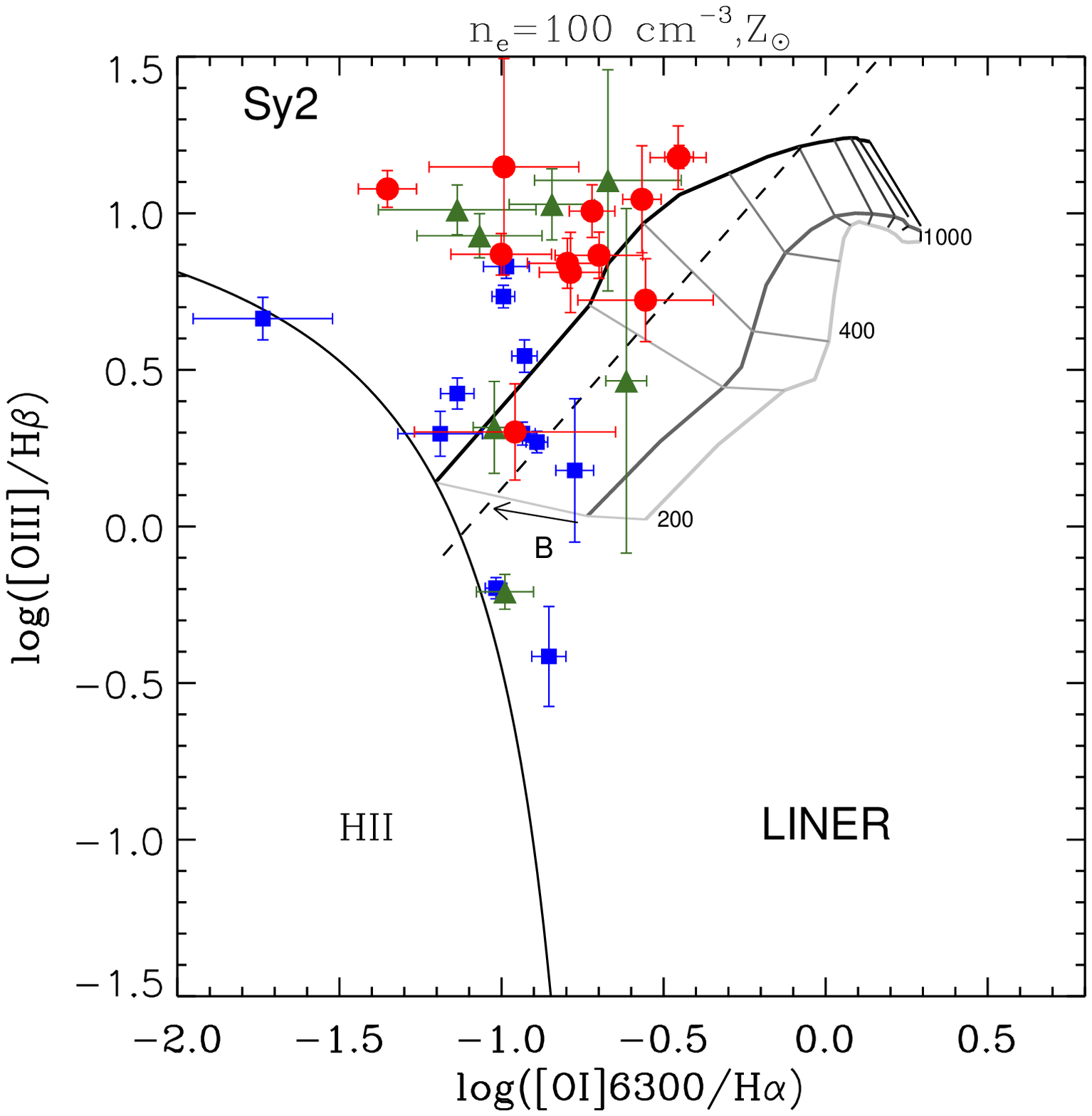,width=8.cm}\\
\hspace{-1.0 cm}\psfig{figure=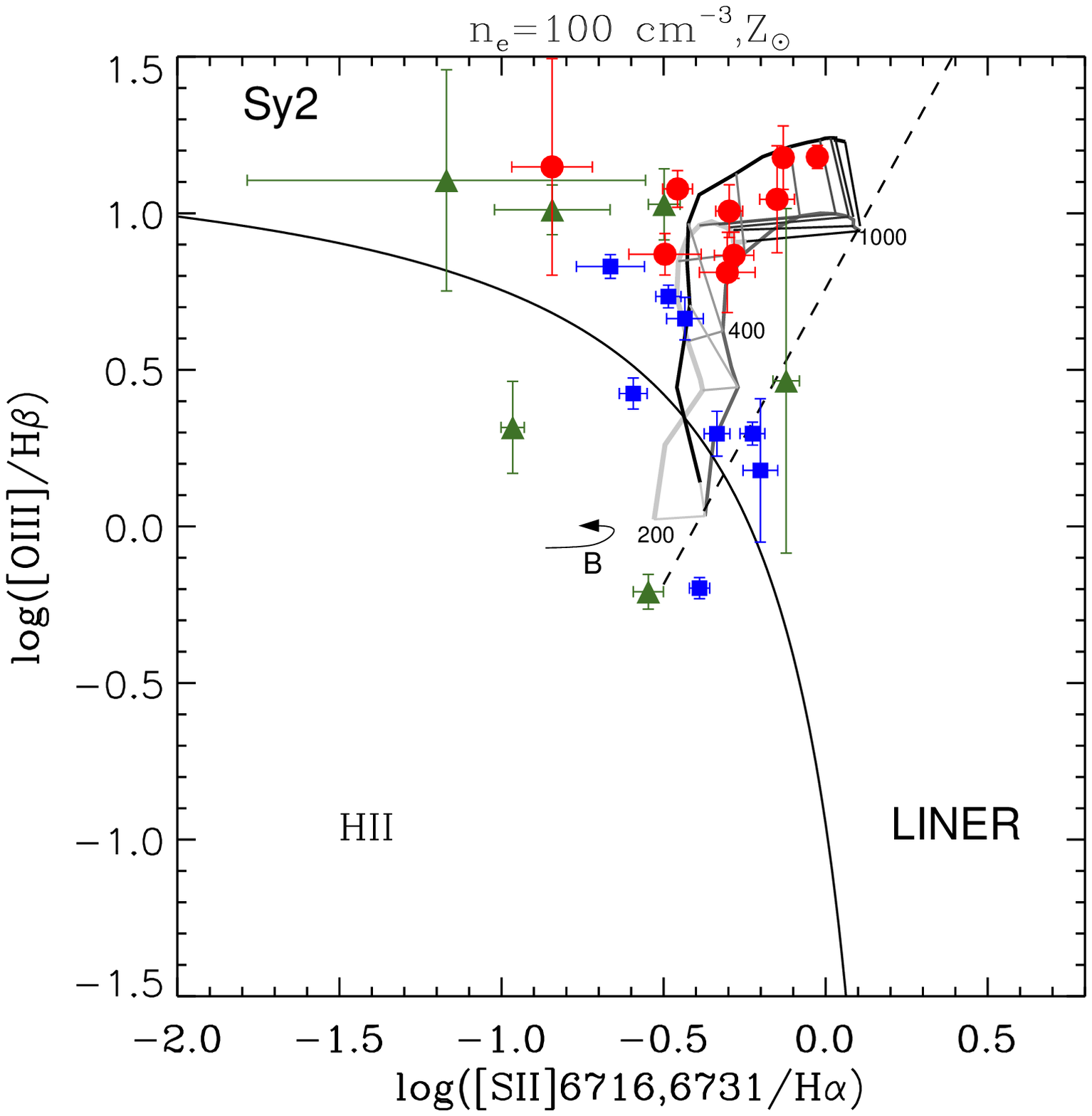,width=8.cm}&
\end{tabular}
\caption[]{Same as Figure \ref{Diagnostics} but showing the \cite{Allen08} grids
  of shock-ionization models. These grids have been generated assuming a
  shock+precursor model with solar metallicity and preshock density of $n_e$=10
  cm$^{-3}$. Gridlines corresponding to 3 values of magnetic field ($B$ = 1, 10
  and 100 $\mu$G) and 9 shock velocity values ($v_s$ = 200, 300, 400, 500, 600,
  700, 800, 900 and 1000 km s$^{-1}$) are shown in the figure. The increasing
  direction of $B$ is indicated with an arrow. In the case of the
  [SII]/H$\alpha$ diagram $B$ increases first to the right and then to left
  which is indicated with a curved arrow. $v_s$ increases always from bottom to
  top with 3 values (200, 400 and 1000 km s$^{-1}$). As in the case of Figure
  \ref{Diagnostics}, the gridlines these are grey-coded from ``light-grey'' to
  black, with light-grey and black corresponding to the lowest and highest values
  of $B$ and $v_s$ respectively.}
\label{Diagnostics_shock}
\end{figure*}

Figure \ref{Diagnostics_shock} shows the \cite{Allen08} models with solar
metallicity (Z$_{\odot}$) and a pre-shock density of 100 cm$^{-3}$. Based on the
results in the previous section, it might be more appropriate to use models with
higher metallicities. However, the highest metallicity used in the
\cite{Allen08} is 2Z$_{\odot}$, and for this value only $n_e$ values of 1 and
1000 are considered. Therefore, we decided to use Z$_{\odot}$, models since they
sample the parameter space better.

One of the largest uncertainties in these models is pre-shock density. Assuming
that we measure the density of the compressed post-shock gas in the
kinematically disturbed emission line components that we detect in the ULIRGs in
our sample, and the measured electron densities of these components are ~1000
cm$^{-3}$ (see section 4.3), then the pre-shock densities could be as low as 10
-- 100 cm$^{-3}$, since the compression factor in the cooled, post-shock gas can
be high \citep[$\sim$10-100, see][]{Dopita95}. Therefore, we have assumed a
pre-shock density of 100 cm$^{-3}$. Finally, \citet{Dopita95,Allen08} found that
the shock+precursor models produce a better fit to the line ratios measured for
their samples of Sy2 and luminous IR galaxies. For this reason, we decided to
use such models for the work presented in this section. The potential effects of
changing these assumptions is discussed later in the section. The gridlines
correspond to 3 values of $B$ from 1-100 $\mu$G (1, 10 and 100 $\mu$G), which
are appropriate for starburst galaxies \citep{Thompson06}, and 9 values of $v_s$
(200, 300, 400, 500, 600, 700, 800, 900 and 1000 km s$^{-1}$). The direction in
which $B$ increases is indicated in the figure with an arrow, while $v_s$
increases from bottom to top with 3 values (200, 400 and 1000 km s$^{-1}$)
indicated in the figure.

As seen in the figure, most of the measured line ratio values do not fall within
the region covered by the grids of shock models; the lack of agreement between
the models and the data is particularly apparent in the diagnostic diagram that
involves the [OI]/H$\alpha$ ratio. However, the models have a relatively large
number of free parameters ($B$, $v_s$, $n_e$ and abundance) plus the additional
possibility of including or otherwise the shock precursor component. Therefore,
it is possible to cover wide ranges of line ratio values using different
assumptions. For example, using twice solar abundance models with a pre-shock
density of 1000 cm$^{-3}$ instead of 100 cm$^{-3}$ substantially extends the
gridlines in the 3 diagrams \citep[see][for details]{Allen08}, providing a
better match to the measured line ratios. We note that the line ratio values
measured for some of the broad kinematic components fall outside the
\cite{Allen08} grids of shock models, or are only reproduced by the most extreme
set of parameter values (e.g. $n_e$ = 1000 cm$^{-3}$ and $B$ = 1000 $\mu$G).

Overall, we find that the line ratios measured for our sample of Sy-ULIRGs are,
in general, better reproduced by photoionization models that include a combination
AGN and starburst photoionization and in which the emitting gas highly enriched in
metals. However, as pointed out by Dopita and collaborators
\citep[e.g.][]{Dopita95,Dopita96,Groves04b,Allen08}, it is not always possible to
decisively distinguish between shock ionization and AGN photoionization 
using only the stronger optical emission lines included in the
diagnostic plots of Figure 6.

\subsection{Physical conditions and reddening}

Potentially, further clues to the nature of the emission line outflows may be
obtained by considering the reddening and densities of the emission line
components.  For example, \citep{Holt03,Holt11} found evidence that the broadest
and most highly shifted emission line components in F13451+1232 (PKS1345+12) are
also the most highly reddened, suggesting that the outflows are stratified and
decelerating, with the fastest components associated with most highly reddened
regions closest to the driving AGN. Similarly, if the most highly disturbed
kinematic components are truly associated with the regions closest to the AGN,
we might expect them to show evidence for higher densities than the less
kinematically disturbed components.

On the outset, it is important to emphasize that estimating the reddening and
densities for the broad, shifted kinematic components is challenging based on
our existing data, because some of the key emission lines (e.g. H$\alpha$,
[SII]$\lambda\lambda$6717,6731) are in blends, and there are potentially
degeneracies involved in the fits that may affect the accuracy of the line ratio
measurements, especially in cases with highly complex, multiple-component line
profiles (see discussion in section 3.1).

To search for reddening gradients, we use the $H{\alpha}/H{\beta}$ ratio,
which, for typical NLR conditions and AGN photoionizing continuum shapes, is
expected to have a value in the range $2.8 < H\alpha/H\beta < 3.1$ in the
absence of reddening \citep{Gaskell84,Osterbrock89}.  Column 9 in Table 3 gives the
$H{\alpha}/H{\beta}$ line ratio values for the various kinematic components of
the galaxies in our sample. At first sight, the modelling results appear to show
evidence for a trend of increasing reddening with linewidth in four of the ULIRG
Seyferts (F13451+1332, F14349+5332, F17179+544, F23060+0505). However, two objects
apparently show the reverse trend (F00188-0856, F15462-0450), and in most cases
there is no clear trend. Clearly, given the large uncertainties in the measured
ratios --- to some extent reflecting the degeneracies in the fits --- the
results are inconclusive.

The results are no more decisive when it comes to estimating the densities using
the [SII](6717/6731) ratio (see the discussion of individual objects in the
Appendix). While it has proved possible to derive accurate densities for the
narrow emission line components in some objects (typical electron densities:
$2.4\times10^2 < n_e < 3.6\times10^3$~cm$^{-3}$), we have failed to measure
accurate densities for the broader emission line components in any
object. However, in the cases of F13305-1739, F13428+5608, F15462-0405 and
F23233+2817, it has been possible to derive lower limits on the electron
densities of the broad components of 9$\times$10$^{3}$, 6$\times$10$^{3}$,
4$\times$10$^{3}$ and 4$\times$10$^{3}$~cm$^{-3}$ respectively.

\subsection{Mass outflow rates and energetics}

AGN feedback is now routinely incorporated into numerical simulations of the
hierarchical growth of galaxies through major galaxy mergers. To reproduce the
correlations between the masses of the black holes and the bulge properties
\citep[e.g.][]{Silk98,diMatteo05} some of these models require that a relatively
large fraction of the available accretion power of the quasars is thermally
coupled to the circumnuclear ISM \citep[$\sim$ 5 -- 10\%, e.g.][but see
  Hopkins et al. 2010]{Fabian99,diMatteo05}.

As described in the introduction, ULIRGs represent just the situation modelled
in many of the merger simulations that include AGN feedback. In order to
investigate whether our observational results are consistent with those of the
merger simulations, it is important to quantify the mass outflow rates and the
kinetic powers of the outflows.

For a spherical outflow, the mass of the ionized gas in the outflow is given by:
\begin{equation} 
M  = n_e m_p V \epsilon 
\end{equation}
\noindent 
where $n_{\rm e}$ is the electron density, $m_p$ the mass of the proton, V is
the total volume and $\epsilon$ is the gas filling factor
\citep{Osterbrock06}. Therefore, for a steady state outflow, the mass outflow
rate ($\dot{M}$) is: 
\begin{equation} 
\dot{M} = n_e m_p v_{out} A \epsilon
\end{equation}
\noindent 
where $v_{out}$ is the velocity of the outflow and $A$ is surface area of the
outflowing region. Also, filling factor $\epsilon$ is related to the H$\beta$
luminosity, the electron density $n_e$ and the emitting volume V by:
\begin{equation} 
\epsilon = \frac{L({\rm H}\beta)}{\alpha_{\rm{H}\beta}^{eff} h\nu_{\rm{H}\beta}
  n_e^2 V}
\end{equation}
\noindent 
hence, the mass outflow rate is related to the H$\beta$ luminosity by:
\begin{equation} 
\dot{M}  = \frac{L({\rm H}\beta) m_p v_{out} A}{\alpha_{\rm{H}\beta}^{eff}
  h\nu_{\rm{H}\beta} n_e V}.
\end{equation}
\noindent
For a spherical outflow geometry this reduces to:
\begin{equation} 
\dot{M}  = \frac{3L({\rm H}\beta) m_p v_{out}}{\alpha_{\rm{H}\beta}^{eff}
  h\nu_{\rm{H}\beta} n_e r} 
\end{equation}
\noindent
where r $is$ the radius of the spherical volume. In addition, the kinetic
power of the outflow ($\dot{E}$) is related to the velocity dispersion ($\sigma
\approx$ FWHM/2.355), mass outflow rate and outflow velocity by:
\begin{equation} 
\dot{\rm E} = \frac{\dot{M}}{2}{(V_{out}^2 +
  3\sigma^2)} .
\end{equation}

The main uncertainty in calculating the mass outflow rates, and hence the
kinetic powers of the outflows, is related the electron density. Unfortunately,
as we saw in the last section, due to degeneracies in the fits to the [SII]
lines, the densities are relatively unconstrained for the broad, shifted
components in most of the objects in our sample. In this case, the densities
could range from the relatively low values typically determined for the
spatially-resolved NLR in some objects ($n_{\rm e} \sim 100$ cm$^{-3}$: Taylor
et al. 2003, Robinson et al. 2000) up to the much higher densities directly
measured for the outflow components in F13451+1232 using the transauroral line
ratios \citep[$2\times10^4 < n_{\rm e} < 5\times10^5$ cm$^{-3}$:][]{Holt11} ---
a range of almost four orders of magnitude. Other sources of uncertainty include
the H$\beta$ luminosities ($L_{\rm H\beta}$), which are affected by uncertain
reddening corrections (see the last section), and the uncertain radii ($r$) of
the outflows (see section 3.4). Each of these adds a least a further order of
magnitude uncertainty when calculating the properties of the outflows.

In order to gain an impression of the range of possible mass outflow rates and
kinetic powers for the warm outflows in the Sy-ULIRGs, we have calculated upper
and lower limits on these quantities, based on the emission line kinematics from
Table 2 and assuming that $\Delta V = V_{out}$ for each of the kinematic
components. To derive the lower limits we have used the high density estimated
by \citet{Holt11} for the broadest component in F13451+1232 ($n_{\rm e} =
5\times10^5$~cm$^{-3}$), the upper radius limits given in column 4 in Table
\ref{weighted}, and H$\beta$ luminosities that have not been corrected for
reddening; whereas for the upper limits we have assumed a density of 100
cm$^{-3}$, a radius of 0.1~kpc, and the reddening-corrected H$\beta$
luminosities. Note that low densities, small radii and high H$\beta$
luminosities favour highly massive and energetic outflows. In the case of
F13451+1232, the density, reddening corrected H$\beta$ luminosity, and radius of
the outflow are know relatively accurately from the studies of
\cite{batcheldor07} and \cite{Holt11}; for this object we have used the results
from the latter studies to calculate the outflow properties.

\begin{table*}
\centering
\begin{tabular}{llllllllll}
\hline\hline
Sample  &Selection criteria &Redshift range &L$_{\rm [OIII]}$ range (W)&
Resolution (km s$^{-1}$)\\
\hline
Sy-ULIRGs (this paper) &S$_{60}>$1 Jy, L$_{IR}>10^{12}$ L$_{\odot}$, &$z < 0.175$ &$2\times10^{33}$ -- $4\times10^{35}$ &$\sim$230\\
&emission line ratios & & \\
\cite{Heckman81}& representative of & z$<$0.076 & 3.5$\times$10$^{32}$ -- 1.7$\times$10$^{35}$ & $\sim$130\\
& local Seyferts & &\\
\cite{Whittle85a}& representative of & z$<$ 0.066 & 2.3$\times$10$^{32}$ -- 3.3$\times$10$^{35}$ & 40$-$80\\
& local Seyferts & & \\
\cite{Veilleux91a} &  $F[OIII] \geq$ 1$\times$10$^{-13}$ & z$<$0.028 & 5.8$\times$10$^{32}$ -- 1.8$\times$10$^{35}$ & $\sim$10  \\
& erg s$^{-1}$cm$^{-2}$& & \\
\cite{Nelson95}& representative of & z$<$ 0.043 & 2.1$\times$10$^{31}$ -- 1.7$\times$10$^{35}$ & 80$-$230\\
& local Seyferts &&\\
\end{tabular}
\caption []{Details of the properties of the local Seyfert comparison
  samples. The [OIII]$\lambda$5007 flux values for the sources within these
  comparison samples have been taken from: \cite{Dahari88}, \cite{Gu06},
  \cite{Moustakas06}, \cite{Massaro09},\cite{Tremblay09}, \cite{Gonzalez09},
  \cite{LaMassa10}, \cite{Greene10} and Marvin Rose (private communication).}
\label{comp_samples_a}
\end{table*}

As expected, with the exception of F13451+1232, the estimates of the outflow
properties are highly uncertain, and encompass several orders of magnitude in
both mass outflow rate and kinematic power for individual kinematic components
in individual objects. Across our sample we find upper limiting mass outflow
rates in the range $1 < \dot{M} < 1.5\times10^3$~M$_{\odot}$ yr$^{-1}$ and upper
limiting kinetic powers in the range $5\times10^{41} < \dot{E} <
2\times10^{45}$~erg s$^{-1}$.  To put these numbers into context, we can compare
them with the bolometric radiative powers of the AGN, which provide an
indication of the available accretion power of the material being accreted onto
the super-massive black holes. We calculate the bolometric luminosities of the
individual AGN by multiplying the infrared luminosities from Table 1 by the
fractional contributions of the AGN to the infrared light from mid-IR
spectroscopic study of the QUEST sample by \cite{Veilleux09}\footnote{Note that,
  for objects in our sample that were not included in the \cite{Veilleux09}
  QUEST sample, we assume that the AGN contribute 50\% of the infrared light --
  a typical value for the QUEST sample objects.}, obtaining values in the range
$2\times10^{45} < L_{\rm BOL}^{\rm AGN} < 6\times10^{45}$ erg s$^{-1}$.
Therefore, considering the extreme high end of the upper limits, the warm
outflows in Sy-ULIRGs can have kinetic powers that are comparable with the
radiative luminosities of the AGN, However, at the lower end of the range of
upper limits -- including F13451+1232 with its more precise $\dot{E}$ estimates
\citep[see][]{Holt11} -- the kinetic powers of the outflows are only a small
fraction of the bolometric luminosities ($<$0.5\%) --- well below the levels
required by some of the models \citep[e.g.][]{Silk98,Fabian99,diMatteo05}, but
perhaps consistent with the two stage feedback mechanism described by
\cite{Hopkins10}.

The discrepancy with the models grows even stronger if we consider the range of
lower limiting mass outflow rates and kinetic powers encompassed by our sample:
$4\times10^{-5} < \dot{M} < 4\times10^{-3}$~M$_{\odot}$ yr$^{-1}$ and
$4\times10^{35} < \dot{E} < 2\times10^{39}$~erg s$^{-1}$.  If these values
proved correct, the AGN outflows would have a negligible impact on the overall
evolution of their host galaxies.

\subsection{Comparison with the emission line kinematics of other samples of Seyfert
galaxies and other types of AGNs}

The hydrodynamical models predict that the AGN-induced outflows should be
particularly strong in the final stages of the type of major, gas rich mergers
represented by ULIRGs. Therefore it is interesting to consider whether the
degree of kinematic disturbance in ULIRGs with optical Seyfert nuclei is unusual
compared with samples of non-ULIRG AGN.

\subsubsection{Comparison with previous kinematic studies of the NLR in Seyfert galaxies} 

For our first comparison, we use the \cite{Heckman81},\cite{Whittle85a},
\cite{Nelson95} and \cite{Veilleux91a} studies of the narrow emission ine
kinematics in samples of local Seyfert and radio galaxies. In general, although
these samples are likely to be representative of nearby Seyfert galaxies, their
selection criteria are not well defined.  The exception is the
\cite{Veilleux91a} sample, for which the objects have $\delta \gsim$
-25$^{\circ}$ and F([OIII]\la5007) $\geq$ 1$\times$10$^{-13}$erg s$^{-1}$
cm$^{-2}$ \citep[see][for a discussion of the potential biases due to their
  selection criteria]{Veilleux91a}. The general properties of these samples are
compared with those of the Sy-ULIRGs in Table \ref{comp_samples_a}. We note
that, although these samples encompass a similar range of [OIII] emission line
luminosity to our Sy-ULIRG sample, they tend to contain a higher proportion of
lower luminosity objects.
 
The previous studies of nearby Seyfert galaxies investigated the kinematics of the NLR 
using a parameter scheme based
on cuts across the profiles at different heights. We will concentrate here in the
so called asymmetry ($AI20$) and kurtosis ($K1$) parameters as defined by
\cite{Heckman81} and \cite{Whittle85a} as follows:
\begin{equation} 
AI20  = \frac{WL20 - WR20}{WL20 +WR20}
\end{equation}
\noindent 
where $WL20$ and $WR20$ are the widths to the right and left of the line center
defined at the 80\% intensity level ($C80$), and:
\begin{equation} 
K1  = 1.524\times\frac{W50}{W20} 
\end{equation}
\noindent 
where $W50$ and $W20$ are the widths at 50\% and 20\% intensity level (i.e $W50
\equiv FWHM$ of the line). The kurtosis $K1$ parameter is defined such that for a
Gaussian $K1=1.0$. Line profiles with $K1$ values lower or higher than 1 are
called platykurtic (``peaky'') or leptokurtic (``stubby'') respectively.

Table \ref{EL_parameters_tab} shows the values of $AI20$, $K1$, $W20$ and $W50$
for the Sy-ULIRGs in our sample. Since the previous studies considered here
concentrate mostly on the [OIII]$\lambda$5007 emission line, the values in the
table have been estimated using that line. When more than one kinematic narrow
component is clearly visible in the emission line profile of a source (e.g. IRAS
F14394+5332), the wavelength of peak of the intensity (i.e. $C80$) is defined
using the kinematic component at rest frame. In addition, in the case of the Sy1
IRAS F15130-1958, we used our modelling results to subtract the FeII emission
prior to calculating the parameters.

\begin{table}
\centering
\begin{tabular}{llcccccc}
\hline\hline
Object & W20 & W50 & AI20 & K1\\
IRAS   & (km s$^{-1}$) & (km s$^{-1}$) &&\\
(1)    & (2)& (3) & (4) & (5)\\   
\hline 
F00188-0856 &   1207$\pm$32  & 509$\pm$56  &	 0.24$\pm$0.05  & 0.64$\pm$0.10\\
F01004-2237 &   1779$\pm$30  & 249$\pm$42  &	 0.71$\pm$0.03  & 0.21$\pm$0.09\\
F12072-0444 &   1002$\pm$30  & 544$\pm$36  &	 0.31$\pm$0.04  & 0.83$\pm$0.14\\
F12112+0305 &   743$\pm$18   & 388$\pm$36  &	-0.20$\pm$0.06  & 0.79$\pm$0.09\\
F13305-1739 &   1786$\pm$31  & 1072$\pm$29 &     0.11$\pm$0.02  & 0.91$\pm$0.05\\
F13428+5608 &   949$\pm$36   & 540$\pm$36  &     0.00$\pm$0.04  & 0.88$\pm$0.05\\
F13451+1232 &   706$\pm$31   & 1553$\pm$36 &	 0.34$\pm$0.02  & 0.68$\pm$0.03\\
F14394+5332 &   2091$\pm$42  & 1488$\pm$30 &     0.65$\pm$0.04  & 1.08$\pm$0.05\\
F15130-1958 &   2029$\pm$30  & 1156$\pm$42 &     0.28$\pm$0.05  & 0.87$\pm$0.14\\
F15462-0450 &   2208$\pm$46  & 1659$\pm$40 &     0.68$\pm$0.05  & 1.14$\pm$0.09\\
F16156+0146 &   989$\pm$29   & 297$\pm$30  &	 0.36$\pm$0.03  & 0.46$\pm$0.13\\
F17044+6720 &   640$\pm$21   & 296$\pm$29  &	 0.04$\pm$0.03  & 0.70$\pm$0.03\\
F17179+5444 &   1202$\pm$29  & 645$\pm$29  &	 0.07$\pm$0.03  & 0.82$\pm$0.06\\
F23060+0505 &   1315$\pm$32  & 721$\pm$30  &	 0.38$\pm$0.03  & 0.83$\pm$0.10\\
F23233+2817 &   1459$\pm$34  & 834$\pm$29  &	 0.31$\pm$0.05  & 0.87$\pm$0.14\\
F23327+2913 &   1151$\pm$40  & 639$\pm$40  &	 0.27$\pm$0.02  & 0.84$\pm$0.11\\
F23389+0300 &   2294$\pm$36  & 571$\pm$29  &	 0.07$\pm$0.01  & 0.38$\pm$0.05\\
\end{tabular} 
\caption{The emission line parameters for the Sy-ULIRSGs in our sample. Col (1):
  galaxy name. Col (2): width at 20\% of the peak intensity level. Col (5):
  width at 50\% of the peak intensity level. Col (4): the asymmetry $AI20$. Col
  (5): the kurtosis parameter $K1$. All these parameter values have been
  quadratically corrected for the instrumental profile.}
\label{EL_parameters_tab}
\end{table}

\begin{figure}
\psfig{figure=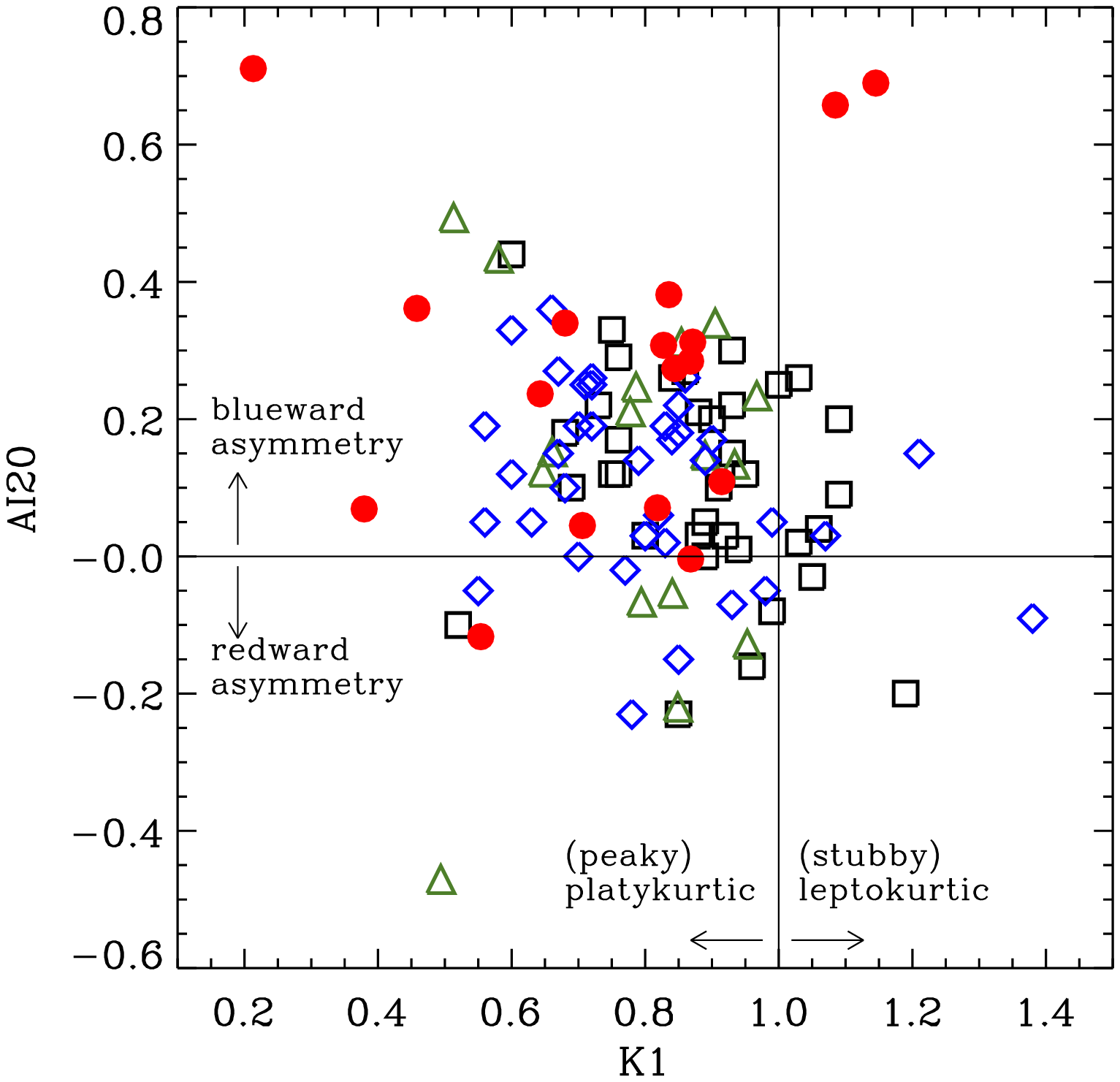,width=8.cm}
\psfig{figure=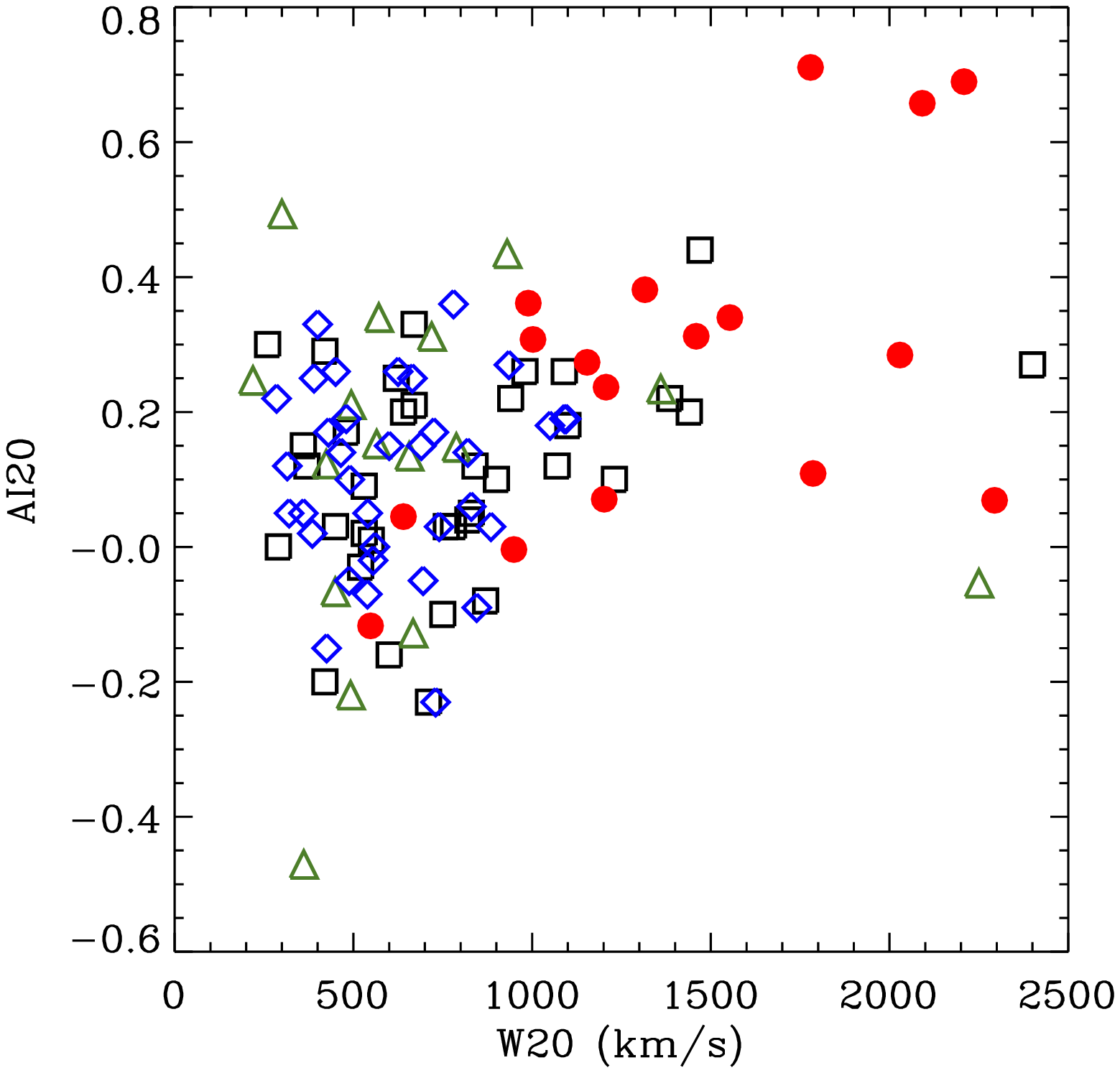,width=8.cm}
\caption[]{Upper panel: the asymmetry $AI20$ plotted against the kurtosis
  parameter $K1$ for the different samples of Sy galaxies. Open black squares:
  \cite{Heckman81}. Open blue diamonds: \cite{Whittle85a}. Open green triangles:
  \cite{Veilleux91a}. Filled red circles: our sample of Sy-ULIRGs. $AI20$ values
  higher (lower) than 0 indicate blueward (redward) asymmetries, while K1 values
  lower (higher) than 1 indicate ``peaky'' (``stubby'') profiles. for a
  perfectly symmetric profile AI20 = 0, while K1 = 1 for a Gaussian
  profile. Lower panel: the asymmetry $AI20$ plotted against W20. }
\label{AI20_K1}
\end{figure}

Figure \ref{AI20_K1} shows $AI20$ plotted against $K1$ and
$W20$. \cite{Nelson95} do not provide the center at 20\% intensity level in
their paper and therefore it is not possible to estimate the $AI20$ parameter
for the objects in their sample. At first sight, the figure shows the well-known
tendency for the [OIII]$\lambda$5007 in Seyfert galaxies to have blueward
asymmetries \citep{Heckman81,Whittle88}. In addition, the upper panel of the
figure shows that, although the values of the different parameters found for the
Sy-ULIRGs are consistent with the main body of points for the 4 comparison
samples, a significant proportion (at least 5 of 17, 30\%) are outliers in these
plots. Moreover, the lower panel in Figure \ref{AI20_K1} shows that Sy-ULIRGs
have, in general, broader emission lines that the objects in the comparison
samples. It is also apparent in the figure that all three of the objects
with the most asymmetric profiles are Sy-ULIRGs.

\begin{table*}
\centering
\begin{tabular}{llcccccc}
\hline\hline
Sample & $AI20$ & $K1$ & $W20$  & $W50$ \\
       & &&(km s$^{-1}$) & (km s$^{-1}$)\\
\hline
\cite{Heckman81}  & 0.11$\pm$0.03 & 0.88$\pm$0.02 & 798$\pm$69 & 455$\pm$40\\
N=36              \\
\hline 
\cite{Whittle85a} & 0.10$\pm$0.02 & 0.80$\pm$0.03 & 632$\pm$40 & 329$\pm$24\\
N=36              \\
\hline 
\cite{Veilleux91a}& 0.12$\pm$0.06 & 0.78$\pm$0.04 & 700$\pm$122 & 374$\pm$73\\
N=16              \\
\hline 
\cite{Nelson95}   & ...  & 0.87$\pm$0.02 & 682$\pm$47 & 388$\pm$28\\
N=77              \\
\hline 
all samples       & 0.10$\pm$0.01 & 0.85$\pm$0.02 & 689$\pm$41 & 384$\pm$16\\  
\hline
This paper        & 0.27$\pm$0.06 & 0.74$\pm$0.06 & 1423$\pm$131& 713$\pm$102\\
N=17              \\
\hline 
\end{tabular} 
\caption{Mean and the standard error of the mean ($\sigma$/$\sqrt{N}$) for 
the kinematic parameters measured in the
various samples. The entry in the table labelled as ``all'' correspond to
  the values considering the 4 comparison samples together.}
\label{param_mean}
\end{table*}

To investigate the possible differences between non-ULIRGs Seyfert galaxies and
Sy-ULIRGs, Table \ref{param_mean} shows the mean values and estimated
uncertainties for $AI20$, $K1$, $W20$ and $W50$. We note that our observations
of Sy-ULIRGs have a lower spectral resolution than those of most of the
comparison samples of Seyferts. Lower spectral resolution will generally result
in larger line widths and lower asymmetry indices
\citep[see][]{Veilleux91b}. Since the $AI20$ values for the Sy-ULIRGs are higher
on average than those for the comparison samples, despite the possible
resolution effects, the results in the table provide strong evidence that, in
general, the emission lines of the Sy-ULIRGs are more asymmetric that those of
non-ULIRGs Seyfert galaxies. Furthermore, the $W20$ values (and to a lesser
extent the $W50$ values) for the Sy-ULIRGs are substantially higher than those
of the objects in the comparison samples. Therefore, even allowing for
resolution effects\footnote{In reality, correcting for resolution has a
  negligible effect on the $W20$ measurements. For example, correcting a typical
  value $W20$ measured for the Sy-ULIRGs of $W20 = 1000$~km~s$^{-1}$ for the
  230~km~s$^{-1}$ instrumental width results in a decrease in $W20$ of only
  2.5\%.}, our results also suggest that the Sy-ULIRGs have broader emission
lines compared to non-ULIRGs Seyfert galaxies.

To further investigate the statistical significance of any differences between
the kinematic properties of the Sy-ULIRGs and the non-ULIRG Seyferts we have
performed a variety of statistical tests. In the first instance, we used the 2D
KS test of \cite{Peacock83} to investigate the significance of the differences
between the positions of the Sy-ULIRGs and the comparison samples in Figure
\ref{AI20_K1}. The results are shown in Table \ref{2D_KS:AI20_K1}. We find that,
especially considering the comparison with all the Seyfert galaxy samples
combined, the differences are significant at the $\gsim$3$\sigma$
level\footnote{Those objects included in more than one of the comparison samples
  are only considered once for this comparison. In particular, we used the set
  of parameter values that are derived with the dataset for which the spectral
  resolution most resembles the spectral resolution of our spectroscopic data.}

We have also compared the distributions of $AI20$, $K1$, $W50$ and $W20$
separately using a 1D two sample KS test. The results are consistent with those
of the 2D KS test of \cite{Peacock83} and are shown in Table
\ref{1D_KS_AI20_K1}. We find that, although the emission line in Sy-ULIRGs are
not particularly more ``peaky'' or ``stubby'' than those of the comparison
samples, they tend to be more asymmetric, and they are certainly broader: the
difference between the W20 and W50 parameters for the Sy-ULIRGs in our sample
and the objects in the comparison samples is significant at the $>3\sigma$
level.

At this stage is important to note that, while the comparison samples include a
significant fraction of broad line Seyfert 1 galaxies, our sample of Sy-ULIRGs
includes only one such object. This could be a problem if the NLR kinematics
depend on Seyfert type (e.g. because of orientation effects). In this context,
the results of \cite{Heckman81} show that galaxies with a significant BLR
component have, on average, narrower [OIII] lines that those that do not have
such components. However, the more comprehensive studies of \cite{Whittle85a}
and \cite{Nelson95}, found no evidence for differences between the line width
distributions of Sy1 and Sy2 galaxies of similar [OIII] luminosity and bulge
properties. To explore any potential biases due to the different proportions of
Sy1 objects in the Sy-ULIRG and comparison samples, we have repeated the
analysis described above but this time including only in those objects
classified as Sy2 for each individual sample. The results are consistent with
the previous results described in this section, although with slightly less
significance in the case of the $AI20$ parameter.

To summarise, these results provide evidence that the emission line kinematics
of the Sy-ULIRGs are more highly disturbed than those of the non-ULIRGs Seyferts
in the comparison samples. Indeed, three of the objects in our sample show more
extreme emission line kinematics than {\it any} of Seyfert galaxies in the
comparison samples. However, we emphasize that the size of our Sy-ULIRGs sample
is relatively small. Therefore, the results from the K-S test must be
interpreted carefully. Observations of larger samples of Sy-ULIRGs will be
required to put this result on a firmer statistical footing.

\begin{table}
\centering
\begin{tabular}{llcccccc}
\hline
Sample & & $AI20$ vs $K1$ & $AI20$ vs $W20$\\
(1)  & (2)  & (3)  & (4)\\
\hline
\cite{Heckman81}&D    &0.52 & 0.56  \\  
                &P(\%)&0.6  & 0.2 \\
\hline
\cite{Whittle85a}&D    &0.46 & 0.70  \\ 
                 &P(\%)&2    & $<$0.1 \\
\hline
\cite{Veilleux91a}&D    &0.36 & 0.70  \\
                  &P(\%)&22   & $<$0.1 \\
\hline
\cite{Nelson95}	&D    &-      & 0.62 \\  
                &P(\%)&-    & $<$0.1 \\
\hline
All             &D    &0.43 & 0.66  \\	          
                &P(\%)&0.6    & $<$0.1 \\
\end{tabular} 
\caption{The results of the 2D KS test of Peacock 1983, comparing our sample of
  Sy-ULIRGs with the comparison samples. Col 1: sample name. Col 2: 2D
  statistics. Col 3 and 4: the two parameters considered for the 2D KS test.}
\label{2D_KS:AI20_K1}
\end{table}

\begin{table}
\centering
\begin{tabular}{llcccccc}
\hline
Sample & &$AI20$ &	 $K1$ &	$W50$ &	$W20$\\
(1)&(2)&(3)&(4)&(5)&(6)\\
\hline\hline
\cite{Heckman81}   & D	        &0.47	&0.43	&0.48	&0.63\\
                   & P(\%)      &0.6	&2	&0.5	&$<$0.1\\
                  \hline
\cite{Whittle85a} & D	        &0.53	&0.20	&0.62	&0.77\\
                  & P(\%)	&0.15	&65	&$<$0.1	&$<$0.1\\
\hline
\cite{Veilleux91a}& D	        &0.33	&0.18	&0.64	&0.75\\
                  & P(\%)	&24	&93	&0.1	&$<$0.1\\
\hline
\cite{Nelson95}   & D	        &...	&0.33	&0.57	&0.66\\
                  & P(\%)	&...	&7.5	&$<$0.1	&$<$0.1\\
\hline           
all               & D           &0.47   & 0.27  & 0.56  & 0.68\\
                  & P(\%)       &0.2    & 17   &$<$0.1 & $<$0.1\\
\end{tabular} 
\caption{The results of the 1D two sample KS tests, comparing $AI20$, $K1$, $W50$
  and $W20$ for our sample of Sy-ULIRGs with those of the four comparison
  samples.}
\label{1D_KS_AI20_K1}
\end{table}

\subsubsection{Comparison with the emission line kinematics of other types of AGN}

Whittle (1985) found evidence -- albeit weak -- for correlations between the
emission line kinematics and the luminosities of the associated of AGN. Given
that the Seyfert galaxy comparison samples considered in the previous section
contain a higher proportion of objects with low [OIII] luminosities than our
Sy-ULIRG sample, it is therefore important to make comparisons with samples than
include higher luminosity objects. For this purpose we use the following three
AGN samples: the Palomar-Green (PG) sample of nearby quasars (Boroson et al.,
1982; Rose et al., 2012), the \cite{Jin12} sample of unobscured Seyfert 1 nuclei
and quasars, and the Rose et al. (2012) sample of nearby AGN selected from the
2MASS sample on the basis of their red near-IR colours. A major advantage of
these samples compared to those used in the previous section is that the [OIII]
emission line profiles have been fitted using the same multiple Gaussian fitting
technique that we have used for the ULIRGs in this paper. The general properties
of these samples are compared with those of the Sy-ULIRGs in Table
\ref{comp_samples}.

\begin{table*}
\centering
\begin{tabular}{llllllllll}
\hline\hline
Sample  &Selection criteria &Redshift range &L$_{\rm [OIII]}$ range &
Resolution & References\\
&&& (W) &(km s$^{-1}$)\\
\hline
Sy-ULIRGs &S$_{60}>$1 Jy, L$_{IR}>10^{12}$ L$_{\odot}$, &$z < 0.175$
&$2\times10^{33}$ -- $4\times10^{35}$ & $\sim$230&This paper \\
&emission line ratios & & & \\
PG Quasars &UV excess, broad lines &$z < 0.5$  &$5\times10^{33}$ -- $5\times10^{36}$ & $\sim$350&\cite{Boroson82}\\
&$M_B < -23$ & & & \\
Unobscured type I &Type 1, X-ray bright &$z<0.4$ &2$\times$10$^{33}$ -- 3$\times$10$^{35}$  &$\sim$160& \cite{Jin12}\\
&Unobscured & & & \\
2MASS AGN &(J-K) $>$ 2 &$z<0.29$ &3$\times$10$^{33}$ -- 6$\times$10$^{35}$  &
$\sim$550&Rose et al. 2012\\
\end{tabular}
\caption []{Details of the properties of the higher luminosity AGN comparison
  samples.}
\label{comp_samples}
\end{table*}

\begin{figure}
\psfig{file=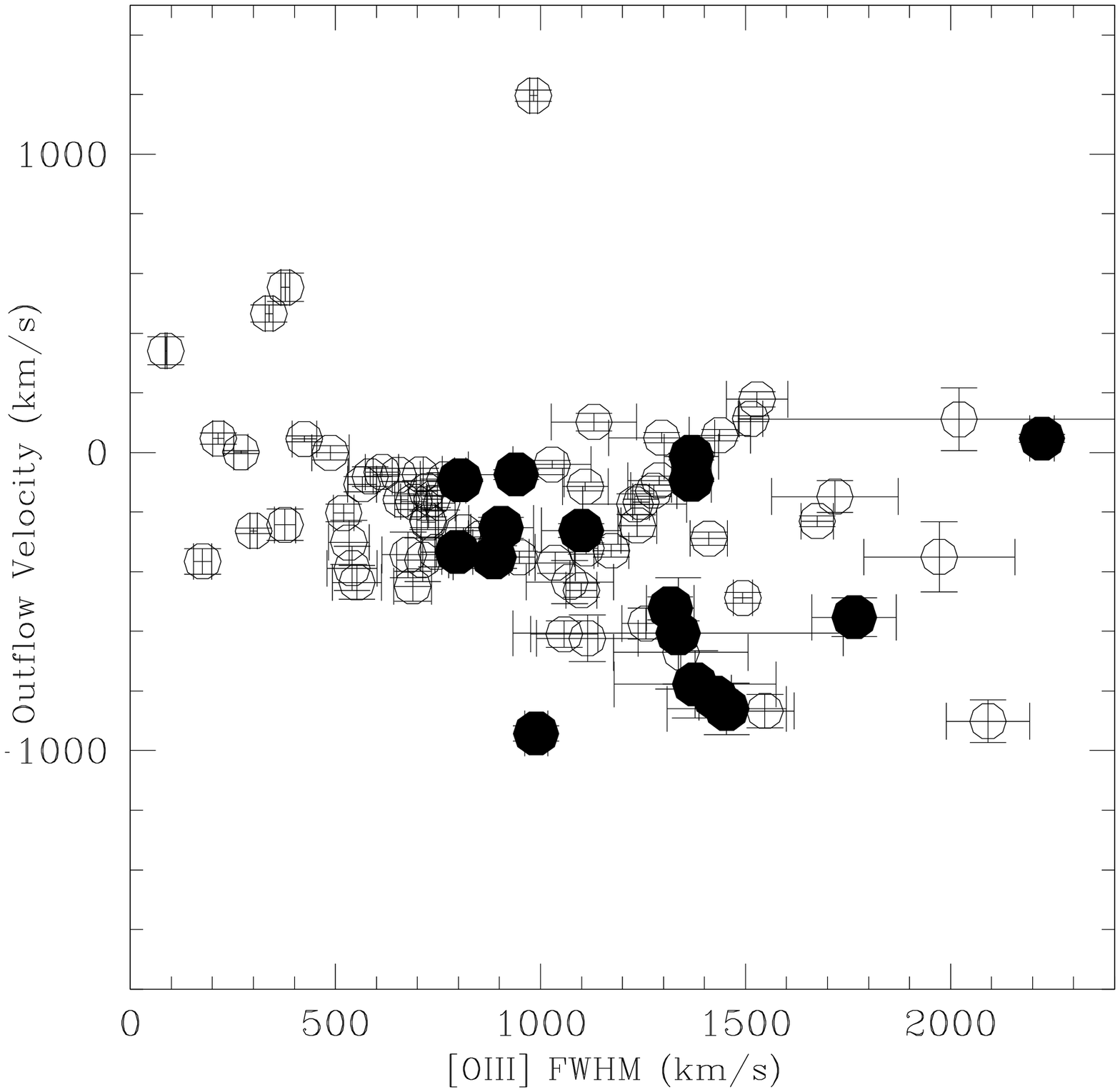,width=6.8cm}\\
\psfig{file=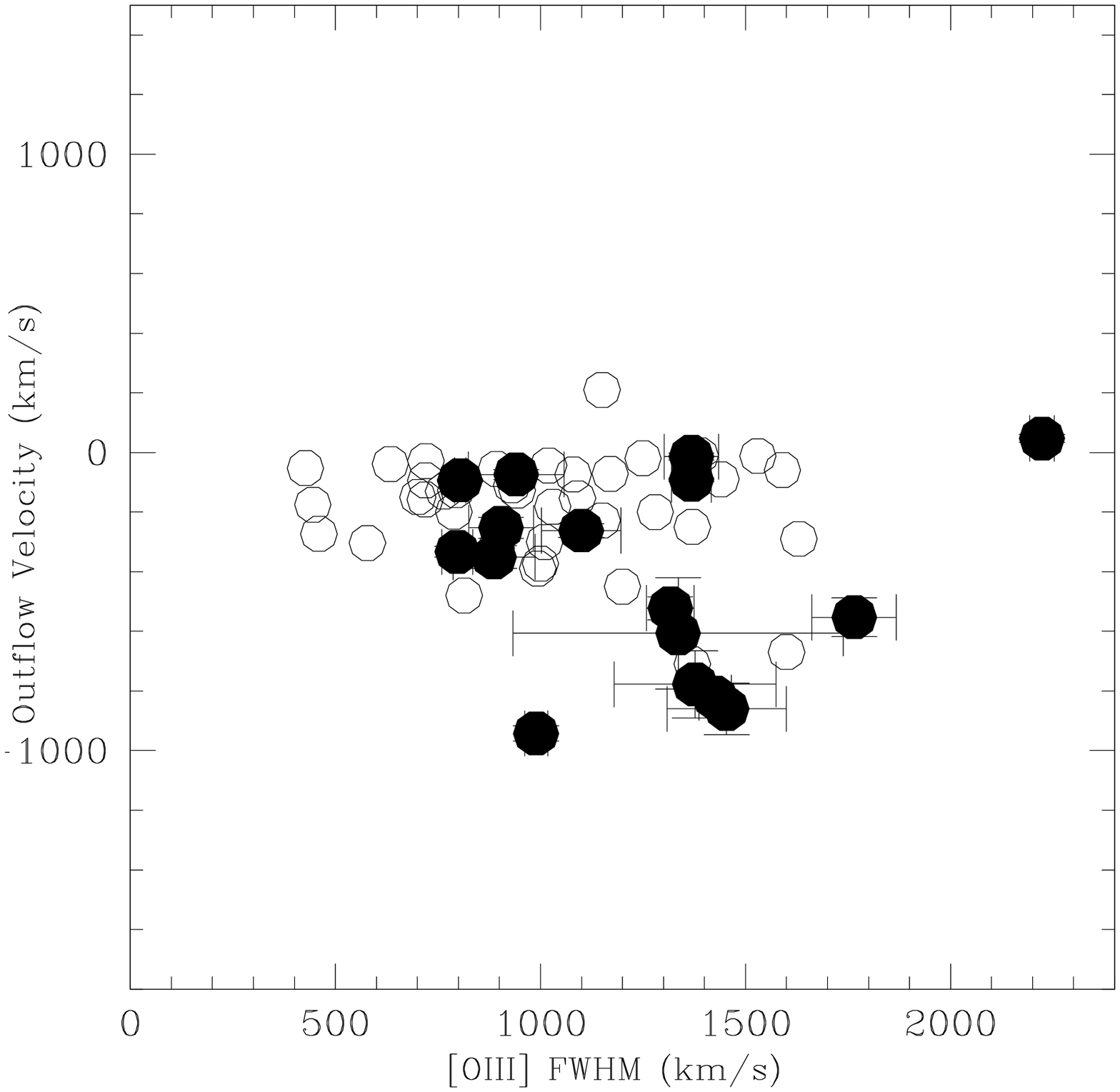,width=6.8cm}\\
\psfig{file=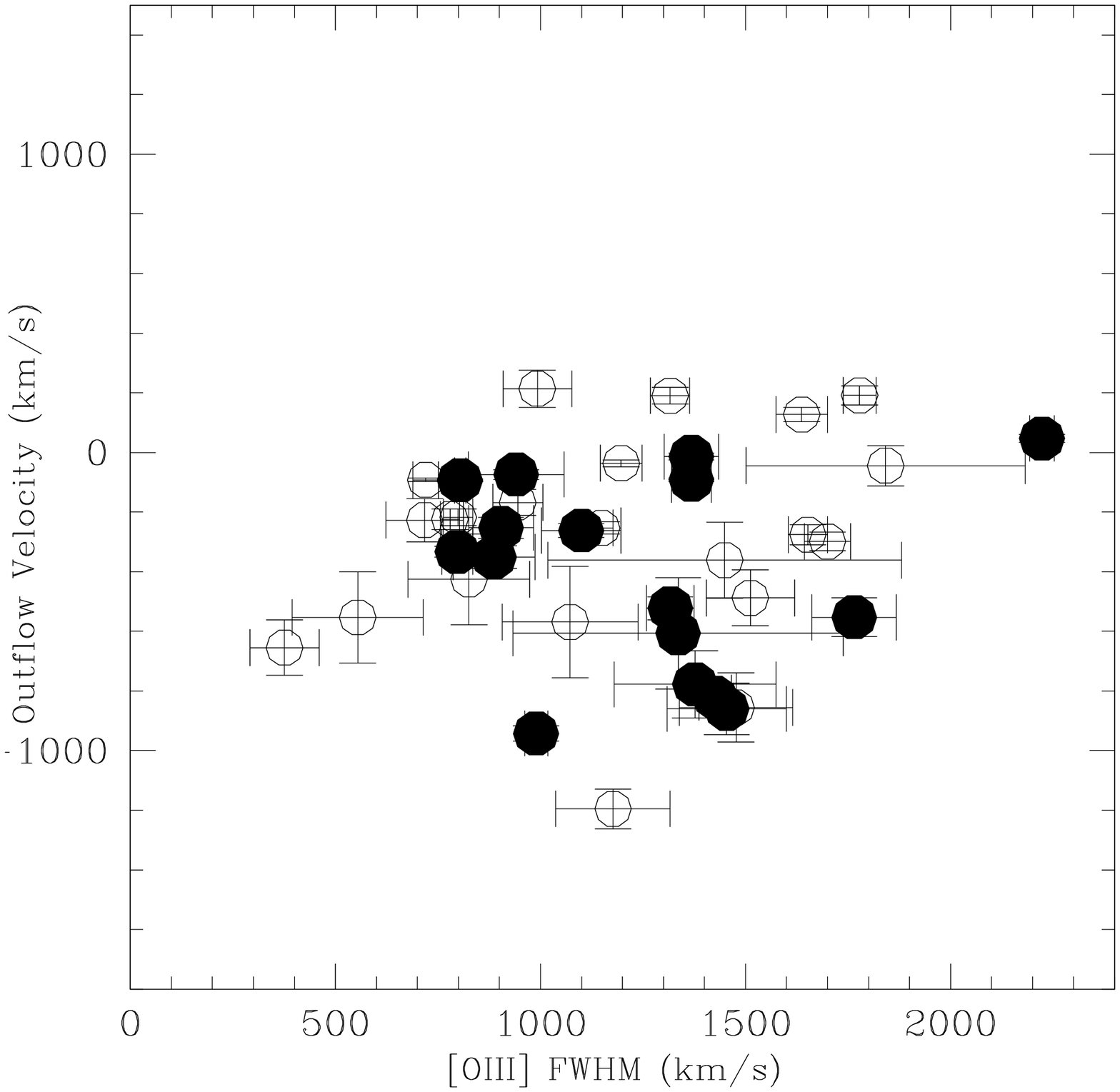,width=6.8cm}\\
\caption[]{Flux weighted mean shifts plotted against the flux weighted mean FWHM
  for the ULIRGs in our sample (black-filled circles) and the PG QSOs (top), the
  \cite{Jin12} sample of unobscured Seyfert 1 nuclei and quasars (middle) and
  the Rose et al. (2012) sample of nearby AGN selected from the 2MASS sample on
  the basis of their red near-IR colours (bottom). The figure show that the
  proportion of objects that show extreme emission line kinematics in the sense
  that they are outliers in the plot is higher for the Sy-ULIRGs than it is
  from the comparison samples.}
\label{Sample_comparison}
\end{figure}

We use the velocity shifts and FWHM of the broad and intermediate emission line
components as our main indicators of the degree of kinematic disturbance. A
complication with this approach is that individual objects --- including many of
the ULIRGs discussed in this paper --- may have two or more broad and/or
intermediate emission line components. In such cases we use the flux weighted
mean shift and FWHM of the broad and intermediate emission line components
(defined in Section 3.3); this yields single estimates of the shift and FWHM for
each object and allows fair comparison of the degree of kinematic disturbance
between the various samples. Note that all the velocity shifts and FWHM are
measured relative to the rest frames of the host galaxies, and narrow emission
line components (by our definition ${\rm FWHM} < 500$ km s$^{-1}$) are not
included in the analysis.

Figure \ref{Sample_comparison} shows the velocity shift plotted against the FWHM
for the Sy-ULIRGs and the three comparison samples. Again, the figure shows
the well known tendency for the kinematically disturbed components to be
blueshifted relative to the host galaxy rest-frames. Also, there is a tendency
for the objects with the largest blueshifts to have the largest FWHM \citep[see
  also][]{Spoon09b}, although the converse in not true, indeed the object with
the largest FWHM --- the Sy-ULIRG F23389+0300 --- shows a slight
redshift. Consistent with the results of the previous section, we find that,
although a significant proportion of the Sy-ULIRGs fall within
the main envelope of points defined by the three comparison samples ($\Delta V >
-500$ km s$^{-1}$, $FWHM <$ 1500 km s$^{-1}$), 50\% are outliers on these plots
with $\Delta V < $-500 km s$^{-1}$ and/or $FWHM >$ 1500 km s$^{-1}$; this proportion
is higher that found in the PG quasar (15\%), unobscured type I (12\%), and 2MASS
(34\%) samples. We further note that detailed examinations of the spectra of the objects in the
high luminosity comparison samples reveals no objects with the type of highly structured,
extremely blueshifted, multiple component [OIII] emission profiles that are
observed in the Sy-ULIRGs F13451+1232 and F14394+4332.

We again use the \cite{Peacock83} 2D KS test to investigate the significance of the
differences between the positions of the Sy-ULIRGs and the comparison
samples in Figure \ref{Sample_comparison}. The results are shown in Table
\ref{Sample_comparison_2D}. From this it is clear that we can reject the null
hypothesis that the Sy-ULIRGs, PG quasars and unobscured type I AGN objects are
drawn from the same parent population at the $\sim$2\% ($\sim 2\sigma$) level of
significance, but do not find a statistically significant difference between the
Sy-ULIRGs and the local 2MASS AGN population.

We have also compared the distributions of FWHM and velocity shift separately
using a 1D two sample KS test. The results shown in Table
\ref{Sample_comparison_1D} reveal that the most significant differences are
between the distributions of FWHM for the Sy-ULIRG and the PG Quasars, with the
difference significant at the 0.3\% ($\sim 3\sigma$) level. This difference is
consistent with that found between the line widths of the Sy-ULIRG and non-ULIRG
Seyfert comparison samples in the previous section\footnote{Note that, in this
  case, there is no issue with differences between the spectral resolutions of
  the samples, because the FWHM values have been corrected for instrumental
  resolution.}. The differences between the 1D distributions of velocity shift
for the Sy-ULIRGs, PG quasars and Unobscured Type I samples are less significant
and, consistent with the results of the 2D KS test, the 1D tests provide no
evidence for significant differences between the FWHM and velocity shifts when
comparing the Sy-ULIRGs with the local 2MASS sample.

\begin{table}
\centering
\begin{tabular}{llcll}
\hline\hline
Comparison sample &N  &2D KS statistic (D) &P(\%) \\ 
\hline
PG Quasars &85 &0.45 &1.9  \\
Unobscured type I &30 &0.47 &2.4  \\
2MASS AGN &30  &0.32 &38 \\
\end{tabular}
\caption{The results of the 2D KS test of Peacock 1983, comparing our sample of
  Sy-ULIRGs with the three comparison samples.}
\label{Sample_comparison_2D}
\end{table}

\begin{table}
\centering
\begin{tabular}{lllll}
\hline\hline
&FWHM & &Shift & \\
Comparison sample &D &P(\%) &D &P(\%) \\ 
\hline
PG Quasars &0.48 &0.3 &0.34 &7.6  \\
Unobscured type I &0.43 &1.9 &0.40 &4  \\
2MASS AGN &0.28 &38 &0.24 &60 \\
\end{tabular}
\caption []{The results of the 1D two sample KS tests, comparing the
FWHM and velocity shift distributions for our sample of Sy-ULIRGs
with those of the three comparison samples.} 
\label{Sample_comparison_1D}
\end{table}

Together, the results of this section show a tendency for the Sy-ULIRGs to
display a higher degree of kinematic disturbance than the PG quasars and
unobscured type I AGN, but again observations of larger samples of Sy-ULIRGs
will be required to strengthen the statistics.

\section{Conclusions and future work}

Most of the recent investigations of outflows in ULIRGs have focussed on
observations of the neutral and molecular gas
\citep[e.g][]{Rupke05c,Fischer10,Rupke11,Sturm11}. These have revealed massive,
highly energetic outflows in several objects, supporting the view that the
feedback effect associated with the outflows can have a significant impact on
the evolution of the host galaxies. However, there remain questions about
whether these molecular and neutral outflows are driven by the circum-nuclear
starbursts or by the AGN that are found in the nuclei of some of the
sources. Indeed many such studies have failed to find significant differences
between the outflows in ULIRGs with and without powerful AGN \citep[e.g.][]{Rupke05c}.

Given that (a) we find that a significantly higher fraction of the ULIRGs with
optical Seyfert nuclei show evidence for nuclear outflows than those lacking
such nuclei, and (b) the outflow components have line ratios consistent with
AGN- rather than starburst-photoionization, it is highly likely that the nuclear
outflows in the Sy-ULIRGs are indeed driven by the AGN. We also find that
the Sy-ULIRGs present outflows that are are more extreme than those in
samples of local non-ULIRG Seyfert galaxies.

Clearly there is now evidence that the AGN associated with ULIRGs drive warm
outflows. What remains uncertain is whether these outflows are sufficiently
energetic and massive to affect the star formation histories and evolution of
the host merger remnants. In section 3.8 we estimated the ranges of mass outflow
rates and kinetic powers that are consistent with our observations. At the upper
end of the range, the mass outflow rates and kinetic powers of the warm outflows
in Sy-ULIRGs are comparable with those typically estimated for the neutral
outflows detected using the optical NaID lines \citep[$6 < \dot{\rm M} <
  400$~M$_{\odot}$yr$^{-1}$; $3\times10^{40} < \dot{\rm E} < 10^{44}$ erg
  s$^{-1}$:][]{Rupke05b,Rupke05c,Rupke11} and the molecular outflows using the
hydroxyl molecule OH \citep[up to 1.2$\times$10$^{3}$ M$_{\odot}$
  yr$^{-1}$]{Sturm11}, and would be likely to have a major impact on the host
galaxies. On the other hand, if the lower limiting estimates were to prove more
typical of the true outflow properties, the warm outflows would be insignificant
compared with the neutral and molecular outflows, and therefore unlikely to
significantly affect the evolution of the host galaxies.

Despite the high degree of uncertainty, there is cause for optimism that the
mass outflow rates and kinetic power estimates will be substantially improved in
the near future. In particular, using the transauroral [OII] and [SII] line
ratios it will be possible to make density and reddening estimates that are
considerably more precise \citep[see][]{Holt11}; the outflow radius estimates
will benefit from HST narrow-band \citep[e.g.][]{batcheldor07} and AO-assisted
near-IR imaging; and the reddening (and hence H$\beta$ luminosities) will be
further refined by combining optical and near-IR measurements of the hydrogen
recombination line fluxes.

\section*{Acknowledgments}

JRZ acknowledges financial support from the spanish grant
AYA2010\_21887-C04-04. The William Herschel Telescope is operated on the island
of La Palma by the Isaac Newton Group in the Spanish Observatorio del Roque de
los Muchachos of the Instituto de Astrofisica de Canarias. MR acknowledges
support from STFC and PPARC in the form of PhD studentships. We thank Todd
Boroson for providing the reduced spectra of the PG quasar sample. This research
is supported in part by STFC grant St/J001589/1.

\bibliographystyle{mn2e}
\bibliography{JRZRefs}

\clearpage

\begin{center}
{\bf APPENDIX: Description of individual objects}\\
\end{center}

\noindent
{\bf F00188--0856.} \cite{Veilleux99} classified this compact object as a LINER at optical
 wavelengths. However, based on the \cite{Kewley06}(hereafter Kwl06)
 classification scheme, Yuan et al. (2010) classified it as ``Sy2:'', meaning
 that F00188--0856 has a Sy2 spectral type in 2 of the 3 BPT diagrams use for
 its optical classification (see Yuan et al., 2010 for details). Therefore, we
 included this object in the sample discussed in this paper.

F00188-0856 is one of the two cases (along with F23327+2913), for
which the [OIII]\lala4959,5007 and H$\beta$ emission lines have a low equivalent
width, and are potentially affected by noise and residual structure in the
underlying continuum. Therefore these lines are not, in this case,
suitable for deriving a model to fit to the other lines in the 
spectrum. For this object, we used the
approach of first fitting the H$\alpha$+[NII] complex and then modelling the other
emission lines in the spectrum with the same components and the same
velocity widths and shifts as H$\alpha$. In addition, due to the presence of a
telluric absorption feature that coincides with the location of the
[SII]\lala6716,6731, it was not possible to model this emission
blend\footnote{There are cases for which a telluric absorption feature affects
  the profiles of some lines (mainly \sulphurtwow~and the
  \ha+[NII]\lala6548,6583 complex) but it is still possible to model the
  emission lines. However, the flux values derived from the modelling will be
  relatively unconstrained. An example of the typical effect of a telluric
  absorption feature on the profile of the \ha+[NII]\lala6548,6583 complex in
  shown in Figure \ref{23060-telluric}.}. The results of the H$\alpha$+[NII]
modelling are shown in Figure \ref{OIII-profiles}.

The best fitting model comprises two components: a narrow component with FWHM =
269 $\pm$ 5 km s$^{-1}$ plus an intermediate component with FWHM = 904 $\pm$ 75
km s$^{-1}$ blueshifted by -253 $\pm$ 36 s$^{-1}$ with respect to the narrow
component. Using these results is possible to measure the line ratios associated
with each individual component and then study the corresponding ionization
mechanisms using the already mentioned Kwl06 classification scheme. We find that
the line ratios associated with the narrow component are consistent with a
composite spectral type (CP) or LINER, while those corresponding to the
intermediate component are consistent with Sy2 ionization, but lie close to the
Sy2/LINER limiting region.

\vspace*{0.2cm}\noindent
{\bf F01004--2237.} The optical spectrum of this single nucleus system shows a mixture of HII and
Sy2 features \citep{Veilleux99}. However, it was classified as a ``Sy2:'' in the
work of Yuan et al. (1010) and therefore, included it in our sample.

Figure \ref{OIII-profiles} shows the [OIII] model for this galaxy, which
includes three components: an unresolved, narrow component, an intermediate
component (FWHM = 850 $\pm$ 71 km s$^{-1}$) blueshifted by -230 $\pm$ 32 km
s$^{-1}$ with respect to the narrow component and a broad (FWHM = 1590 $\pm$ 92
km s$^{-1}$ ) component blueshifted by -999 $\pm$ 80 km s$^{-1}$ with respect to
the narrow component. Note that the broad, shifted components dominate the 
[OIII] emission line flux in this case.

Due to the number of components, the degeneracy between
the different models that adequately fit the emission line profiles is a
relatively important issue for this object. For example, when the intensities of
the different components were left as a free parameter during the modelling of
the \ha+\nitrogen complex, the \nitrogen$\lambda$6583/\ha~line ratio
corresponding to the broad component was $\sim$4. As shown in Yuan et
al. (2010), their Figure 2, this line ratio is always smaller than
$\sim$3. Therefore, during the modelling, we set the value of the
\nitrogen$\lambda$6583/\ha~line ratio to 3, obtaining a good fit to the [NII]+H$\alpha$
blend.

In addition,  the \sulphurtwo6716/6731 line ratio was higher than 1.42 (the low
density limit) for the narrow component, but lower than 0.44
(the high density limit) for the two broad components (i.e., the results were
unphysical) when making free fits to
the [SII] blend. Therefore, during the modelling, we also forced the values of these
line ratios to be 1.42 and 0.44 for the narrow and the two broad components
respectively. Then, the value of one of these line ratios was changed in steps
on 0.02 while the values of the other two remained unchanged until the fit was
deemed not adequate. We estimate a lower limit for the intensity ratio
\sulphurtwo6716/6731 of 1.2 for the narrow component while acceptable fits are
obtained for any line ratio value within the allowed range (0.44 $\leq$
\sulphurtwo6716/6731 $\leq$ 1.42) in the cases of intermediate and the broad
components.

Regarding to the ionization mechanisms associated with each of the individual
components, we find that the line ratios corresponding to the narrow and
intermediate components are consistent with HII ionization or fall close to the
HII/Sy2 limiting region in the diagrams. On the other hand, the line ratios
associated with the broad component are consistent with NLAGN and Sy2 ionization
in two of the BPT diagrams, while they fall close to the HII/Sy2 limiting region
in the [SII] diagram. Given these results, it is clear why the spectral
classification of this object based on line ratios derived from integrated line
fluxes \citep[e.g.][]{Veilleux99}, was ambiguous in previous studies; our
analysis shows the clear presence of an AGN associated with the broad,
blueshifted emission line components. Note that the approach used to model the
\sulphurtwo6716/6731 doublet does not significantly change these results.

The nuclear emission line kinematics of this object have been discussed in the
past by \cite{Armus88}, \cite{Lipari03}, and \cite{Farrah05}. While the
spectra of \cite{Armus88} and \cite{Lipari03} show the same blue wings to the
[OIII] emission lines that we detect in the spectrum presented in this study,
\cite{Farrah05} report that they fail detect the blue wings in their HST/STIS
spectrum (0.2'' slit width) that isolates the compact knot in the nucleus of the
galaxy. This is interesting because it suggests that the AGN outflow may be
significantly extended in this object. However, the substantially worse spectral 
resolution of the HST/STIS spectra of \cite{Farrah05} compared with
the spectra presented in this paper may have prevented 
the detection of the shifted features.

\vspace*{0.2cm}\noindent
{\bf F12072--0444.} As in the case for F00188--0856, the presence of a telluric absorption
feature coinciding with the location of the [SII]\lala6716,6731 blend prevents any
attempt to model this feature for this Sy2 galaxy. The best fitting model to
all the other observed emission lines comprises 3 components: a narrow component (FWHM
= 275 $\pm$ 8 \kms) and intermediate component with FWHM of 525 $\pm$ 28 and
blueshifted -276 $\pm$ 30 \kms with respect to the narrow component and a broad
component with FWHM of 1343 $\pm$ 81 \kms and blueshifted -446 $\pm$ 35 \kms. The
line ratios measured for all of the three components are consistent with
NLAGN or Sy2 ionization.

\vspace*{0.2cm}\noindent
{\bf F13305--1739} F13305--1739 has by far the highest emission line luminosity of all
the objects in our sample (see Table 1); its [OIII] emission line luminosity
would lead to it being classified as a type II quasar according the the criteria
of \cite{Zakamska03}, and it shows a rich emission line spectrum from the
UV ([MgII]$\lambda$2798) to the red end (\sulphurtwow) of its
spectrum. Figure \ref{OIII-profiles} shows the \oxythr~model that successfully
fits all the other emission lines detected in the optical spectrum of this
source. Clearly, the broad/shifted components dominate the [OIII] emission line
flux in this case.

\begin{figure*}
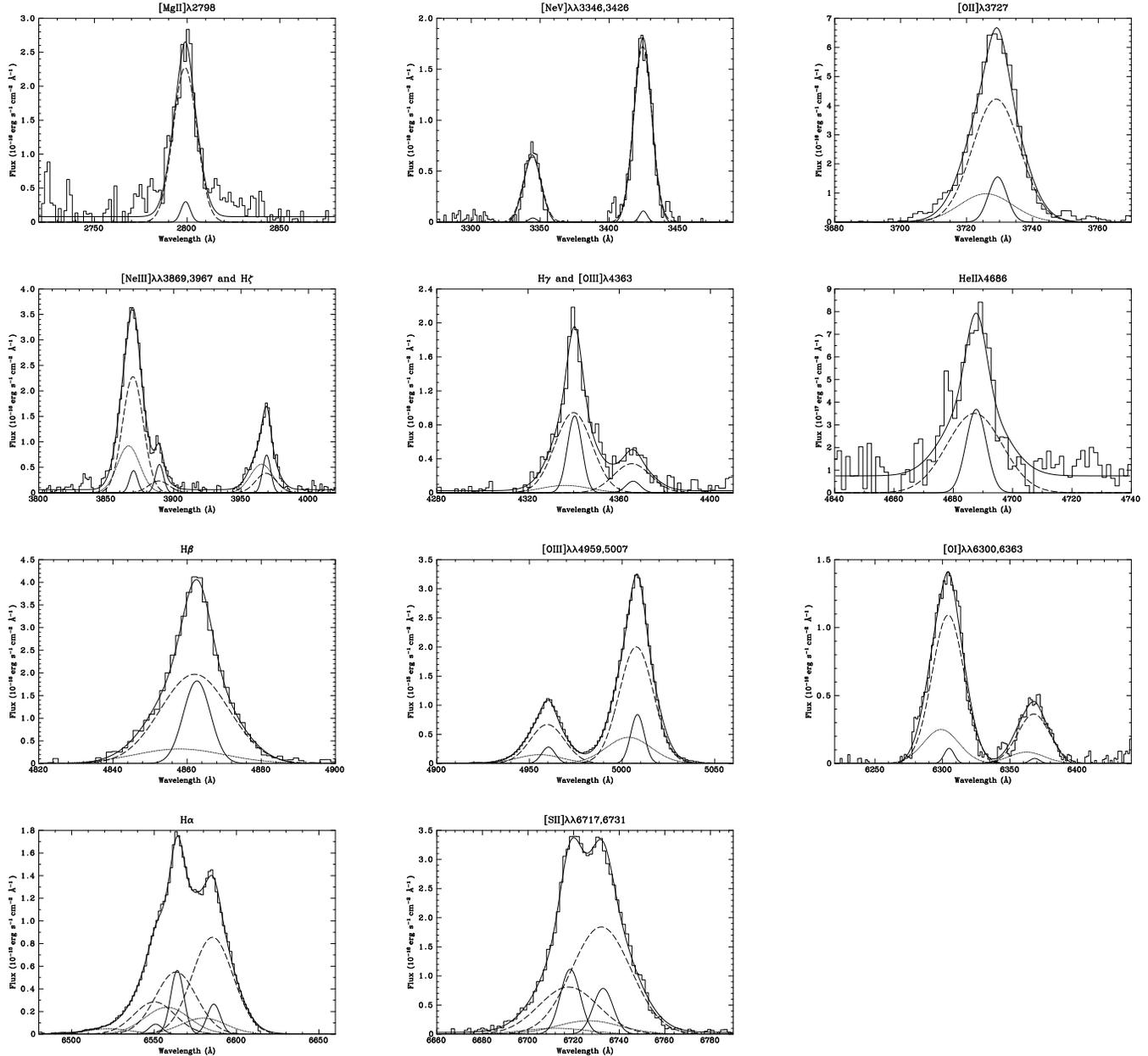

\begin{tabular}{ccc}
\hspace{-0.5 cm}\psfig{figure=IRASF13305_MgII_profile,width=4.cm,angle=-90.}&
\psfig{figure=IRASF13305_NeV_profile,width=4.cm,angle=-90.}&
\psfig{figure=IRASF13305_OII_profile,width=4.cm,angle=-90.}\\
\hspace{-0.5 cm}\psfig{figure=IRASF13305_NeIII_profile,width=4.cm,angle=-90.}&
\psfig{figure=IRASF13305_Hgamma_profile,width=4.cm,angle=-90.}&
\psfig{figure=IRASF13305_HeII_profile,width=4.cm,angle=-90.}\\
\hspace{-0.5 cm}\psfig{figure=IRASF13305_HB_profile,width=4.cm,angle=-90.}&
\psfig{figure=IRASF13305_OIII_profile,width=4.cm,angle=-90.}&
\psfig{figure=IRASF13305_OI_profile,width=4.cm,angle=-90.}\\
\hspace{-0.5 cm}\psfig{figure=IRASF13305_Ha_profile,width=4.cm,angle=-90.}&
\psfig{figure=IRASF13305_SII_profile,width=4.cm,angle=-90.}\\
\end{tabular}
\caption{The results obtained by using the \oxythr~model to fit all the optical
  emission lines detected in the spectrum of F13305--1739. The resulting model
  is overplotted (bold line) on the extracted spectrum (faint line). The
  different kinematic components are also plot in the figure. The solid, dashed
  and dotted line correspond to the narrow component and the two broad
  components respectively (referred as N, B1 and B2 in the text). The figure
  shows that B1 (dashed line) dominated the flux contribution for the majority
  of the emission lines detected in the optical spectrum of the source.}
\label{13305-profiles}
\end{figure*}

We use the case of F13305--1739 as an example of our modelling technique, and
Figure \ref{13305-profiles} shows the modelling results obtained for all the
emission lines using the so called \oxythr~model. This model comprises three
comp: a narrow component (FWHM=435$\pm$25 \kms), a first broad component (B1)
with FWHM of 1276$\pm$29 \kms blueshifted by only -36 $\pm$ 14 \kms with respect
to the narrow component and a second, broad component (B2) with FWHM of
1685$\pm$286 \kms and blueshifted by -281$\pm$29 \kms with respect to the narrow
component. Figure \ref{13305-profiles} shows that, in general, the B1 component
dominates the flux contribution for the majority of the emission lines detected.

As in the case of F01004--2237, degeneracy is an important issue in the case of
F13305--1739. When the intensities of the different components were left as a
free parameter during the modelling the \sulphurtwo6716/6731 line ratio was
higher than 1.42 (the low density limit) for the narrow component, but but this
ratio was lower than 0.44 (the high density limit) for the two broad components
(i.e., the results were unphysical). Using the same approach described before,
we estimate a lower limit for the intensity ratio \sulphurtwo6716/6731 of 1.2
for the narrow, component and an upper limit of 0.6 for the broad components.

In terms of the ionization mechanisms, the line ratios measured for all three
kinematic components are consistent with Sy2 ionization. Note that the approach
used when modelling the \sulphurtwow~doublet does not significantly change these
result.


\vspace*{0.2cm}\noindent
{\bf F13428+5608 (Mrk273).} This well-known ULIRG is the closest in our sample of Sy-ULIRGs
($z=0.037$), and in optical images shows an impressive tidal tail that extends
over 30 kpc towards the south of the galaxy. A double nucleus structure is
visible at near-IR \citep{Armus90,Majewski93,Knapen97,Scoville00} and radio
\citep{Ulvestad84,Condon91,Cole99} wavelengths, with a nuclear separation of
$\sim$ 700 pc, although the two nuclei are unresolved in our spectroscopic data.

As shown in Figure \ref{OIII-profiles}, the best fitting model for this galaxy includes
two components: a narrow component with a FWHM of 449$\pm$7 \kms plus a broad
component with a FWHM of 1369$\pm$62 \kms that is redshifted by 14$\pm$12 \kms with
respect to the narrow component. 

We find that, when the intensities of the two components are left as free
parameters during the modelling of the \ha+\nitrogen~complex, the
\nitrogen$\lambda$6583/\ha~line ratio measured for the broad component (FWHM
= 1369 km s$^{-1}$) is significantly higher than 3, which, based on the results
of Yuan et al. (2010), can be considered as an upper limit for this line
ratio. In addition, the \sulphurtwo6716/6731~line ratio obtained for the broad
component is outside the allowed range (i.e. \sulphurtwo6716/6731 $<$ 0.44, the
high density limit). Therefore, as discussed before, we set the value of the
\nitrogen$\lambda$6583/\ha~line ratio to 3 and the \sulphurtwo6716/6731 line
ratio at the high density limit. Then we gradually increased the value of the
\sulphurtwo6716/6731~line ratio until the overall fit was poor,
which occurred at the value of 0.8.

In terms of the ionization of the gas, the line ratios measured for the 
narrow component lie close to the
limiting regions between CP and narrow line AGN (NLAGN) ionization, and LINER
and Seyfert2 galaxies in the Kwl06 classification scheme. On the other hand, the
line ratios measured for the broad component are consistent with Sy2
ionization. Note that the approach used when modelling the
\ha+\nitrogen~complex and the \sulphurtwow~doublet does not change this result.

\vspace*{0.2cm}\noindent
{\bf F13451+1232.} A detailed discussion of the spectacular emission line kinematics this object
can be found in \cite{Holt03} and \cite{Holt11}.

\vspace*{0.2cm}\noindent
{\bf F14394+5332.} F14394+5332 shows the most spectacular [OIII] emission line kinematics of all
the objects considered in this paper (see Figure 1), with the clear detection of
three distinct narrow components, as well as an underlying broad
component. Based on our analysis of the spatially extended [OIII] emission line
kinematics in this object, the redmost narrow component in the integrated
spectrum represents the galaxy rest frame. In this case, the two other narrow
components are blueshifted by -700 and -1360 km s$^{-1}$ relative to the rest
frame, and the underlying broad components are also blueshifted; clearly this is
a case in which the compenents that are broad and/or blueshifted dominate the
[OIII] emission line flux. The multi-component nature of the [OIII] lines in
this object was also detected by \cite{Lipari03}, albeit at lower S/N.

\begin{figure*}
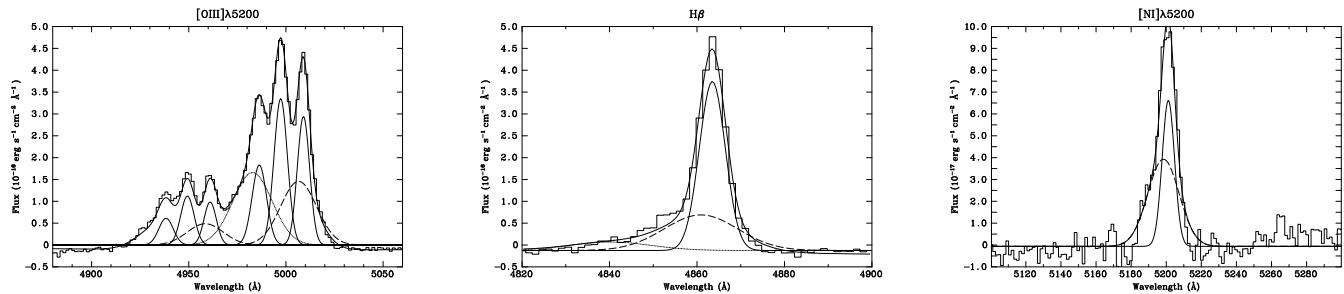

\begin{tabular}{ccc}
\hspace{-0.5 cm}\psfig{figure=IRASF14394_profile_OIII.ps,width=4.cm,angle=-90.}&
\psfig{figure=IRASF14394_profile_Hb.ps,width=4.cm,angle=-90.}&
\psfig{figure=IRASF14394_profile_NI5200.ps,width=4.cm,angle=-90.}\\
\end{tabular}
\caption{The modelling results obtained for [OIII], H$\beta$ and
  [NI]$\lambda$5200. The resulting model is overplotted (bold line) on the
  extracted spectrum (faint line). The different components are also plot in the
  figure. The solid, dashed and dotted line correspond to the narrow components
  (3 in the case of the [OIII] emission line referred as N1, N2 and N3 in Table
  2) and the two broad components respectively (referred as B1 and B2 in Table
  2). The figure shows that the blueshifted narrow lines (N2 and N3) are not
  clearly detected in the lower ionization lines such as H$\beta$ or
  [NI]$\lambda$5200.}
\label{IRAS14394_emission}
\end{figure*}

The profiles of the [OIII] emission lines in this object are strikingly
different from those of most of the other emission lines. This is illustrated by
the comparison between the H$\beta$, [OIII] and [NI] emission lines shown in
Figure \ref{IRAS14394_emission}. In particular, we do not clearly detect the
blueshifted narrow lines in {\it any} of the lower ionization lines such as
[OII]$\lambda$3727, H$\beta$, [NI]$\lambda$5200,
[OI]$\lambda\lambda$6300,6363\footnote{N.B. The blue wing of the
  [OI]$\lambda$6300 line may be significantly affected by telluric atmospheric
  B-band absorption.}, H$\alpha$, [NII]$\lambda$$\lambda$6548,6584 and
[SII]$\lambda$$\lambda$6717,6731, apart from possible hints of the first shifted
narrow component in the form of a ``shelf'' in the blue wings of
[NI]$\lambda$5200 and [OI]$\lambda$6300. On the other hand, the
[NeIII]$\lambda$3869 line has a profile that is consistent with that of
[OIII]$\lambda$5007, although the lower S/N, lower effective velocity
resolution, and possible residual features from the subtraction of the (complex)
underlying stellar continuum, preclude an accurate comparison between the line
profiles. We further note that the [NeV]$\lambda$3425 and HeII(4686) features
show evidence for broad, blueshifted features, but the S/N is again too low to
allow a detailed comparison with the [OIII] line profile.

The fact that the discrete blueshifted [OIII] emission lines components are not
clearly detected in the other lines, many of which have similar {\it total}
emission line fluxes, immediately implies that the gas emitting these
blueshifted narrow components has a high ionization state. Nonetheless, the low
ionization lines do show some evidence for kinematic disturbance in the form of
blue wings. Overall, we find that we can obtain an acceptable fit to all the
low ionization and Balmer lines in the nuclear spectrum using a model that
includes a narrow component with the same kinematic properties as those of the
redmost (rest frame) narrow component detected in the [OIII] line, plus two
broad components blueshifted shifted by -156 and -1574 km s$^{-1}$ with FWHM of 1272 and
1401 km s$^{-1}$ respectively. We also find that we can fit the [OIII]
profiles with the same two broad components plus the three narrow components
described above; this is fit is shown in Figure 1, and all the corresponding fit
components are listed in Table 2.

To get some idea of the ionization of the kinematically disturbed emission line
components, we have compared the sum of the fluxes of all the blueshifted
components of [OIII] (2 narrow $+$ 2 broad) with the sum of the fluxes of the
blueshifted components fitted to the lower ionization lines (2 broad). The
resulting line ratios are labelled ``B'' in Table 3. Interestingly, despite the
fact that the discrete blueshifted narrow components in [OIII] must be emitted
by gas with a very high ionization state, the overall B line ratios are low
ionization in character, with [OII]/[OIII]$\sim$ 1 and [OIII]/H$\alpha$ $\sim$
1, although more consistent with AGN than starburst heating\footnote{Note that
  integrated H$\beta$ flux for the kinematically disturbed components is likely
  to be strongly affected by the underlying H$\beta$ absorption. Therefore B ratios
  measured relative to H$\beta$ are unlikely to be reliable}. On the other hand,
the line ratios for the red narrow component fall squarely in the HII/AGN region
of all three diagnostic diagrams.

Clearly, the extraordinary kinematical and ionization properties of this source
warrant further investigation with higher resolution and S/N optical
spectroscopy.

\vspace*{0.2cm}\noindent
{\bf F15130--1958.} Three components were required to adequately model the profile of the emission
lines observed in the optical spectrum of this single nucleus ULIRG. A first,
intermediate component (I1, FWHM = 545$\pm$69 \kms) component, a second,
intermediate component (I2, FWHM = 700$\pm$188 \kms) blueshifted by -351$\pm$280
\kms with respect to I1, and a broad (FWHM = 1630$\pm$41 \kms) component
blueshifted by -725$\pm$131 \kms with respect to the narrow component.

As in previous cases, when only the shifts between the different components and
their widths of are fixed during the modelling of the \ha+[NII] complex, the
results obtained lead to unphysical \ha/\hb, \nitrogen$\lambda$6583/\ha~line
ratio values for I2, and \sulphurtwo6716/6731~line ratios for I2 and the broad
component. Therefore, during the modelling, the \hb~flux of the  I2 component was
set such the the \ha/\hb~line ratio associated with this
component is equal or higher than that of I1 (\ha/\hb~= 3.74 and 4.01 for I1 and
I2 for the best fitting model). In addition, the \nitrogen$\lambda$6583/\ha~was
set to 3 and the \sulphurtwo6716/6731 line ratio was set to 1.42, 0.44 and 0.44
for I1, I2 and the broad component respectively. The values of each of the
\sulphurtwo6716/6731~line ratios was then changed in steps on 0.02, while the
values of the other two remained unchanged, until the fit was deemed 
poor. Hence, we estimate a lower limit for the intensity ratio
\sulphurtwo6716/6731~of 1.2 for I1, while acceptable fits are obtained for any
line ratio value within the allowed range (0.44 $\leq$ \sulphurtwo6716/6731
$\leq$ 1.42) in the cases of I2 and the broad component.

Finally, we note that the line ratios of I1 are consistent with NLAGN ionization
or fall close to the LINER/Sy2 limiting region in the optical diagrams, whereas
the values measured for the I2 and broad components are consistent with Sy2
ionization.

\vspace*{0.2cm}\noindent
{\bf F15462-0450.} This is the only Seyfert 1 galaxy included in the sample discussed in this
paper. Figure 1 shows the best fitting model obtain from fitting simultaneously
the H$\beta$, [OIII]$\lambda$$\lambda$4959,5007 and the broad
FeII$\lambda$$\lambda$4924,5018\footnote{Note that, in order to model the broad
  FeII lines, we assumed that they have the same profile as the broad H$\beta$
  line.} emission lines, typical of Sy1 galaxies. Due to the large number of
emission lines close together in wavelength and bearing in mind that multiple
(and broad) kinematic components contribute to each individual emission line,
clearly overcoming the degeneracy issue is not easy in the case of F15462--0450.

Overall, we find that the Balmer lines (H$\delta$, H$\gamma$,H$\beta$ and
H$\alpha$) are adequately fitted using three kinematic components: a narrow
component of width 147 $\pm$ 47 \kms~at rest frame, plus a first broad component
of width 1509 $\pm$ 47 \kms~redshifted 169 $\pm$ 8 \kms~with respect to the
narrow component and a second broader component of width 3673 $\pm$ 137
\kms~redshifted 292 $\pm$ 21 with respect to the narrow component.

On the other hand, the forbidden lines ([NeV]\lala3346,3425,
[OII]\lala3726,3728, [NeIII]\lala3868,3968, [OIII]\lala4959,5007,
[NII]\lala6548,6583~and [SII]\lala6716,6731) are adequately fitted using two
kinematic components that correspond to those presented in Table 2, and labelled
as the [OIII] model for this galaxy: a narrow component at rest frame,
consistent with that found for the Balmer lines (i.e. FWHM = 147 $\pm$ 47 \kms and
referred as `N' in Table 2) plus a broad component (FWHM = 1426 $\pm$ 47 \kms)
blueshifte by -822 $\pm$ 25 \kms with respect to the narrow component (referred to as `B'
in Table 2). Note that the broad component dominates the [OIII] emission line flux
in this object.

Note that the broad component associated to the forbidden lines (B) it is still
required to adequately model the profile of the H$\beta$ emission line. Such
component is represented by the most redshifted, dashed-line Gaussian in the
H$\beta$ fit in Figure 1. Most likely, this component is also present in the
other Balmer lines. However, the dominant flux contribution associated to the
emission from the Broad Line Region (BLR) as well as the relatively low S/N in
the cases of the H$\delta$, H$\gamma$ emission lines prevents a clear detection
of this kinematic component for the other Balmer lines. In addition, due to the
dominant BLR emission, it is not possible to constrain the flux contribution of
the broad component to the [NII]\lala6548,6583~emission lines with any accuracy.

Given the fact that the broad component present in the forbidden lines is not
clearly detected in the H$\alpha$ emission line, no attempt to study the
ionization mechanisms was made for that component. Interestingly, the line
ratios measured for the narrow component detected in {\it all} lines are
consistent with HII ionization.

\vspace*{0.2cm}\noindent
{\bf F16156+0146.} For the study presented here we used the 5kpc-I aperture described in the work
of \cite{Rodriguez-Zaurin09}. This aperture samples the north-western nucleus of
this double nucleus system, which is associated with the AGN activity. This is
one of the four objects in our sample (along with F13451+1232, F14394+5332 and
F23389+0300) for which is not possible to find one model that fits all the
emission lines observed in the optical spectrum. However, it is possible to
model all the emission lines using two different approaches. A first approach in
which a different model is used for each of the observed optical spectrum, and a
second one in which we use two models: one that adequately fits the
\oxythrw~emission lines, and a second model that fits all the other optical
emission lines well.

\begin{figure}
\psfig{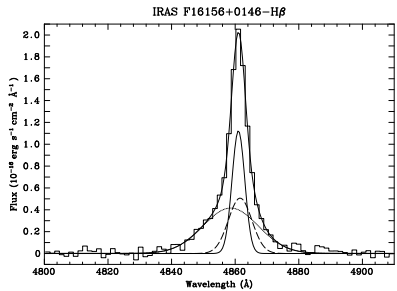}
\caption{In the case of F16156+1046 it was not possible to model {\it all} the
  emission lines using a single model. However, with the exception of the
  [OIII]\lala4959,5007, it was possible to model the rest of the observed
  emission lines with the model shown in the figure. This model comprises an
  unresolved component at rest frame (solid line), a narrow component (FWHM =
  446 $\pm$ 86 \kms) redshifted 33 $\pm$ 29 \kms with respect to the unresolved
  component (dashed line) and a third, broad component (FWHM = 1225 $\pm$ 108
  \kms) that is blueshifted by -150 $\pm$ 48 \kms relative to the narrow,
  unresolved component (dotted line). The resulting model (bold line) is
  overplotted on the extracted spectrum.}
\label{16156_Hb}
\end{figure}

The best fitting model for \oxythrw, which is the same for the two approaches,
is shown in Figure \ref{OIII-profiles} and comprises three components: an
unresolved narrow component, an intermediate component (FWHM = 804 $\pm$ 20
\kms) blueshifted -186 $\pm$ 10 \kms with respect to the narrow component and a
broad (FWHM = 1535 $\pm$ 59 \kms) blueshifted by -374 $\pm$ 26 \kms with respect
to the narrow component. The second model, shown in Figure \ref{16156_Hb} for
the H$\beta$ line comprises another 3 components: an unresolved component, a
narrow component (FWHM = 446 $\pm$ 86 \kms) redshifted by 33 $\pm$ 29 \kms with
respect to the unresolved component, and a third, broad component (FWHM = 1225
$\pm$ 108 \kms) that is blueshifted by -150 $\pm$ 48 \kms relative to the narrow,
unresolved component.

Overall, the fact that we do not find one adequate model for all emission lines
suggest that, instead of a discrete number of kinematic components, there may be
a continuous range of density, ionization and kinematics through the Narrow Line
Region (NLR), leading to each line having a different profile. We did not
attempt to estimate line ratios associated with the different components in the
case of F16156+0146.

\vspace*{0.2cm}\noindent
{\bf F17044+6720.} This object was originally classified as a LINER by \cite{Veilleux99}. However,
optical spectroscopic classification of this source was recently reviewed by
Yuan et al. (2010), who found that the line ratios of F17044+6720 are typical of
a Sy2 galaxy.

Figure \ref{OIII-profiles} shows the \oxythrw~model for this galaxy comprising
two components: a narrow component of width FWHM = 290 $\pm$ 5 \kms plus a broad
component of width FWHM = 1765 $\pm$ 100 \kms that is blueshifted by -553 $\pm$ 65
\kms with respect to the narrow component. Due to the presence of a telluric
absorption feature it was not possible to model the \sulphurtwow~emission lines
for this galaxy. The line ratio values measured for the narrow component are
consistent with CP ionization, or lie close to the limiting region between LINER
and Sy2, while the line ratios associated with the broad component are
consistent with NLAGN, or fall close to the LINER/Sy2 ionization.

\vspace*{0.2cm}\noindent
{\bf F17179+5444.} Three components are required to model the optical spectrum of this single
nucleus Sy2 galaxy. Figure \ref{OIII-profiles} shows an \oxythr~model including
a narrow component with a FWHM of 356 $\pm$ 22 \kms, an intermediate component
with a FWHM of 515 $\pm$ 51 \kms~that is redshifted by 246 $\pm$ 67 \kms with
respect to the narrow component, and a third, broad component with a FWHM of
1562 $\pm$ 41 \kms that is blueshifted with respect to the narrow component by
-119 $\pm$ 61 \kms. The line ratios measured for the narrow component are
consistent with CP or LINER ionization, while the line ratios of the shifted
components are, in general, consistent with Sy2 ionization.

\begin{figure}
\psfig{figure=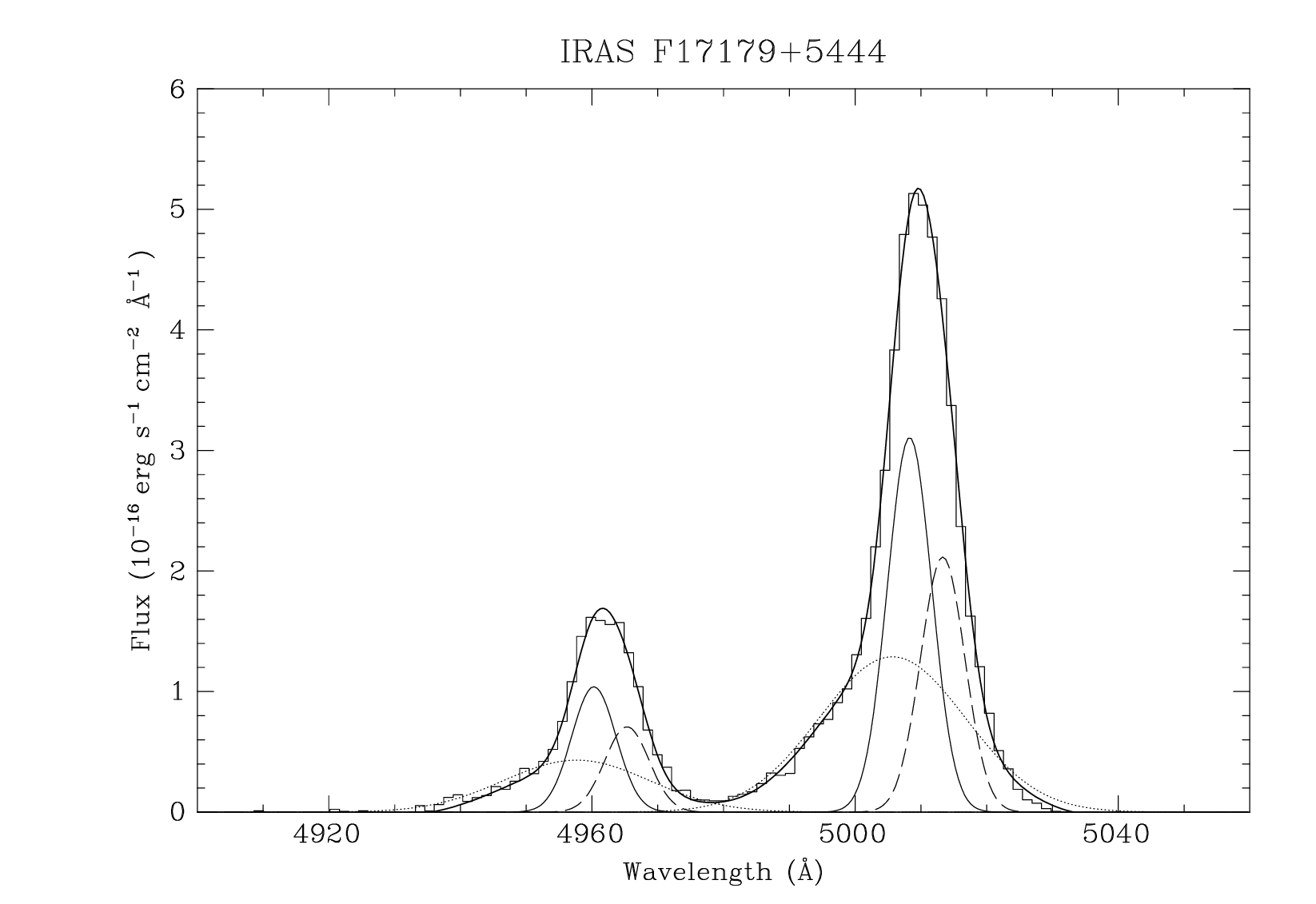,width=5.5cm,angle=-90.}
\caption{A second, possible [OIII] model that adequately fits all the observed
  emission lines in the case of F17179+5444. The different components shown in
  figure are: a narrow component (N1, FWHM = 386 $\pm$ 14 \kms), a second narrow
  component (N2) with the same width redshifted 301 $\pm$ 21 \kms with respect
  to N1 and a broad component (FWHM = 1543 $\pm$ 34 \kms ) blueshifted by -157
  $\pm$ 33 \kms with respect to N1. This model is interpreted as unresolved
  rotation}
\label{17179_rotation}
\end{figure}

\begin{figure*}
\begin{tabular}{cc}
\hspace{-0.5 cm}\psfig{figure=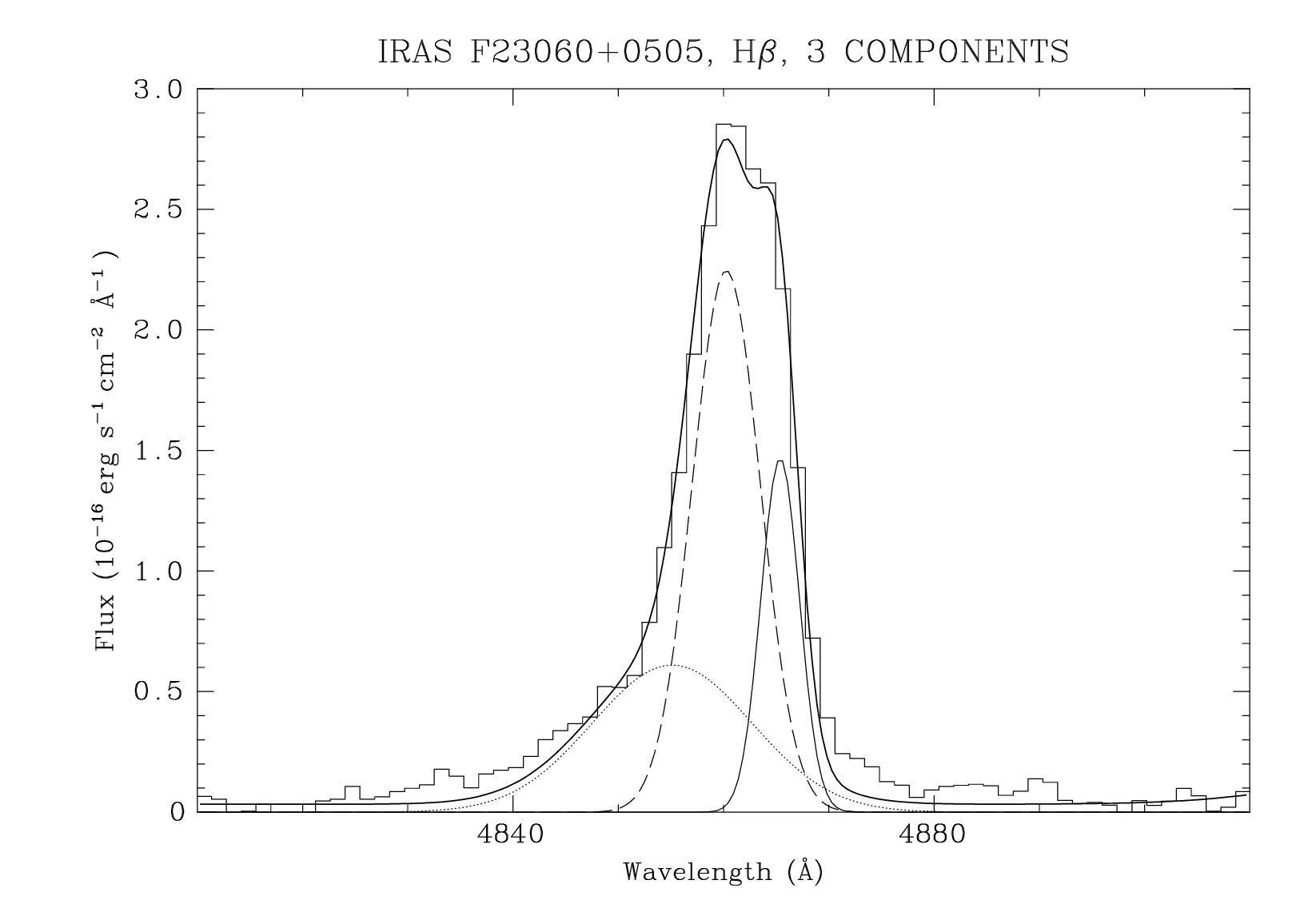,width=5.5cm,angle=-90.}&
\psfig{figure=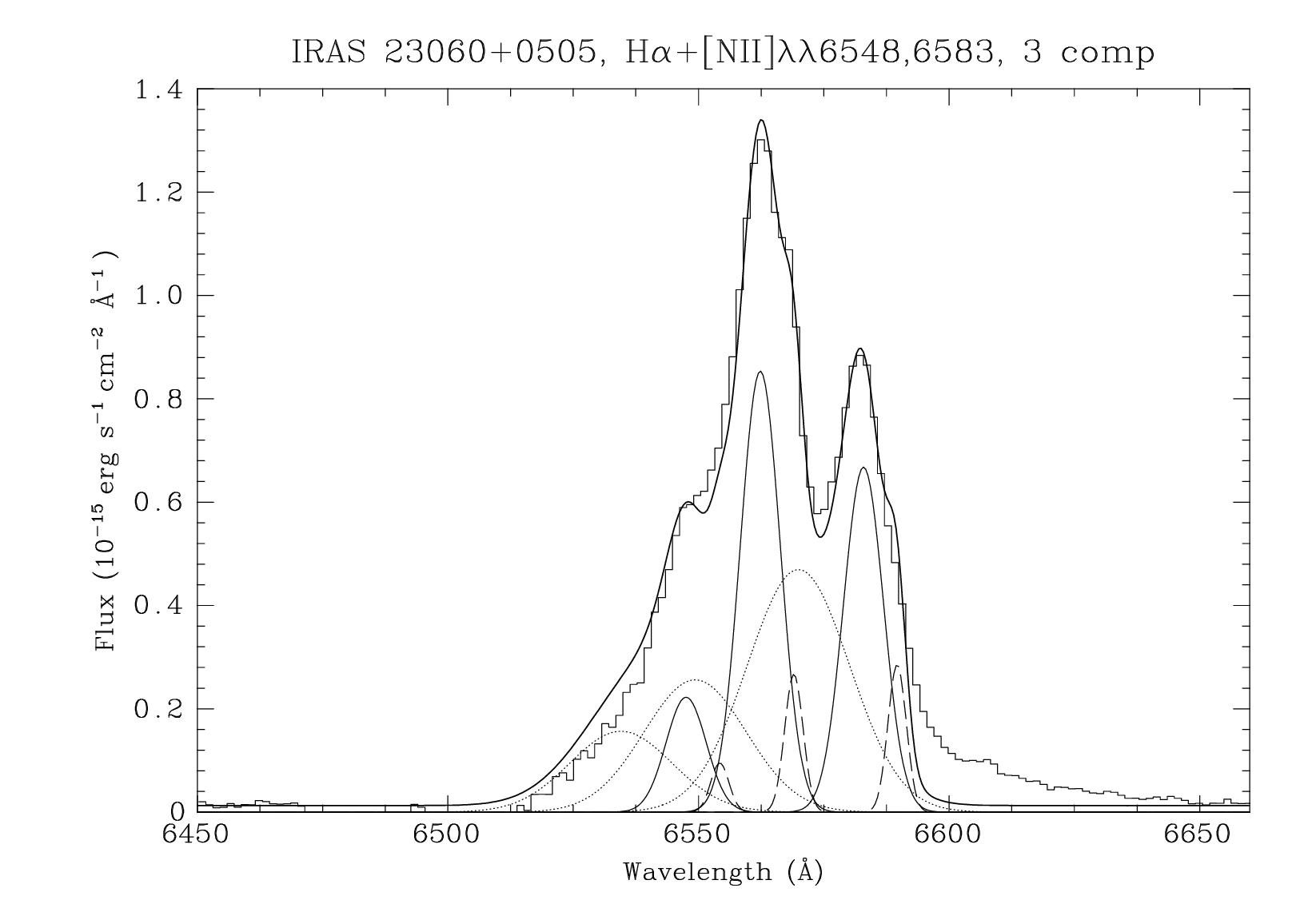,width=5.5cm,angle=-90.}\\
\hspace{-0.5 cm}\psfig{figure=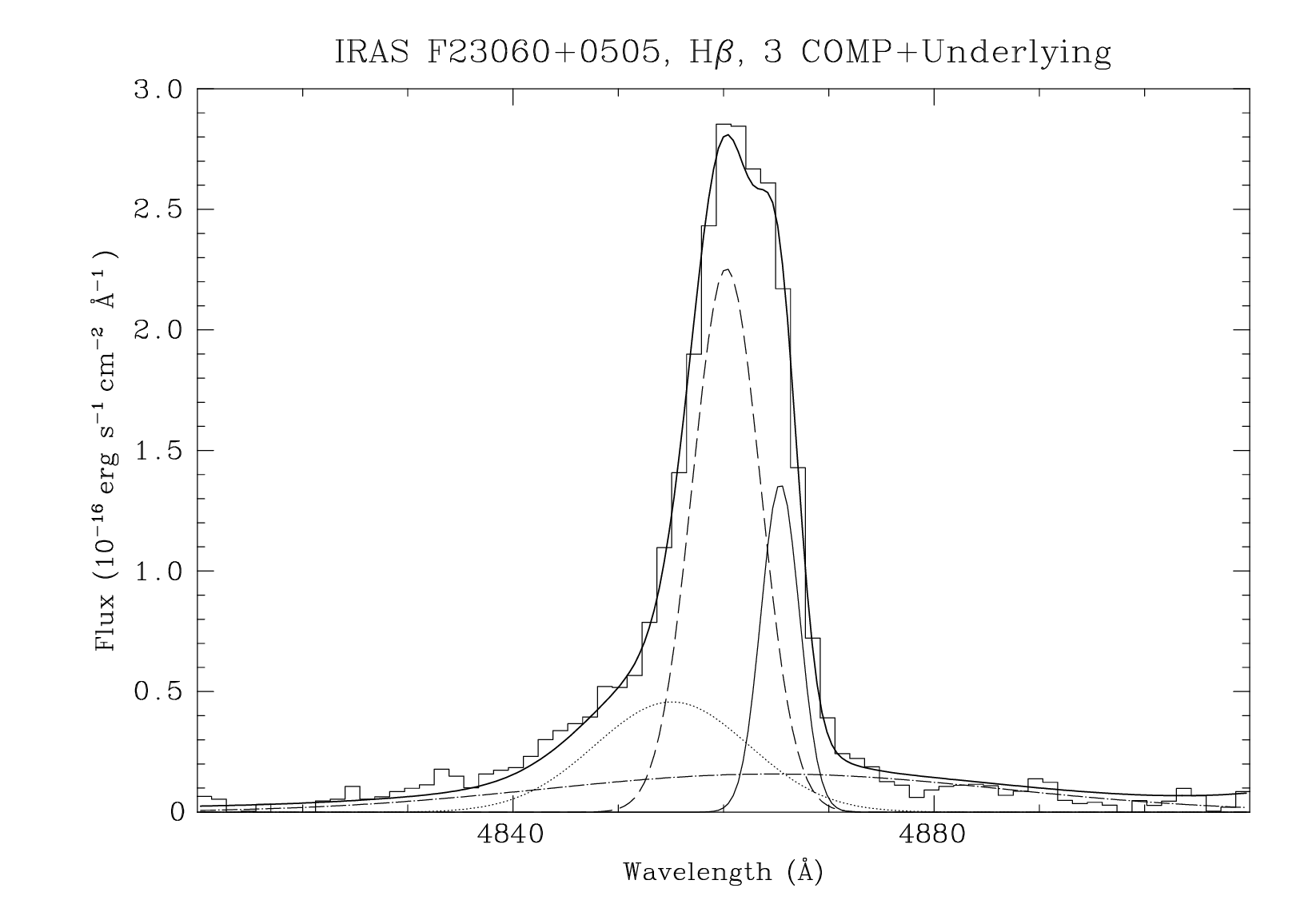,width=5.5cm,angle=-90.}&
\psfig{figure=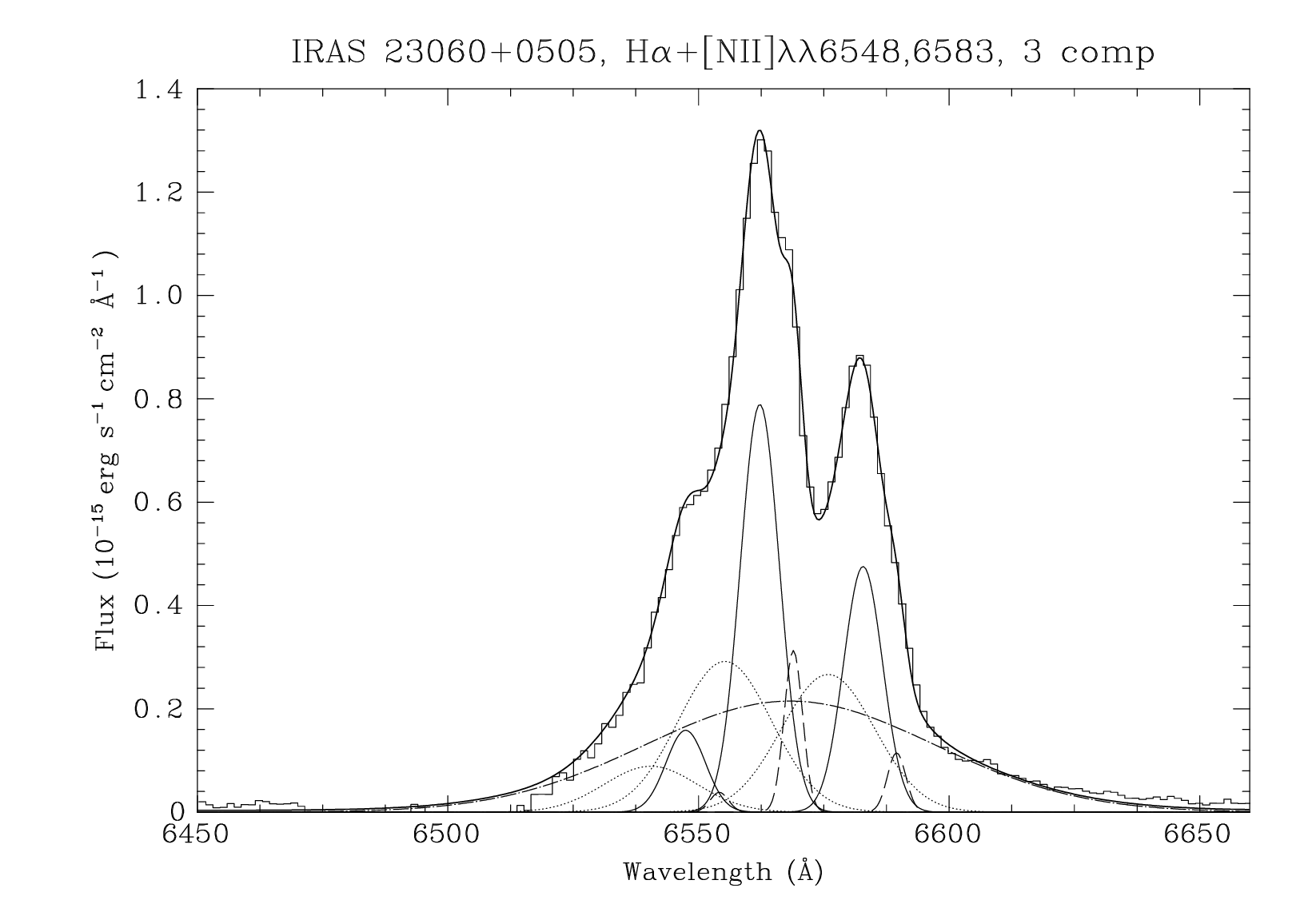,width=5.5cm,angle=-90.}\\
\end{tabular}
\caption{Fits to the optical emission lines in F23060+0505. Top panel: \hb and \ha+[NII]\lala6549,6583~fit including the two narrow
  components (N1 and N2) plus the broad components corresponding to the [OIII]
  model for this galaxy. Lower panel: \hb~fit including the already mentioned
  three components plus an underlying broad component (representing the
  BLR). When only 3 components are included in the modelling, the blue, and
  paricularly the red wings of the line profiles are not adequately fitted. The
  problem disappears when including the underlying broad component.}
\label{23060-Hb}
\end{figure*}

However, for this galaxy, it is important to add a caveat about the uniqueness of
the best fitting model. As described above, there are two close narrow components that
have relatively similar widths (356 $\pm$ 22 \kms and 515 $\pm$ 51
\kms). Therefore, it is possible that this narrow line splitting represents
unresolved rotation, or traces a large-scale bipolar outflow in the gas
\citep[see, for example][]{Holt08}. To test this scenario we set the widths of
the two narrow components to have the same value during the modelling of the
\oxythrw~emission lines; the result is presented in Figure
\ref{17179_rotation}. The model, that also successfully fits all the observed
emission lines, comprises three components: a narrow component (N1, FWHM = 386
$\pm$ 14 \kms), a second narrow component (N2) with the same width redshifted
301 $\pm$ 21 \kms with respect to N1 and a broad component (FWHM = 1543 $\pm$ 34
\kms ) blueshifted by -157 $\pm$ 33 \kms with respect to N1.

Note that when only the widths and shifts of the different components
were constrained for the two best fitting models described before, the
\ha/\hb~line ratios associated with the intermediate and the N2 component were
lower than 2.88, the theoretical lower limit assuming case B recombination
theory in absence of reddening. A similar issue is found for the broad component
in the case of the \sulphurtwow~doublet, where the \sulphurtwo6716/6731~line
ratio is found to be higher than 1.42, the low density limit. This is due to
the fact that the flux contribution in the case of the \sulphurtwow~emission
lines is dominated by the narrow components at rest frame (for the two best
fitting models considered). Therefore, the intensities of the broad components
are relatively unconstrained.

To overcome these problems the \hb~flux contribution of the intermediate and the
N2 was forced to have an intensity such as the \ha/\hb~line ratio associated
with this component is higher or equal to that of the narrow component at rest
frame. In addition, following the approach used for other objects, the
\sulphurtwo6716/6731~line ratio of the shifted components were set to 0.44
during the modelling and then slightly increased in steps of 0.02 until the fit
became poor. As we mentioned before, the flux contribution of these
components is relatively small, and adequate fits are obtained for any value of
this ratio within the allowed range (0.44 $\geq$\sulphurtwo6716/6731$\geq$1.42).

In terms of line ratios we find similar results for the two models described
above. The values measured for the narrow component is consistent
consistent with  CP and LINER ionization. On the other hand, the line ratios of
the I and B components are consistent with Sy2 ionization. We note
that telluric atmospheric absorption features
affect, to some extent, the profile of the [OI]$\lambda$6300 emission line and the
blue wing of the \sulphurtwow~doublet. Therefore, the line ratios for the broad components
of the latter lines are relatively
unconstrained for either of the models considered here.


\vspace*{0.2cm}\noindent
{\bf F23060+0505.} Strong emission lines are observed from the UV end (MgII$\lambda$2798) to the
red end (\sulphurtwow) of the spectrum of this compact galaxy. The best fitting
model presented in Figure \ref{OIII-profiles} comprises 4 components: a narrow
(N1, FWHM = 376 $\pm$ 6 \kms), a second, unresolved, narrow component (N2)
redshifted by 306 $\pm$ 4 \kms with respect to the narrow component at rest
frame, a broad component (FWHM = 1001 $\pm$ 23 \kms) blueshifted by -310 $\pm$
14 \kms with respect to the narrow component at rest frame and a very broad
component (FWHM = 2148 $\pm$ 125 \kms) that is blueshifted -1073 $\pm$ 122 \kms
with respect to the narrow component in the rest frame.  As in the previous
case, it is possible that the observed narrow line splitting represents
unresolved rotation, or traces a large-scale bipolar outflow in the gas.

The very broad component is only detected in high ionization emission lines
(\oxythrw, HeII$\lambda$4686 and NeV$\lambda$3426). The other observed emission
lines, with the exception of \hb~and the~\ha+[NII]\lala6549,6583 complex, are
adequately modeled using a model including the two narrow components plus the
broad component. In the case of the \hb~line and the \ha+[NII]~complex the
profiles the blue and red wings are not adequately fitted with a model
comprising the two narrow plus a broad component (see Figure
\ref{23060-Hb}). Accounting for a fourth, blueshifted component does not improve
the quality of the fit. In fact, when a forth, broad component consistent with
that detected in \oxythr~is included during the modelling, DIPSO returns
negative flux values for this component. Therefore, we used the approach of
including a fourth, unconstrained component (i.e. shift, width and flux of this
component are free parameters) during the modelling of the \hb~and~\ha+[NII]
emission lines, and the results are shown in Figure \ref{23060-Hb}. This fourth
component has a FWHM of 2345$\pm$ 368 \kms~and is redshifted by 188 $\pm$ 100
\kms~with respect to the narrow component at rest frame; its H$\alpha$/H$\beta$
ratio of 18$\pm$2.7 implies a reddening of E(B-V)=1.68$\pm$0.14 assuming a
\cite{Calzetti00} extinction law. At the same time, the nuclear spectrum of the
object shows an unusually red continuum compared with other local ULIRGs
(Rodriguez Zaurin et al.  2009, see also Hill, G. et al. 1987). Overall, the
properties of the nuclear spectrum of this source are consistent with it
containing a moderately reddened Seyfert 1 nucleus. This is consistent with the
clear detection at near-IR wavelengths by \cite{veilleux97} of a broad
Pa$\alpha$ line of similar width (FWHM$\sim$1954 km s$^{-1}$) to the broad
Balmer lines detected in our optical spectrum.

Note that a telluric absorption feature might affect the blue wing of the
\ha+[NII] complex, as shown in Figure \ref{23060-telluric}. Nonetheless, we have
successfully fitted the \ha+[NII] blend using a BLR component that has the same
width and shift as that fitted to the H$\beta$ line.

\begin{figure}
\psfig{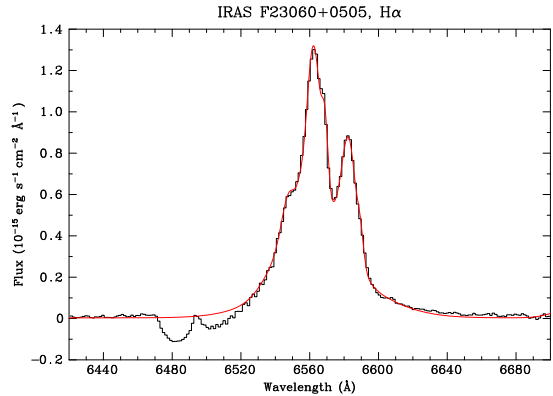}
\caption{Figure using the \ha+[NII] complex to illustrate the effects of the
  telluric absorption feature in the case of F23060+0505. Although the telluric
  feature affects the shape of the profile for the complex, it is still possible
  to model the \ha~and the \nitrogen$\lambda\lambda$6549,6583 emission lines
  using the [OIII] model. The red, solid line is our best fitting model.}
\label{23060-telluric}
\end{figure}

Regarding the ionization mechanisms, we find that the line ratios measured for
the three components observed in all the emission lines are consistent with Sy2
ionization. As in some previous cases, when only the shifts between the
different components and their widths are fixed, the \sulphurtwo6716/6731 line
ratio associated with N2 was higher than 1.42 (the low density limit), while
this value was lower than 0.44 (the high density limit) for the broad
component. Following the approach described before, we set the values of this
line ratio to be 0.44 and then change the value for each component individually
in steps on 0.02 until the fit was deemed not adequate. Acceptable fits are
obtained for any line ratio value within the allowed range (0.44 $\leq$
\sulphurtwo6716/6731 $\leq$ 1.42). Finally, we note that the degeneracy issue
associated with the \sulphurtwow~doublet does not change the results related to
the ionization mechanism, i.e. Sy2-like line ratios are always found for all the
different components.

\vspace*{0.2cm}\noindent
{\bf F23233+2817.} The \oxythr~model for this very compact object comprises 4 components: a narrow
component (FWHM = 179 $\pm$ 4 \kms), an intermediate component (FWHM = 640 $\pm$
24 \kms) blueshifted by -161 $\pm$ 15 \kms with respect to the narrow component,
a broad component (FWHM = 1511 $\pm$ 118) blueshifted -447 $\pm$ 41 \kms with
respect to the narrow component, and a very broad component (FWHM = 2706 $\pm$
422 \kms) that is blueshifted by -3433 $\pm$ 904 \kms with respect to the narrow
component. As shown in Figure \ref{OIII-profiles}, the flux contribution of the
very broad component is relatively small. Therefore, the uncertainties
associated with the corresponding shift and width are large. As in the previous
object, this component is only required to obtain an adequate fit in the case of
the \oxythrw. All the other observed lines can be modelled with the first three
components listed before.

As in most of the cases, when only the shifts between the different components
and their widths of are fixed, the \sulphurtwo6716/6731 line ratio associated
with the intermediate and broad components are outside the allowed
range. Therefore, following the approach described above we set the ratio to
0.44 for the intermediate and broad components and then increase the value of
each of these two components in steps of 0.02 until the fit was considered poor. 
In this way, we estimate an upper limit for the 
\sulphurtwo6716/6731 ratio of 1.0 for these two components.

Regarding the ionizing mechanisms, we find that the line ratios measured for the
narrow component are consistent with NLAGN ionization, or fall close to the
HII/Sy2 limiting region in the diagrams. On the other hand, the line ratios
measured for both the intermediate and the broad components are consistent with
Sy2 ionization.  Note that that a Sy2 spectral type is found for these two
components independently of the \sulphurtwo6716/6731 line ratio used during the
modelling.

\vspace*{0.2cm}\noindent
{\bf F23327+2913.} Yuan et al. (2010) recently classified this double nucleus system as Sy2:,
although it was originally classified as LINER by \cite{Veilleux99}. For the
work presented here we concentrate on the 5kpc-I aperture described in
\cite{Rodriguez-Zaurin09}. This aperture covers the emission from the southern
source, and is the only aperture for which the corresponding extracted spectrum
shows evidence for outflows.

The main problem when trying to model the emission lines in this object is that
the \oxythrw and H$\beta$ emission lines have low equivalent widths, and are
potentially affected by noise and residual structure in the underlying
continuum. Therefore these lines are not suitable for deriving a model to fit to
the other lines in the spectrum. As in the case of F00188--0856, we use the
approach of first fitting the H$\alpha$+[NII] complex and then model the other
emission lines observed in the spectrum with the same components and the same
velocity widths and shifts as H$\alpha$. The results of the H$\alpha$+[NII]
modelling are shown in Figure \ref{OIII-profiles}.

The best fitting model comprises 2 components: a narrow component (FWHM = 109
$\pm$ 2 \kms) and an intermediate component (FWHM = 805 $\pm$ 13 \kms)
blueshifted by -92 $\pm$ 4 \kms with respect to the narrow component. The narrow
component makes a negligible contribution to the flux of the \hb~emission
line. In fact, the \hb~emission line can be modeled using only the intermediate
component. 
Therefore, it
is not possible to study the ionization mechanism associated with
that component. Regarding to the intermediate component, we find that the line
ratios measured for that component are consistent with NLAGN or HII
ionization.

\vspace*{0.2cm}\noindent
{\bf F23389+0300.} Two 5 kpc apertures (5kpc-I and 5kpc-II) were used during the investigation of
the stellar populations of this double nucleus system presented in
\cite{Rodriguez-Zaurin09,Rodriguez-Zaurin10} However, aperture used for this
study is the 5kpc-I aperture, centered on the northern source, which includes
the AGN nucleus \citep[e.g.][]{Nagar03}

The [OI]\lala6300,6393 emission lines, which are stronger powerful than the
observed \oxythrw~emission lines, are particularly impressive in this source. 
In terms of the modelling, it was not possible to
find one model that adequately fits all the detected emission lines. In this
particular case, each emission line was fitted using an individual model. For
completeness, Figure \ref{OIII-profiles} shows the modelling result for the
\oxythr~emission lines. The model shown in the figure comprises a narrow
component (FWHM = 289 $\pm$ 7 \kms) plus a very broad component (FWHM = 2223
$\pm$ 30 \kms) that is redshifted by 47 $\pm$ 13 \kms with respect to the narrow
component.

As in the cases of F13451+1232, F14394+5332 and F16156+0146, the fact that we do
not find one adequate model for all emission lines suggest that, instead of a
discrete number of kinematic components, there may be a continuous range of
density, ionization and kinematics through the Narrow Line Region (NLR), leading
to each line having a different profile. No attempt has been made to estimate
line ratios of the different kinematic components for this galaxy.

\end{document}